\documentclass[11pt]{article}

\usepackage{amsmath, bm, amssymb, amsthm}
\usepackage{array}
\usepackage{comment}
\usepackage{dsfont}
\usepackage{enumerate}
\usepackage[margin=1in]{geometry}
\usepackage{graphicx}
\usepackage{hyperref}
\usepackage{latexsym}
\usepackage{lineno}
\usepackage{longtable, float}
\usepackage{lscape}
\usepackage{mathtools}   
\usepackage{multirow}
\usepackage{rotating}
\usepackage[doublespacing]{setspace}
\usepackage{caption}
\usepackage{subcaption}
\usepackage{tabularx}
\usepackage{ragged2e} 
\usepackage{thmtools}
\usepackage{threeparttablex}
\usepackage{threeparttable}
\usepackage{xcolor}
\let\origappendix\appendix 
\renewcommand\appendix{\clearpage\pagenumbering{roman}\origappendix}
\usepackage{booktabs}
\usepackage{colortbl}
\usepackage{makecell}
\usepackage{pdflscape}
\usepackage{tabu}
\usepackage[normalem]{ulem}
\usepackage{wrapfig}
\usepackage{mwe}

\hypersetup{
   pdffitwindow=true,pdfstartview={FitH},colorlinks=true,linkcolor=red,citecolor=blue,filecolor=magenta,urlcolor=cyan,breaklinks=true}

\theoremstyle{definition}

\theoremstyle{definition}

\theoremstyle{definition}

\theoremstyle{remark}

\theoremstyle{remark}

\DeclareMathOperator{\EX}{\mathbb{E}}

\usepackage{pgfplots}
\pgfplotsset{width=10cm,compat=1.9}
\usepgfplotslibrary{external}
\usetikzlibrary{through,calc,arrows,intersections,patterns}
\usetikzlibrary{decorations.pathreplacing}
\usetikzlibrary{decorations.markings}
	\tikzset{->-/.style={decoration={markings,mark=at position #1 with {\arrow{>}}},postaction={decorate}}}

\usepackage[round,authoryear]{natbib}
\title{The Effects of the Pandemic on Market Power and Profitability\footnote{We thank Dick Startz, Ted Frech, Javier Birchenall, Gonzalo Vazquez-Bare, Francis X.\ Diebold, Pablo Kurlat, Hashem Pesaran and  the applied economics reading group and applied lunch seminar at UC Santa Barbara, and the LACEA-LAMES 2022 attendants for valuable comments in an earlier version of this manuscript.}}
\author{Juan Andres Espinosa-Torres\footnote{Department of Economics, University of Southern California. Address: 3620 S. Vermont Ave. Kaprielian Hall (KAP) 363 Los Angeles, CA 90089-0253. Email: \href{mailto:juanaesp@usc.edu}{juanaesp@usc.edu} }\\
Jaime Ramirez-Cuellar\footnote{Office of the Chief Economist, Microsoft, Redmond, WA, USA. Email: \href{mailto:jaimeramirez@microsoft.com}{jaimeramirez@microsoft.com}.}
}
\date{
\today
\\
First version: May 27, 2022
}

\begin{document}

\maketitle
\vspace{-1cm}
\begin{abstract}
    We explore firm-level markup and profit rates during the COVID-19 pandemic for a panel of 3,611 publicly traded firms in Compustat and find increases for the average firm. We offer conditions to give markups and profit rate forecasts a causal interpretation of what would have happened had the pandemic not happened. Our estimations suggest that had the pandemic not happened, markups would have been 4\% and 7\% higher than observed in 2020 and 2021, respectively, and profit rates would have been 2.1 and 6.4 percentage points lower. We perform a battery of tests to assess the robustness of our approach. We further show significant heterogeneity in the impact of the pandemic on firms by key firm characteristics and industry. We find that firms with lower than forecasted markups tend to have lower stock-exchange tenure and fewer employees.
    \\
    \noindent {\it Keywords: COVID-19, markups, profitability, firm dynamics, market power.} 
    \\
    \noindent \textit{JEL classification: D22, D43, E3, L1.}
\end{abstract}

\section{Introduction}

The COVID-19 pandemic deeply disrupted both supply and demand in a myriad of markets. In order to understand these disruptions, we analyze markup and profit rates from 2020 to 2021. This analysis is relevant for two reasons: First, markup rates summarize the interactions between supply and demand in a market and can therefore provide insight into the effects of the pandemic on markets. Second, it is uncertain whether the pandemic exacerbated previous trends in markup or profit rates, or whether it created the conditions for an increase in market power. The goal of this paper is not only to document recent movements in markups and profitability during the pandemic but also to determine whether pre-pandemic trends can predict these movements or whether the pandemic itself triggered the conditions for an increase in market power.

This paper makes three contributions to the recent literature on the market power and profitability of publicly-traded firms in the US during the COVID-19 pandemic. First, we document both the increasing markup and profit rates, and more volatile profit rates. Specifically, after average markup rates decreased from 1.61 to 1.49 between 2016 and 2019, they rebounded to 1.54 in 2021. Additionally, the average profit rate reached its all-time high at 21.6\% in 2021, at least since the start of historical records in 1955. We present our findings using both yearly and quarterly firm-level data.

Second, to further understand the rising trends in market power and profitability during the pandemic, we evaluate whether previous trends in firm-level markup and profit rates predict these increases. To do so, we use separate firm-level Bayesian structural time series models to compute firm-level forecasts of markups and profit rates, using information up to 2019. We then calculate the difference between the observed value and its forecast. Our results indicate that, for the average firm, the observed markup rate was 4.3\% and 6.6\% lower than its forecast in 2020 and 2021, respectively. In contrast, the average firm had a profit rate of 21.6\% in 2021, which is 6.4 percentage points higher than its forecast value. These findings suggest that while markup rates were lower than expected, profit rates were higher than expected.

Third, we analyze the heterogeneous effects of rising market power and profitability during the pandemic. Firms that performed poorly in terms of markups during the pandemic tend to have become recently publicly traded and have fewer employees. As for profitability, there seems to be low effect heterogeneity across firms with different levels of employment, tenure in the stock exchange, or market share. 

Alternatively, when analyzing effect heterogeneity in markups by industry, information, real estate, and chemical manufacturing experienced lower markups than expected, whereas warehousing and entertainment had higher markups than expected. When analyzing heterogeneity in profit rates by industry, transportation, entertainment, and hospitality tended to perform poorly, whereas warehousing and real estate outperform. These findings suggest that the impact of the pandemic on market power and profitability was uneven across firms.  

To support these contributions, we perform a battery of statistical tests to assess the efficacy of the Bayesian structural time series model in producing forecasts of firm-level markup and profit rates. Specifically, we perform three exercises to evaluate the quality of our models. First, we statistically test whether a critical assumption of our Bayesian structural time series (BSTS) models holds for most firms. Second, we evaluate whether these firm-specific models accurately reflect the observed firm’s markup and profit rates before the pandemic. Third, we confirm that our forecasts have adequate forecast quality for a sample before the pandemic.

In addition to evaluating the quality of the BSTS models, we offer robustness checks for our results by considering two alternative modeling choices and frameworks. First, we vary the modeling choices of the BSTS model by choosing alternative hyperparameters of the priors in the Bayesian model. These hyperparameters attribute either more or less variation to the latent components of the structural model. Second, we alternatively implement forecasts using the flexible approach of local linear projections \citep{Jorda2005}. Particularly, when comparing the forecast quality of BSTS and local linear projections models, the later tend to have wide confidence intervals which suggests that local linear projections do not have as much power for this data. Despite these changes, our main results remain unchanged. These robustness checks help to strengthen the reliability and robustness of our findings.

Our results inform the recent literature on the rising markups of publicly-traded firms in the US \citep{Deloecker2012,Hall2018,DeLoecker2020,Autor2020,berry2019increasing}. In particular, \cite{DeLoecker2020} documents the rising nature of markups from 1955 to 2016. We extend their analysis by providing evidence suggesting that previous trends in markup and profit rates do not predict average market power behavior during the pandemic. 
 
Additionally, we uncover the average behavior in markups and profitability and analyze its heterogeneity by key firm characteristics such as size, market share, and tenure on the stock exchange. We further argue that our methods may provide a causal accounting of the effects of the pandemic on markups and profit rates. We support our identification strategy on a similar fact noted in \citet{jorda2022longer}: pandemics are exogenous and unpredictable to the economy and, even more so, to granular firm dynamics. A similar approach in the context of country-level impact of the pandemic uses shocks to economic growth expectations as identifying variation \citep{chudik2021counterfactual}. Our methodology is related to the event study method commonly used in finance and economics \citep{mackinlay1997event}. However, we do not limit our analysis to estimating the impact on the stock price of a given sample of companies. 

Furthermore, a growing literature documents the effects of the COVID-19 pandemic on macroeconomic aggregates, such as inflation \citep{Cavallo2020,Ball2022,binder2022expected,dietrich2022news} and economic output \citep{ludvigson2020covid,mckibbin2021global,baqaee2020nonlinear}, and the consequential public policy response \citep{Guerrieri2022, chudik2021counterfactual, woodford2022effective,bigio2020transfers,auerbach2021inequality}.  \cite{Brodeur2021} provides a comprehensive survey of the rapidly growing body of literature on the economics of COVID-19. None of these papers offers evidence of the disruption of the pandemic on competition dynamics.

This paper has six sections including this introduction. The second section describes the data and outcomes of interest. The third section describes our methodology including the Bayesian structural time series model. The fourth section presents our main results. The fifth section introduces our robustness checks. The sixth section concludes.

\section{Data}

We use yearly (and quarterly) information for 3,611 (3,192) publicly-traded companies from Compustat, which provides a panel of financial statements since 1955 (1980) representing roughly 29\% of US jobs \citep{davis2006volatility}. To make our sample comparable to that in previous literature, we follow the same sample selection steps as in \citet{DeLoecker2020}.\footnote{For a more detailed description of the sample, please refer to Appendix \ref{app:data}.} Furthermore, to evaluate whether our sample is comparable to the sample in previous literature, we compute summary statistics for sales, COGS, capital stock, SG\&A, wage bill, and employment in both 1955--2021 and 1955--2016. Our numbers are indeed quite similar to those in \citet{DeLoecker2020}. Differences arise as Compustat updates and companies revise the corresponding financial statements.\footnote{Table \ref{tab:summary_stats} in the Appendix reports summary statistics for these variables for three samples: 1955--2016, 1955--2021, and the sample in \citet{DeLoecker2020}. We use the same coding routine in Stata as those authors, which is available in their published supplemental material. Although the mean and median of the cost of goods sold differs between our sample and the sample in \citet{DeLoecker2020}, we find a similar series of average markup rates as those authors, and both series have a high positive correlation (0.92).}

\subsection{Outcomes of interest}

\subsubsection*{Markup rates}
We compute markup rates using the production function approach \citep{Hall1988,Deloecker2012,DeLoecker2020}.\footnote{Alternatively, there are two other competing methods to compute markup rates. First, the demand approach by \cite{Berry1995}, which requires information on the demand side and, consequently, is more data-intensive and hard to carry out for an extensive set of industries simultaneously. Second, the approach assumes constant returns of scale, which implies that the markup rate can be estimated from the ratio of sales to costs.} This approach identifies the markup rate by solving the firm's cost minimization problem and two main assumptions: first, the firm is a price taker in the input market. Second, there are two types of factors,  a variable factor, in which firms can flexibly decide how much to use within a given period (typically a year), and a fixed factor, such as capital, which remains constant within the same period. We compute markup rates, $\mu_{ijt}$, at firm $i$, industry $j$, and period $t$ (either quarter or year) as\footnote{We define an industry at the 2-digit level of the North American Industry Classification System (NAICS).}
\begin{gather*}
    \mu_{ijt} = \theta_{jt} \frac{\text{Sales}_{ijt}}{\text{COGS}_{ijt}},
\end{gather*}
where $\theta_{jt}$ is the industry-level output elasticity of variable inputs, and $\text{COGS}_{ijt}$ are costs of goods sold.
  
\subsubsection*{Profit rates}

Our measure of profitability follows standard accounting practice by calculating the percentage of sales that remains after subtracting the costs of goods sold and selling, general, and administrative expenses from total sales. 
\begin{gather*}
    \text{Profit rate}_{it} = \frac{\text{Sales}_{it}-\text{COGS}_{it}-\text{XSGA}_{it}}{\text{Sales}_{it}}, 
\end{gather*}
where XSGA is selling, general and administrative expenses. 

\subsection{Entry and exit of stock exchange listings in Compustat}
 
\begin{figure}[H]
\caption{Entry and Exit Rates}
    \label{fig:entryexitrates}
       \centering  
           \includegraphics[width=0.7\linewidth]{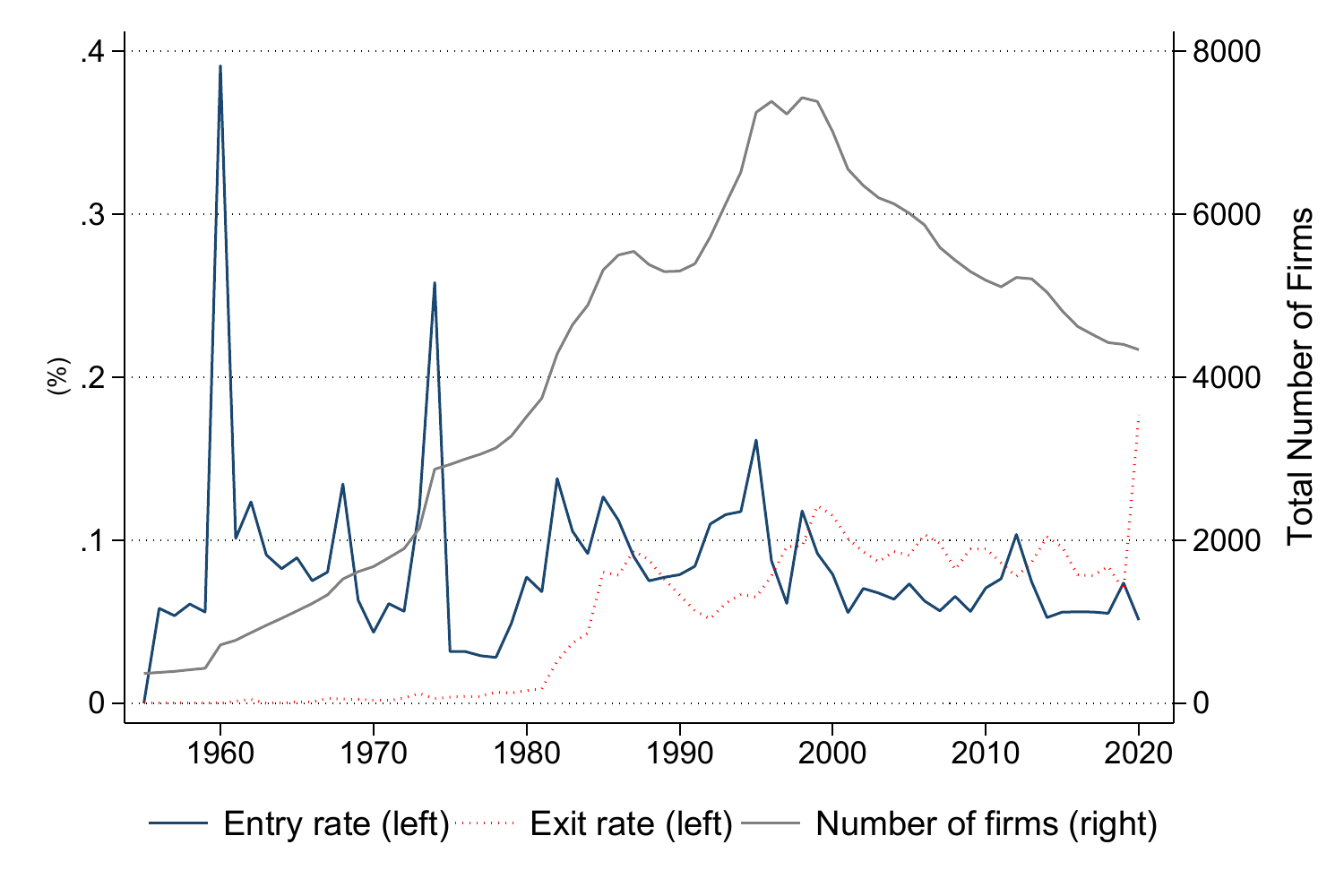}
         
    \caption*{\footnotesize \textit{Notes:} The figure displays entry and exit rates for the firms in Compustat. We classify a firm as entering at year $t$ if $t$ is the first year it appeared in the survey (the firm became publicly traded), and we label a firm as exiting at time $t$ if it is the last period where we observe   the firm in the survey.}
\end{figure}

We further analyze the composition of firms in Compustat's survey to evaluate whether our sample exhibits similar patterns in the share of firms entering or exiting before and during the pandemic. \figurename~\ref{fig:entryexitrates} depicts the yearly entry and exit rates and the total number of companies from 1955--2021.
 
One potential issue with our sample of firms is that the COVID-19 pandemic may have caused some firms to leave Compustat. However, our sample indicates that only 7\% of the firms in Compustat in 2019 left the survey in 2020, which is only slightly above the 1965--2019 exit rate average of 5.7\%. This exit rate increased in the second year of the pandemic, with 17.6\% of firms with data in 2020 leaving the survey in 2021.

\begin{figure}[H]
\caption{Average Markup and Profit Rates for Entrant and Exiting firms.}
    \label{fig:mupientryexit}
       \centering
        \begin{subfigure}{.5\textwidth}
      \centering
           \includegraphics[width=\linewidth]{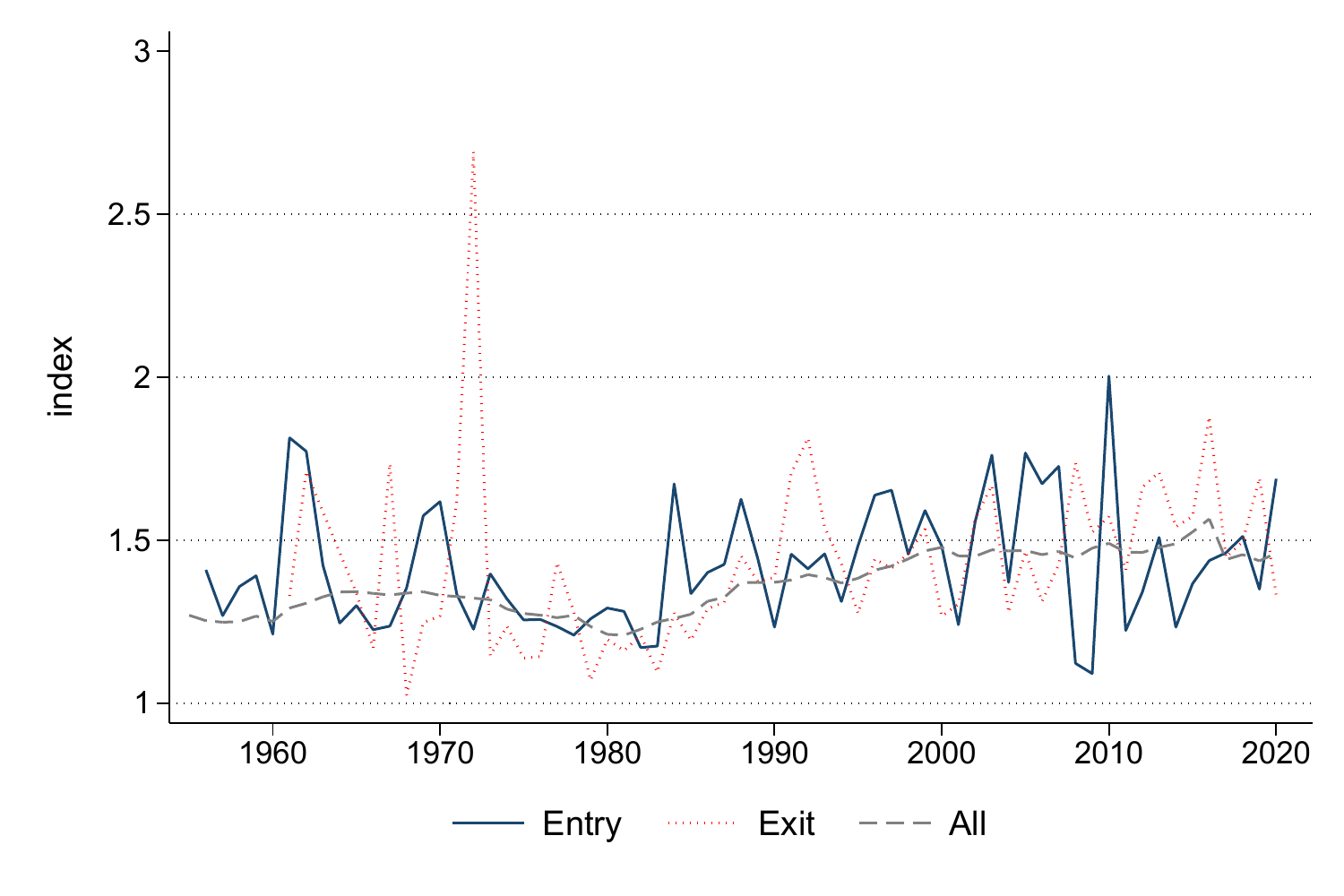}
      \caption{Average Markups}
      \label{fig:Markupsentryexit}
    \end{subfigure}%
        \begin{subfigure}{.5\textwidth}
      \centering
      \includegraphics[width=\linewidth]{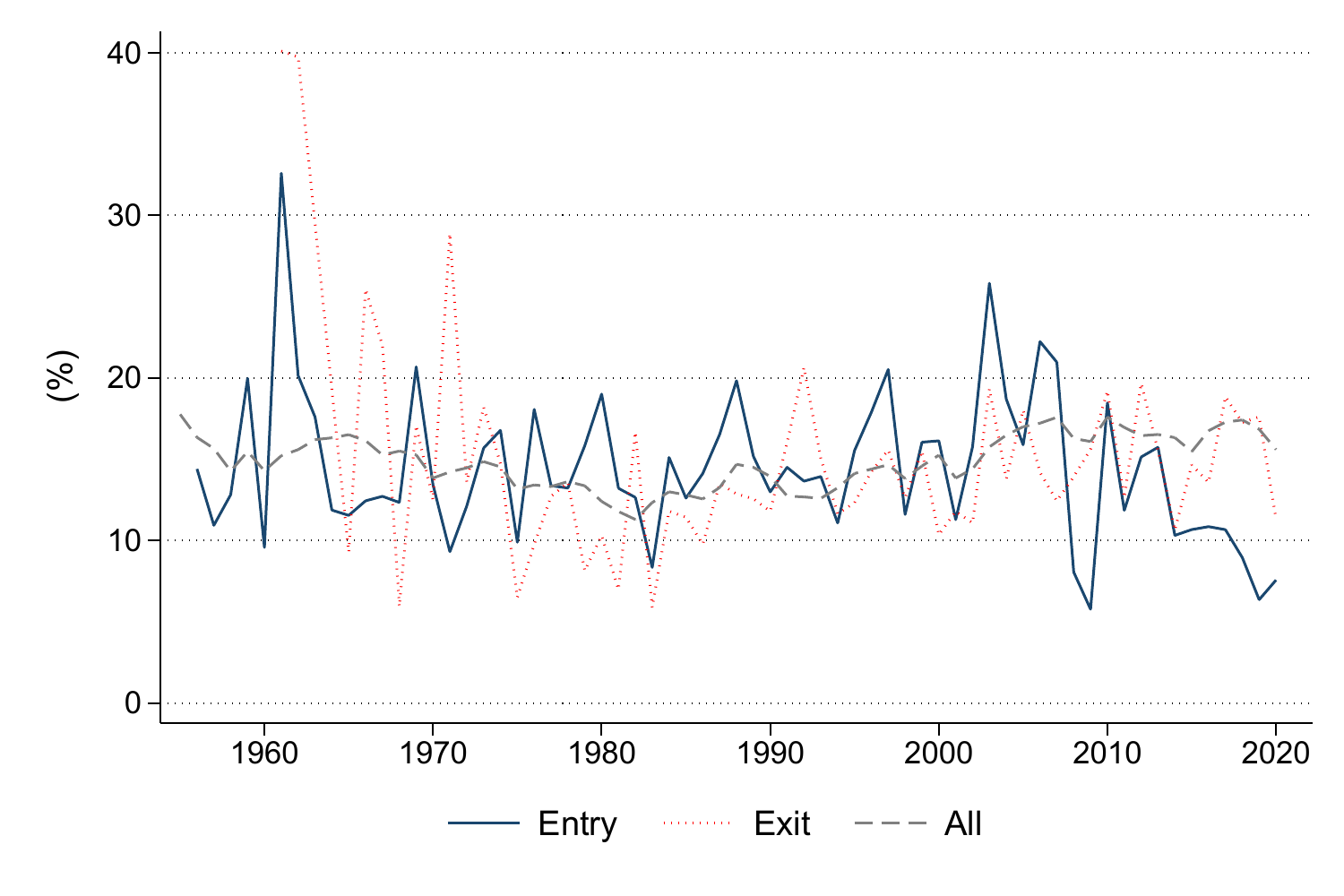}
      \caption{Average Profit Rates}
      \label{fig:profitsentryexit}
    \end{subfigure}%
    \caption*{\footnotesize \textit{Notes:} This figure plots the average markups and profit rates for three samples: firms entering the stock exchange, firms exiting, and all firms in the stock exchange.}
\end{figure}

Furthermore, firms that enter and exit Compustat's survey each year are similar to the firms that remain in the survey, at least in terms of markups and profit rates. \figurename~\ref{fig:mupientryexit} plots the average markup and profit rates for three samples: first, a sample including all the firms in a given year; second, firms entering the stock exchange; and third, firms exiting the stock exchange. The markup rates for all three samples have been increasing since 1980, although entering and exiting firms' markup rates tend to vary more relative to the average markup rate. Likewise, entering and exiting firms tend to have the same profit rates as the average firm. These findings suggest that the firms that leave the survey are not significantly different from those that remain.
 
\subsection{Historical trends in markup and profit rates}
 
\begin{figure}[H]
\caption{Markup and Profit Rates}
    \label{fig:hist_mu_pik_year}
    \begin{subfigure}{.5\textwidth}
      \centering
      \includegraphics[width=\linewidth]{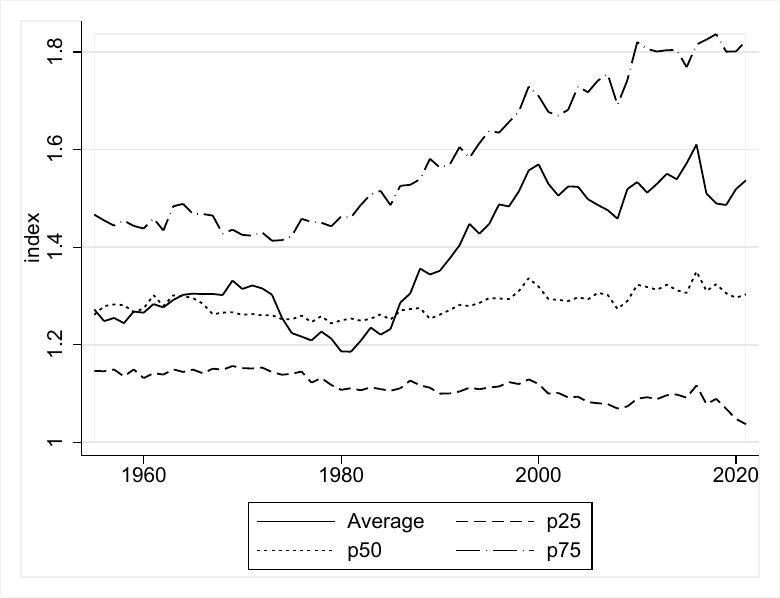}
      \caption{Markup Rates}
      \label{fig:hist_markup}
    \end{subfigure}%
    \begin{subfigure}{.5\textwidth}
      \centering
      \includegraphics[width=\linewidth]{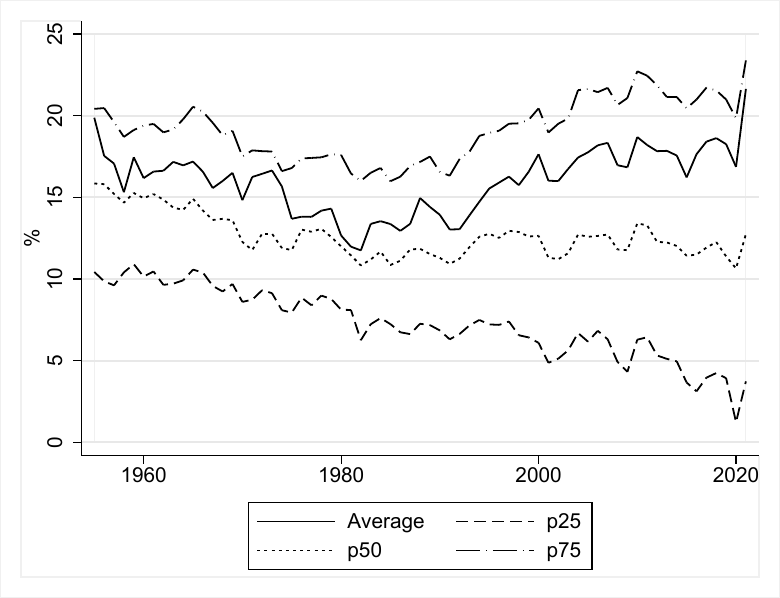}
      \caption{Profit Rates}
      \label{fig:hist_profitRate}
    \end{subfigure}
    \caption*{\footnotesize \textit{Notes:} The figure shows the sales-weighted average, the first, second, and third quartiles for the markup rates, based on \cite{DeLoecker2020}, and profit rates for publicly traded firms collected in Compustat. Markup rates are calculated as the multiplication of industry-level variable-input output elasticities and the firm-level sales to cost of goods sold ratio. Profit rates are total sales minus costs of goods sold and selling, general and administrative expenses as a share of sales.}
\end{figure}

We supplement this section with a historical analysis of markup and profit rates. Figure \ref{fig:hist_mu_pik_year} presents annual markup and profit rates for publicly-traded firms in the US between 1955 and 2021. The average profit rate is at its highest level in history, whereas the average markup rate is below its highest recorded value in 2016.

The markup rate saw a significant decrease between 2016 and 2019, but this trend reversed during the pandemic. In the first year of the pandemic, the sales-weighted average markup rate increased from 1.49 to 1.52. In the second year, it increased further to 1.54, which is slightly lower than the historical maximum observed in 2016. As documented in \citet{DeLoecker2020} through 2016, the increase in markup rates was driven by the rising markups of top firms, which continued from 2016--2021. For example, the third quartile markup rate increased from 1.81 to 1.82 between 2016 and 2021, while the markup rate for firms in the lower quartile decreased from 1.12 to 1.04.
 
Unlike markup rates, profit rates decreased in 2020 before recovering in 2021. For instance, the sales-weighted profit rate increased from 17.6\% to 21.6\% between 2016 and 2021, with a temporary decrease to 16.9\% in 2020. Interestingly, the most profitable firms were less affected than less profitable firms: the third quartile firms by profit rate experienced profit rates of 19.8\% and 23.4\% in 2020 and 2021, respectively, with the latter being their highest profit rate since 195. In contrast, the lower quartile firms reduced their profit rate from 3.1\% to 1.2\% between 2016 and 2020 before rebounding to 3.7\% in 2021.

\begin{figure}[H]
\caption{Markup and Profit Rate Volatility}
    \label{fig:volatility}
    \begin{subfigure}{.5\textwidth}
      \centering
      \includegraphics[width=0.9\linewidth]{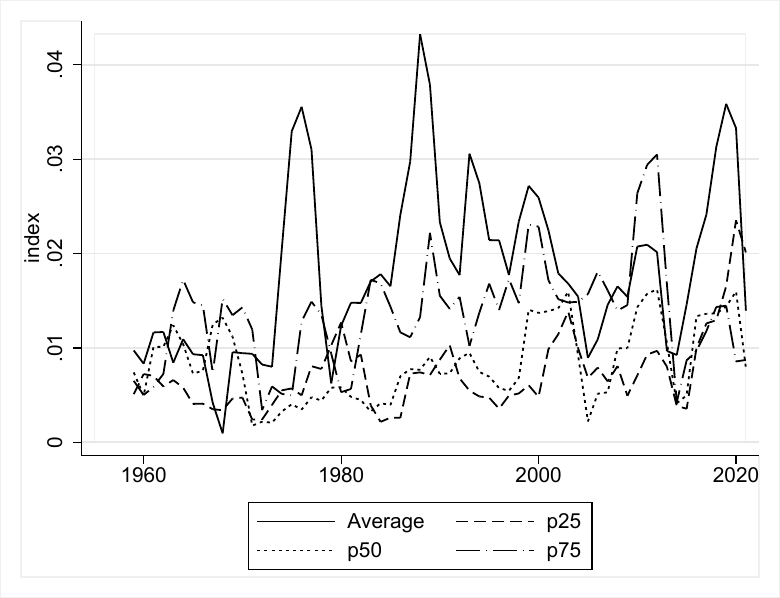}
      \caption{Markups}
      \label{fig:mu_coeff_variat}
    \end{subfigure}%
    \begin{subfigure}{.5\textwidth}
      \centering
      \includegraphics[width=0.9\linewidth]{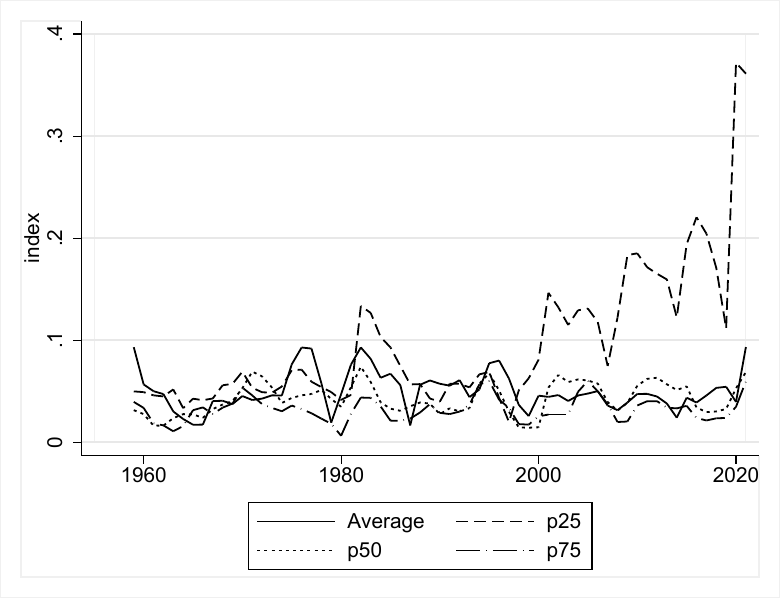}
      \caption{Profit Rates}
      \label{fig:pik_coeff_variat}
    \end{subfigure}
    \caption*{\footnotesize \textit{Notes:} The figure shows the 5-year rolling coefficient of variation for the sales-weighted average, the first, second, and third quartiles for the markup rates and profit rates for publicly traded firms collected in Compustat.}
\end{figure}

In addition to the increase in markup and profit rates caused by the pandemic, our findings indicate that their volatility was higher than usual during the first two years of the pandemic. However, this increased volatility was still within historical levels for most firms. Figure \ref{fig:volatility} shows how the volatility of sales-weighted markup and profit rates changed from 1959 to 2021.\footnote{For a given time series, The 5-year rolling coefficient of variation is calculated as the absolute value of the ratio of the standard deviation and the average, with the average and standard deviation being determined using data from the previous five years.} Consistent with the heterogeneous effects of the pandemic on profit rates, the lower quartile experienced increasing volatility from 2000 to 2021, with a particularly pronounced increase during the pandemic.

\section{Methodology}

    Let $Y_{i,t}$ be an observed outcome of interest, either the markup rate or the profit rate, for a firm $i$ in period $t$. Let $Y_{i,t}(1)$ and $Y_{i,t}(0)$ be the potential outcomes for a firm $i$ in time period $t$ when a pandemic happens and a pandemic does not happen, respectively. We denote $T$ as the last period before the pandemic happens.
    
    Our main parameter of interest is the effect of the pandemic in a firm-level outcome, namely,
    \begin{gather*}
        \tau_{i,T+h} = Y_{i,T+h}(1) - Y_{i,T+h}(0), \quad h=1,2,\dots. 
    \end{gather*}
    We only observe the potential outcome when a pandemic happens after time period $T$ for any firm, so $Y_{i,T+h} = Y_{i,T+h}(1)$ for $h=1,2,\dots$, but we cannot observe the outcome had the pandemic not taken place, $Y_{i,T+h}(0)$.
    
    To approximate the outcome had the pandemic not taken place, we propose to use the forecast based on information up to period $T$ as a counterfactual for each firm $i$, $Y_{i,T+h}(0)$. Specifically, let $I_{T}$ denote all information available up to period $T$; we define our counterfactual as the conditional expectation of $Y_{i,T+h}(0)$ given all observed information up to period $T$, i.e., $\EX[Y_{i,T+h}(0)|I_{T}]$, which potentially includes pre-treatment values of both outcomes and available covariates.

    Using the conditional expectation, we can compute an approximation to the causal effect as the difference between the realized value of our outcome of interest and the counterfactual
    \begin{gather*}
        \widetilde{\tau}_{i,T+h} = Y_{i,T+h}(1) - \EX[Y_{i,T+h}(0)|I_{T}].
    \end{gather*}
    
    To operationalize our counterfactual, we pick a family of models to estimate the conditional expectation of the outcome had the pandemic not taken place. We explain this family of models in Section \ref{sec:forecasting_model}. In this sense, our firm-level identification rests on the specific parametric family we use to estimate the counterfactual. We denote our estimator as  $\widehat{Y}_{i,T+h}(0) \equiv \widehat{\EX}[Y_{i,T+h}(0)|I_{T}]$.  Our final estimator of the firm-level causal effect is
    \begin{gather*}
        \widehat{\tau}_{i,T+h} = Y_{i,T+h} - \widehat{Y}_{i,T+h}(0).
    \end{gather*}
  
    The causal effect estimator, $\widehat{\tau}_{i,T+h}$, is a good approximation to the true causal effect, $\tau_{i,T+h}$, under two conditions: first, the 
    difference between the actual potential outcome without the pandemic and its counterfactual is near zero, $Y_{i,T+h}(0)-\EX[Y_{i,T+h}(0)|I_{T}]\approx 0$, also called conditional expectation error. Second, the difference between the actual forecast and the counterfactual is approximately zero, $\EX[Y_{i,T+h}(0)|I_{T}]-\widehat{Y}_{i,T+h}(0)\approx 0$, which we refer to as forecasting error. We make these conditions explicit in the following decomposition
    \begin{gather}
        \widehat{\tau}_{i,T+h} = \tau_{i,T+h} + (Y_{i,T+h}(0) - \EX[Y_{i,T+h}(0)|I_{T}]) + (\EX[Y_{i,T+h}(0)|I_{T}]-\widehat{Y}_{i,T+h}(0)).
        \label{eq:tau_decomposition}
    \end{gather}

    In practice, it is difficult to assess these two conditions. We can, however, aim to reduce the forecasting error as much as possible by selecting the best model within a comprehensible family of models. 
    
    Furthermore, we can provide conditions to obtain a consistent estimator of the average treatment effect across firms, $ \frac{1}{n}\sum_{i=1}^n \EX[\tau_{i,T+h}]$. In particular, let $Y_1,\dots,Y_n$ be a random sample, and define $\EX[\tau_{T+h}]=\EX[\tau_{i,T+h}]$. In Equation \eqref{eq:tau_decomposition}, summing across firms and dividing by the number of firms, we obtain
    \begin{gather*}
        \frac{1}{n}\sum_{i=1}^n \widehat{\tau}_{i,T+h} = \frac{1}{n}\sum_{i=1}^n \left[ \tau_{i,T+h} + (Y_{i,T+h}(0) - \EX[Y_{i,T+h}(0)|I_{T}]) +  (\EX[Y_{i,T+h}(0)|I_{T}]-\widehat{Y}_{i,T+h}(0))\right].
    \end{gather*}
    By invoking the law of large numbers, the average conditional expectation error should be close to zero as the number of firms in our sample increases,
    \begin{gather*}
        \frac{1}{n}\sum_{i=1}^n (Y_{i,T+h}(0) - \EX[Y_{i,T+h}(0)|I_{T}]) \xrightarrow[]{P} \EX[Y_{i,T+h}(0)] - \EX[\EX[Y_{i,T+h}(0)|I_{T}]] = 0,
    \end{gather*}
    where the equality follows from the law of iterated expectations, and $\xrightarrow[]{P}$ means convergence in probability.
    
    Finally, if our estimator of the conditional expectation is consistent in the sense that
    \begin{gather}
        \frac{1}{n}\sum_{i=1}^n (\EX[Y_{i,T+h}(0)|I_{T}]-\widehat{Y}_{i,T+h}(0)) \xrightarrow[]{P} 0, \label{eq:consistent_forecast}
    \end{gather}
    we can conclude that
    \begin{gather*}
        \frac{1}{n}\sum_{i=1}^n \widehat{\tau}_{i,T+h} \xrightarrow[]{P} \frac{1}{n}\sum_{i=1}^n \EX[\tau_{i,T+h}]=\EX[\tau_{T+h}],
    \end{gather*}
where the equality follows from the random sample assumption.

    \subsection{Forecasting model}
    \label{sec:forecasting_model}

    We use a Bayesian structural time series (BSTS) model forecast to create a counterfactual at the firm level \citep{Brodersen2015}. These models are highly flexible in the sense that they embed a broad class of time-series models. Besides their flexibility, they are also modular: they encompass different features of time-series data such as trend, seasonality, or holiday effects.

    Within the class of BSTS models, assume that the firm-level outcome of interest, $Y_{i,t}$, has a stochastic trend representation (also known as a local-level trend or random walk) as follows
            \begin{gather}
                Y_{i,t} = \delta_{i,t} +\varepsilon_{i,t},
                \label{eq:obs_eq}
                \\
                \delta_{i,t} =   \delta_{i,t-1}  + \eta_{\delta,i,t},
                \label{eq:trend}
               \end{gather}
    where $\delta_{i,t}$ is a trend component at time $t$, and,  $\varepsilon_{i,t}$ and $\eta_{\delta,i,t}$ are mutually independent Gaussian error components with variances $\sigma_\varepsilon^2$ and $\sigma_\mu^2$, respectively. In the state-space representation literature, Equation \eqref{eq:obs_eq} is commonly known as the observation equation, whereas Equation \eqref{eq:trend} is the state equation.
    
   For the quarterly specification, the model also incorporates a seasonal component. Specifically, let $\gamma_{i,t}$ be an additive quarterly component such that we augment the model in Equations \eqref{eq:obs_eq}-\eqref{eq:trend} with the following two equations
            \begin{gather}
                Y_{i,t} = \delta_{i,t} + \gamma_{i,t} +\varepsilon_{i,t},
                \label{eq:obs_eq_quarter}
                \\
                \gamma_{i,t+1} = -\sum_{s=0}^{2} \gamma_{i,t-s} + \eta_{\gamma,i,t}.
                \label{eq:seasonal}
            \end{gather}
    The seasonal specification in Equation \eqref{eq:seasonal} allows for the seasonal component to change over time.\footnote{In contrast to a specification with seasonal indicator variables, which restricts the seasonal effect to be fixed across years, the specification in Equation \ref{eq:seasonal} allows the  seasonal component to change over time \citep[see][Section 2.3.4]{harvey1990forecasting}.} To avoid perfect multicollinearity in the estimation of the seasonal component, the seasonal component is restricted to sum to zero across all seasons and time periods.
   
    To avoid identification or endogeneity issues, we decide not to include any contemporaneous variables on the right-hand side of Equations \eqref{eq:obs_eq} or \eqref{eq:obs_eq_quarter}. In principle, the pandemic may have affected all contemporaneous variables that are relevant to markup and profit rates. The choice to exclude these variables is supported in the bad control problem literature \citep{angrist2008mostly}.
   
\subsection{Estimation}

    \label{sec:estimation}

    To estimate the model in Equations \eqref{eq:obs_eq}-\eqref{eq:trend} (Equations \eqref{eq:obs_eq}-\eqref{eq:seasonal} in the quarterly model), we implement Gibbs sampling. This method jointly estimates unobservables, such as trend and seasonality components, as well as the parameters of the model, such as the error term variances. To ensure that the results are not sensitive to the initial values chosen and given the Markov-chain nature of Gibbs sampling, we choose 10,000 Monte-Carlo iterations.  
    
     Table \ref{tab:priors} shows the priors for the variances of the model and the initial values of the state components of the model. As is common with linear models and following \citet{Brodersen2015}, we set an inverse-Gamma prior for any of the variances $\sigma_{i}^2$ (either $\sigma_{\varepsilon,i}^2$, $\sigma_{\delta,i}^2$, or $\sigma_{\gamma,i}^2$) as $1/\sigma_{i}^2 \sim \mathcal{G}(v/2,s/2)$ for firm $i$. Since the ratio of the hyperparameters $s$ and $v$ is equal to the expected value of $\sigma_{i}^2$, it is common to refer to $s$ as the sum of squares and $v$ as the degrees of freedom. A common choice for the hyperparameters is to set $s/v=0.1s_{y_i}$, where $s_{y_i}$ is the sample standard deviation of the outcome of interest $Y_i$ for firm $i$. As in \citet{Brodersen2015}, we have a prior belief that the error components are small so we set small values for both the degrees of freedom, $v$, and the expected value of the variance, $s/v$. We use weakly informative Gaussian priors for the initial values of the state parameters.

      \begin{table}[ht]
          \centering
          \caption{Priors on Bayesian Structural Model}
          \label{tab:priors}
      \begin{threeparttable}[]
          \begin{tabular}{cccc}
            \hline
            Parameter & Prior density & Hyperparameter 1 & Hyperparameter 2  \\
            \hline
              $\sigma_{\varepsilon,i}$ & Inverse-Gamma & 1 & 0.01 \\
              $\sigma_{\delta,i}$ & Inverse-Gamma & 0.01 & 32 \\
              $\sigma_{\gamma,i}$ & Inverse-Gamma &  0.01 & 0.01 \\
              $\delta_{i,1}$ & Normal &   $\frac{y_0-\overline{y}}{ \hat{\sigma}_y}$     & 1 \\ 
              $\gamma_{i,1}$ & Normal &   0 & 1 \\
             \hline 
          \end{tabular}
        \begin{tablenotes}
         \item \noindent \footnotesize Notes:  $\sigma_{\epsilon,i}$, $\sigma_{\delta,i}$, and $\sigma_{\gamma,i}$ are the variances of the observation equation error (Equation \eqref{eq:obs_eq}), trend, and seasonal components, respectively, for a firm $i$. $\delta_{i,1}$ and $\gamma_{i,1}$ are the initial values of the trend and seasonal state components. When the prior density is inverse-gamma, hyperparameters 1 and 2 can be interpreted as the degrees of freedom and the sum of squared errors, respectively \citep[see][]{Brodersen2015}. When the prior density is normal, hyperparameters 1 and 2 are the mean and standard deviation, respectively. 
        \end{tablenotes}
    \end{threeparttable}

      \end{table}

\subsection{Assumption testing, model fit, and quality of forecasts}

\label{sec:model_fit}

We perform a battery of statistical tests to evaluate whether the firm-level markup and profit rates follow the data-generating process described in Equations \eqref{eq:obs_eq}-\eqref{eq:trend}, which forms the base of our forecasts. Under this data-generating process, the differentiated series of markup and profit rates,  $Y_{i,t}-Y_{i,t-1}$, should be normally distributed for each of the firms in our sample. To test for the normality of the differentiated series, we perform Jarque-Bera tests on these series. Indeed, we do not find evidence to reject the hypothesis that a majority of these series have a normal distribution. Specifically, for yearly (quarterly) markups, we find that 84.6\% (64\%) of the cases among 3,604 (3,192) firms do not reject the null hypothesis of normality at the 5\% level after applying the Bonferroni correction for multiple testing. Similarly, for yearly (quarterly) profit rates, we find that 82.2\% (74\%) of our tests among 3,611 (3,192) firms do not reject the normality assumption after controlling for the Bonferroni correction. We obtain similar results after using other corrections for multiple testing, including the Holm, Hochberg, and Hommel corrections. In summary, we do not find evidence to reject the normality assumption embedded in our forecasting model.

We also evaluate whether the firm-specific models accurately reflect the observed firm's behavior  before the pandemic. To this purpose, we calculate the fraction of periods for each firm when the observed outcome, either markup or profit rate, falls outside the model's 95\% Bayesian credible interval for each period that the firm has data before 2020. \tablename~\ref{tab:significance_bayesian} shows the average of these fractions across firms for each outcome and frequency. On average, 3\% of the observations are not within the Bayesian credible interval for both markup and profit rates. In addition, at least 75\% of the firms have a model that leaves no more than 6.3\% of the periods outside of the 95\% Bayesian credible intervals (see \tablename~\ref{tab:significance_bayesian}, column 3). These results suggest that most of the models correctly pin down the pre-pandemic trends. 

\begin{table}
 
\caption{\label{tab:significance_bayesian} Fraction of Periods that Observed Outcome Falls Outside of Bayesian Credible Intervals}
\centering
\begin{threeparttable}

    \begin{tabular}[ht]{lrrrrrr}
    \toprule
    Outcome (frequency) & Average  & Percentile 75th & Number of firms \\
    \midrule
    Markups (quarterly)      & 0.025 & 0.043 & 3,192 \\  
    Markups (yearly) & 0.028 & 0.053 & 3,611 \\
    Profit rate (quarterly) & 0.029 & 0.045 & 3,192 \\
    Profit rate (yearly) &    0.035 & 0.063 & 3,611\\
    \bottomrule
    \end{tabular}
           \begin{tablenotes}
            \item \noindent \footnotesize Notes:  The table introduces model fitness measures computed for markup rates and profit rates during the pre-pandemic period. For each firm, we compute the fraction of observations of the pre-pandemic  period that do not belong to the estimated 95\% Bayesian credible intervals. The table displays the average and the 75th percentile of these fractions across firms.
           \end{tablenotes}
   \end{threeparttable}

\end{table}

Furthermore, we find that our forecasts have adequate forecast quality. To evaluate our model's forecast quality, we compute forecasts for markup rates and profit rates for 2018--2019 using firm-level models that are estimated with observations up to 2017. We present some common forecast quality statistics in \tablename~\ref{tab:forecast_qual}. Two of these statistics confirm the forecast quality of our models. First, as suggested by Equation \eqref{eq:consistent_forecast}, the objective is to minimize the mean forecast error (ME) to be close to zero, which seems to be the case for the markup rates. Second, for at least half of the firms, the median absolute percentage error (MAPE) is 0.1 and 0.3 for markup and profit rates, respectively. These metrics support the forecast quality of our Bayesian models.

\begin{table}

\caption{\label{tab:forecast_qual} Forecast Quality Statistics of Bayesian Structural Time Series Model}
\centering
\begin{threeparttable}

    \begin{tabular}[ht]{llrrrrcp{1cm}r}
    \toprule
    Variable & Year & ME & RMSE & MAE & MAES & MAPE & Median MAPE & $n$\\
    \midrule
    Markup rate & 2018& -0.008 & 0.827 & 0.373 & 0.215 & 0.204 & 0.107 & 2858 \\
    Markup rate & 2019 & -0.046 & 0.900 & 0.409 & 0.241 & 0.241 & 0.122 & 2872 \\
    Profit rate & 2018 & 19.923 & 147.174 & 28.151 & 3.402 & 2.951 & 0.335 & 2858 \\
    Profit rate & 2019 & 18.482 & 151.506 & 30.590 & 4.464 & 3.610 & 0.380 & 2872 \\
    \bottomrule
    \end{tabular}
            \begin{tablenotes}
             \item \noindent \footnotesize Notes: The table introduces forecast quality statistics computed for 2018--2019 forecasts for markup and profit rates  estimated using firm-level models with samples up to 2017. The statistics computed are:  mean error (ME),  root mean square error (RMSE),  mean absolute error (MAE),  standardized mean absolute error  (MAES),  and the mean absolute percentage error (MAPE). Median MAPE is the median MAPE across firms. $n$ is the number of firms. 
            \end{tablenotes}
    \end{threeparttable}

\end{table}
 
\section{Counterfactual analysis results}

\label{sec:counteractual_analysis}

We introduce the results of the counterfactual effects of the pandemic in the markups and profit rate in this section. We first summarize the findings for the average firm. We then perform a heterogeneity analysis to elucidate which firms fared better during the pandemic.

\subsection{Aggregate data}
\label{sec:results_aggregate}

\begin{figure}[H]
\caption{Comparison of Observed and Counterfactual Markup Rates using Bayesian Structural Time Series Models}
\begin{subfigure}{.5\textwidth}
  \centering
  \includegraphics[width=0.85\linewidth]{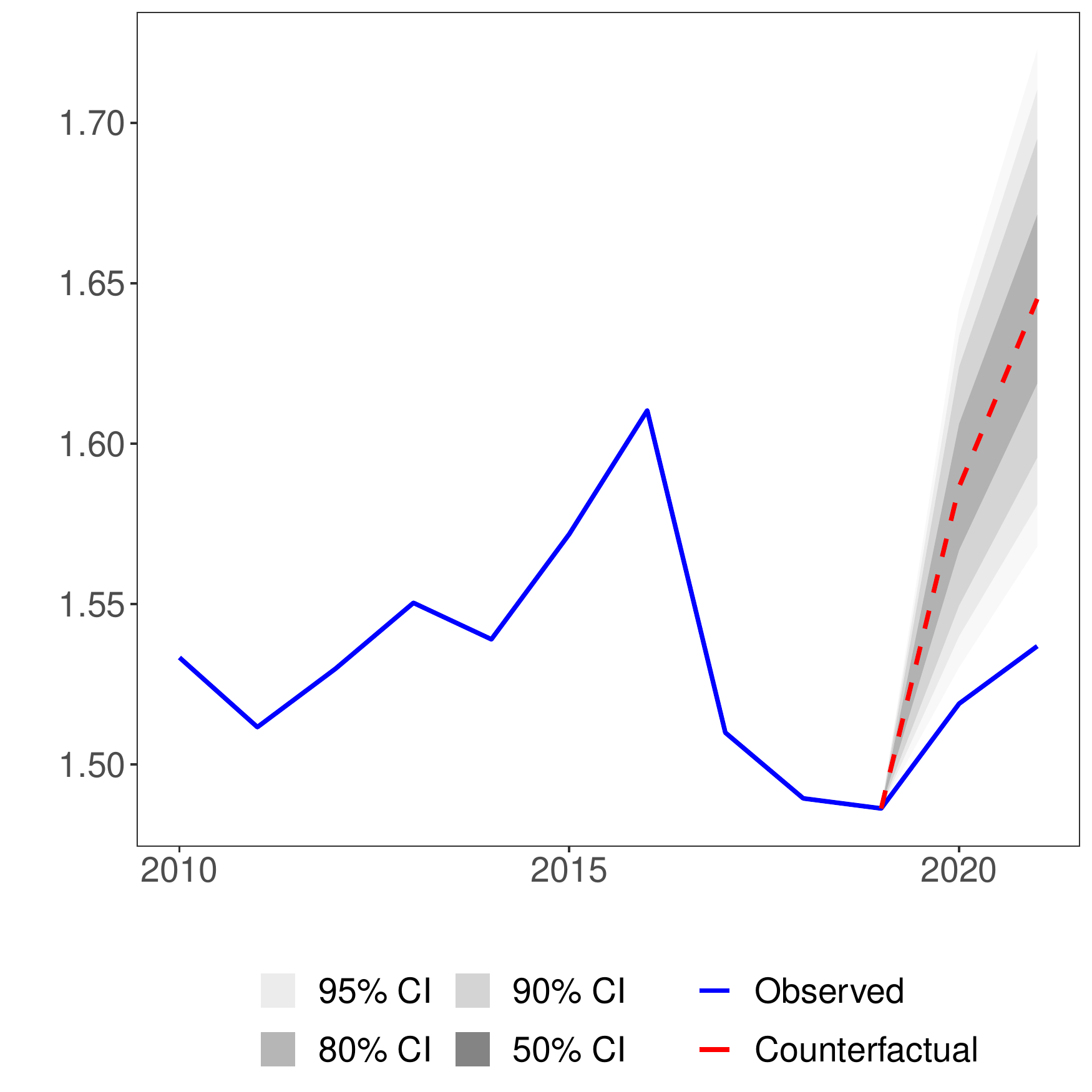}
  \caption{Average}
  \label{fig:Trend_fanchart_mu_year}
\end{subfigure}%
\begin{subfigure}{.5\textwidth}
  \centering
  \includegraphics[width=0.85\linewidth]{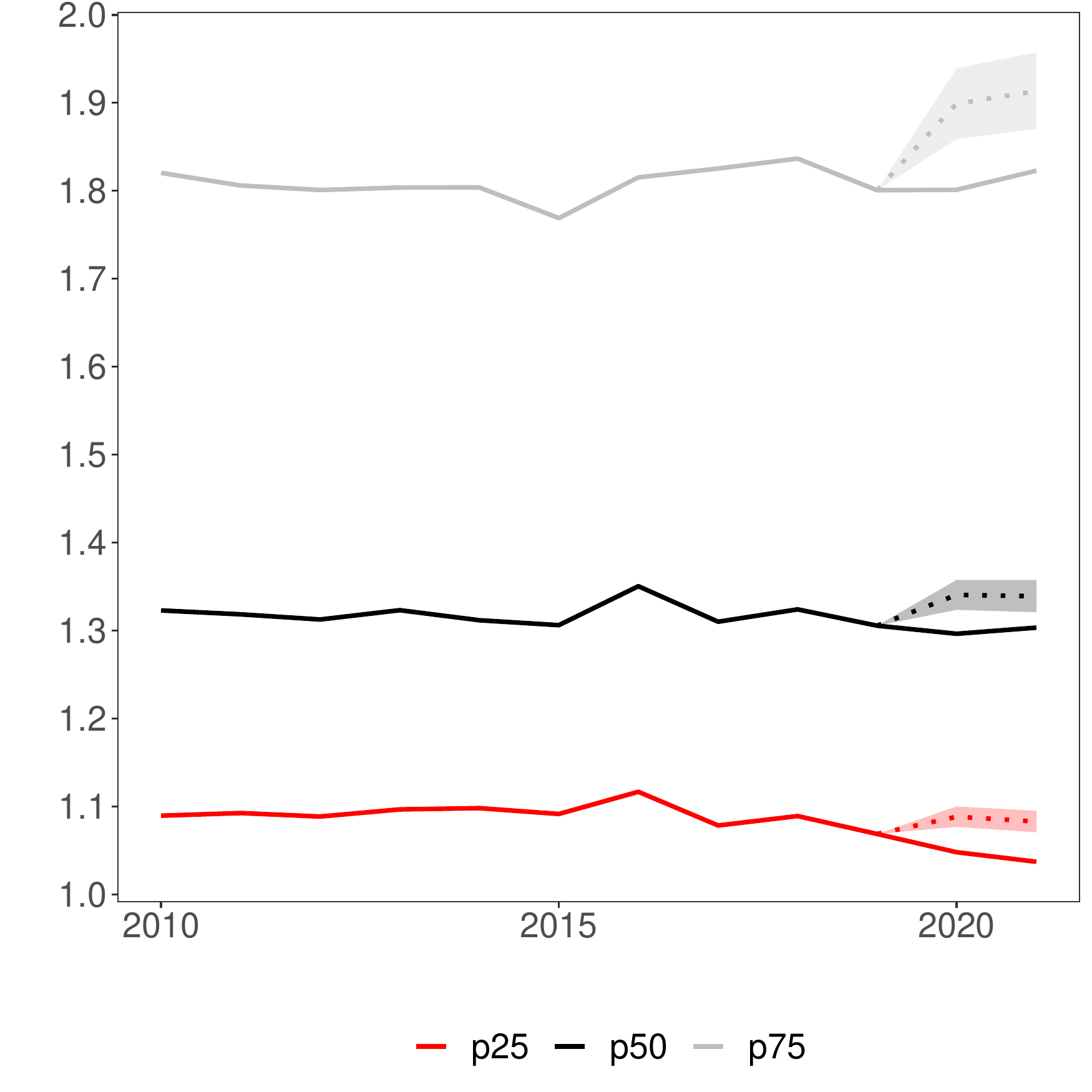}
  \caption{Quartiles}
  \label{fig:Trend_distr_mu_year}
\end{subfigure}
\label{fig:Trend_mu_year}
\caption*{\footnotesize Note: The figure shows four aggregate statistics of the yearly observed and counterfactual markup rates. The statistics are the sales-weighted average (shown in Panel (a)), the first quartile (shown in Panel (b) in red), the second quartile (shown in Panel (b) in black), and the third quartile (shown in Panel (b) in gray). The solid lines represent the observed values, while the dotted lines represent the counterfactual values for 2020-2021, as calculated using firm-level Bayesian structural time series models. The shaded areas represent the  95\%, 90\%, 80\%, and 50\% (equally-tailed) credible interval for each statistic, calculated based on 5,000 posterior simulations on Panel (a). 95\% (equally-tailed) credible intervals are displayed for the statistics on Panel (b).  }
\end{figure}

Most firms had lower markup rates than what could have been observed without the pandemic happening. To illustrate this point, we show in \figurename~\ref{fig:Trend_mu_year} four aggregates of both the observed and counterfactual markup rates, namely, its sales-weighted average and  quartiles in 2000--2021. For instance, for the average firm, the observed markup rate was 4.3\% and 6.6\% lower than its counterfactual value in 2020 and 2021, respectively. These effects are significant as the observed markup rate falls outside the 95\% credible interval of the counterfactual. Furthermore, there is considerable heterogeneity across firms in terms of quartiles of markup rates. For instance, the firms in the third quartile of markup rates had the largest negative effects in 2020 (2021) with a decrease of 5\% (4.6\%) relative to their counterfactual value. In contrast, the median markup firm (quartile 2) experienced less dramatic effects: a 3.3\% (2.6\%) lower  markup rate with respect to their counterfactual in 2020 (2021).\footnote{Similar effects arise when using quarterly data (see Figure \ref{fig:Trend_distr_mu_quarter}  in Appendix \ref{sec:further_analysis}).}

\begin{figure}[H]
\caption{Comparison of Observed and Counterfactual Profit Rates using Bayesian Structural Time Series Models}
\begin{subfigure}{.5\textwidth}
  \centering
  \includegraphics[width=0.85\linewidth]{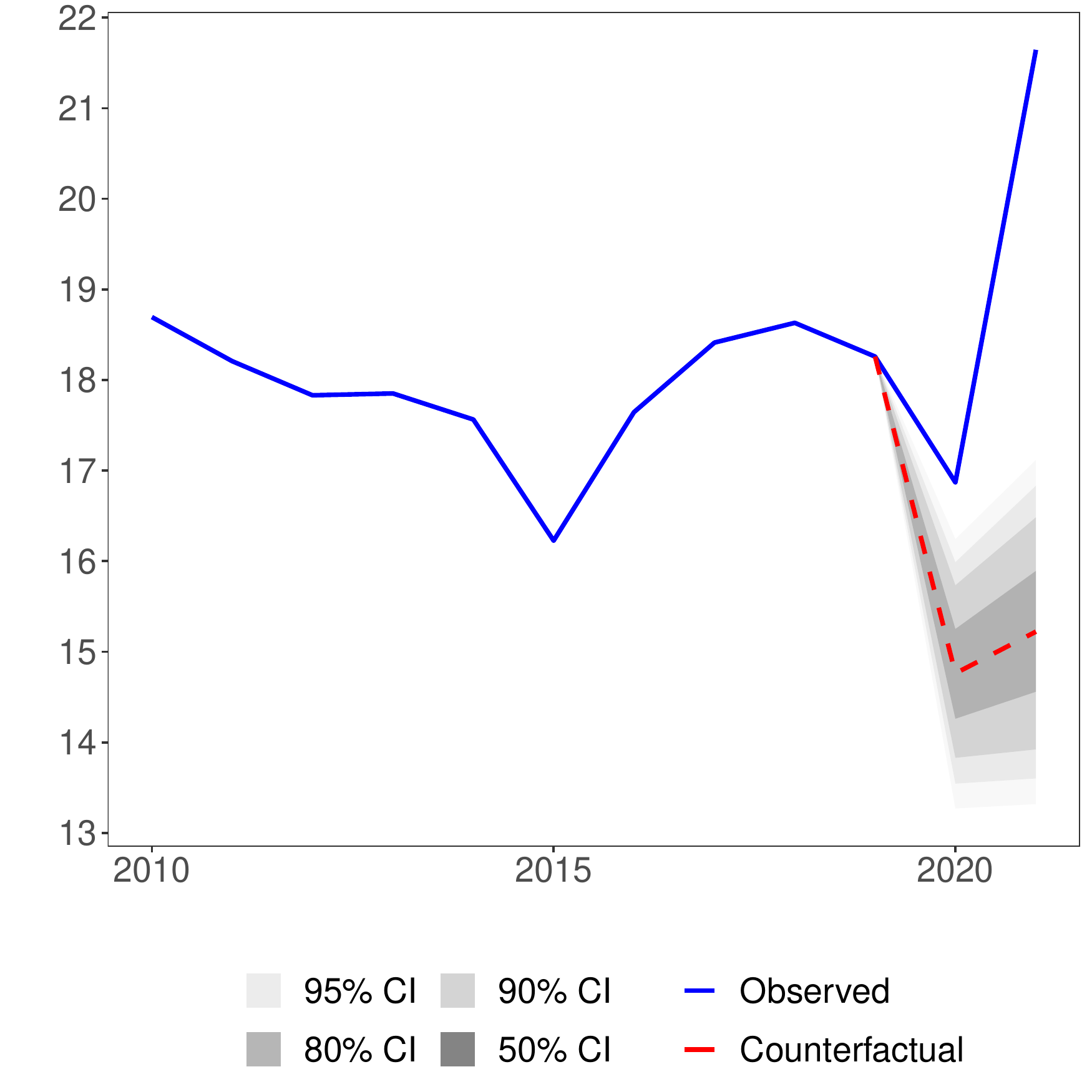}
  \caption{Average}
  \label{fig:Trend_fanchart_pik_year}
\end{subfigure}%
\begin{subfigure}{.5\textwidth}
  \centering
  \includegraphics[width=0.85\linewidth]{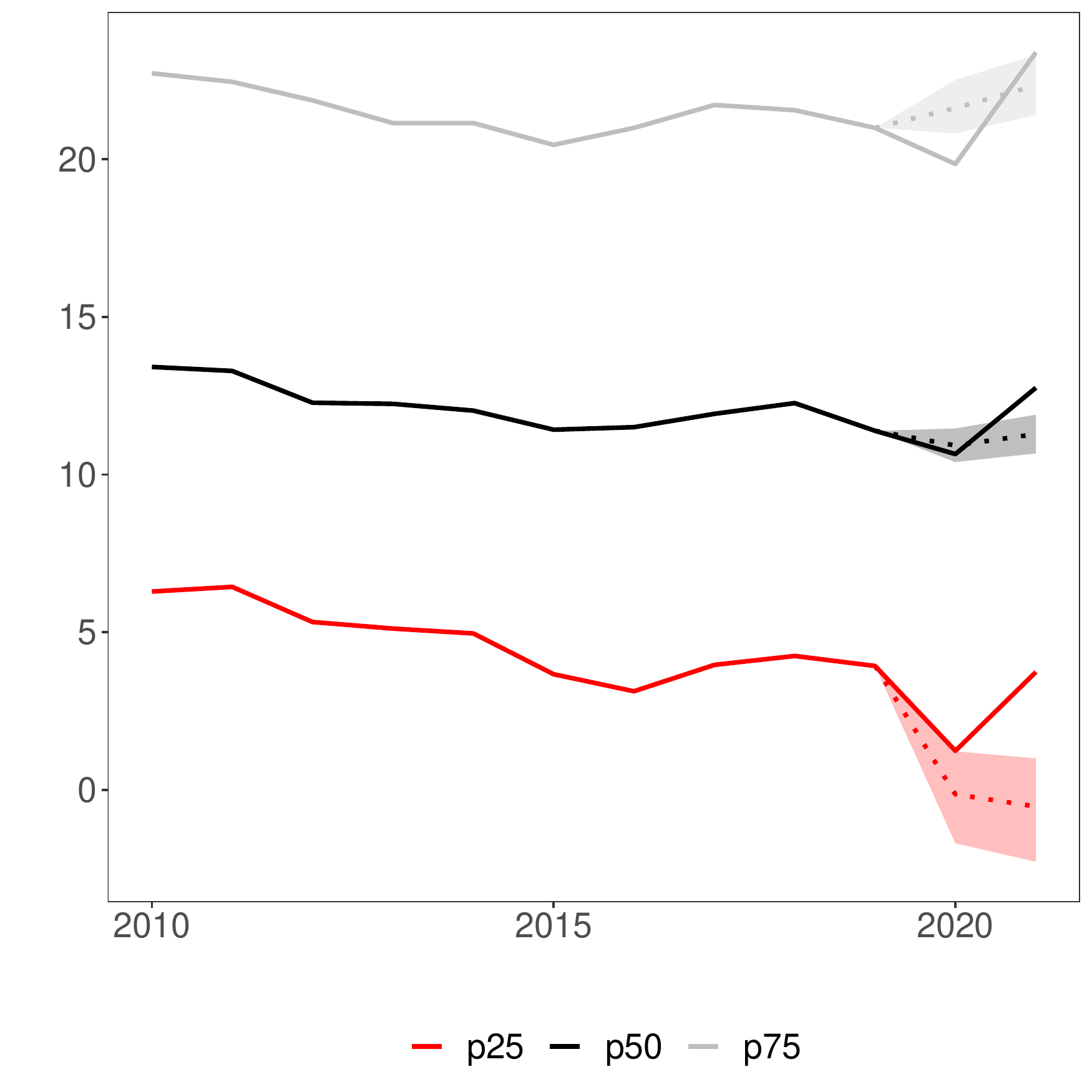}
  \caption{Quartiles}
  \label{fig:Trend_distr_pik_year}
\end{subfigure}
\label{fig:Trend_pik_year}
\caption*{\footnotesize Note: The figure shows four aggregate statistics of the yearly observed and counterfactual profit rates. The statistics are the sales-weighted average (shown in Panel (a)), the first quartile (shown in Panel (b) in red), the second quartile (shown in Panel (b) in black), and the third quartile (shown in Panel (b) in gray). The solid lines represent the observed values, while the dotted lines represent the counterfactual values for 2020-2021, as calculated using firm-level Bayesian structural time series models. The shaded areas represent the  95\%, 90\%, 80\%, and 50\% (equally-tailed) credible interval for each statistic, calculated based on 5,000 posterior simulations on Panel (a). 95\% (equally-tailed) credible intervals are displayed for the statistics on Panel (b).}
\end{figure}

Unlike the observed negative effects of markups, we find that most firms had higher profit rates than their counterfactuals during the pandemic. \figurename~\ref{fig:Trend_pik_year} shows the (sales-weighted) average and quartiles of both the observed and counterfactual profit rates across firms. When analyzing the counterfactual effects of the pandemic on profit rates, most firms had a positive impact. For instance, the average firm had a profit rate of 16.9\% and 21.6\% in 2020 and 2021, respectively, which are 2.1 and 6.4 percentage points higher than the counterfactuals. These effects are significant as the observed markup rate falls outside the 95\% credible interval of the counterfactual. Additionally, firms with higher markup rates experienced lower markup rates than what could have been expected without the pandemic. For instance, the third-quartile firm registered a 19.8\% profit rate in 2020 which is 1.7 percentage points lower than the profit rate these companies would experience had the pandemic not happened. In contrast, firms in the first and second quartiles experienced no effects on profit rates in 2020 while experiencing positive effects in the second year of the pandemic.\footnote{Similar effects arise when using quarterly data  (see Figure \ref{fig:Trend_distr_pik_quarter} in Appendix \ref{sec:further_analysis}).}

Looking at both results for markups and profit rates jointly, it seems that the uncertainty during the pandemic potentially promoted the decrease in markups and the increase in profits\citep{Altig2020}. In the next section, we explore the heterogeneity of these effects across firm characteristics as well as the industries most affected.

\subsection{Heterogeneity analysis}\label{sec:heterogeneity}

We investigate potential heterogeneity in the effects of the pandemic on markup rates and profitability to elucidate whether firms with different levels of key characteristics differ in their response to the pandemic. These characteristics include pre-pandemic levels of cost of goods sold, sales, employment, stock tenure, and market share. Specifically, to summarize the effect heterogeneity in a convenient form, we run the following cross-sectional regression
 
\begin{align*}
    \overline{\widehat{\tau}}_i & =\beta_0+ \beta_1\log \overline{\text{COGS}}_{i} + \beta_2\log \overline{\text{Sales}}_{i} + \beta_3\log \overline{\text{Employment}}_{i} + \beta_4\overline{\text{Stock tenure}}_{i} 
    \\
    & \phantom{space} + \overline{\text{Market share}}_{i} + \delta_{j} + \epsilon_i,
\end{align*}
where $\overline{\widehat{\tau}}_i$ is the average effect in 2020--2021 for a firm $i$ (on either markup or profit rates), the right-hand side variables are firm-level averages for COGS, sales, employment, stock tenure, market share in 2015--2019, and $\delta_{j}$ is an industry-level fixed effect such that firm $i$ belongs to industry $j=j(i)$. Tables \ref{tab:het_markups_yearly} and \ref{tab:het_profit_rates_yearly} show the estimated regression for the yearly specification for markups and profit rates, respectively.
As expected, those firms with higher tenure in the stock exchange have larger positive effects on markup rates since they are more likely to be in a consolidated industry or have survived a merger or an acquisition. Likewise, firms with high sales or low costs have mechanically more positive effects on profit rates. These regressions suggest that market share, number of employees, or tenure in the stock market imply higher effect heterogeneity in profit rates.

We find similar effect heterogeneity results when using the quarterly specification for the markup and profit rates (see Tables \ref{tab:het_markup_rates_quarterly} and  \ref{tab:het_profit_rates_quarterly}, respectively). Likewise, we find similar conclusions when we build scatter plots of firm-level effects against pre-pandemic values of key firm characteristics. In particular, we plot the effect in either markup or profit rates against pre-pandemic values of markup rates, sales, stock exchange tenure, profit rate, market share, and employment---the binscatter plots in Appendix \ref{app:het} illustrate this analysis. 

We provide further evidence of the heterogeneity of the response to the pandemic by tracking the fraction of firms with negative and positive effects of the pandemic in each quarter in 2020--2021 (see Appendix \ref{app:heterogeneity}). In particular, at the beginning of the pandemic, within those firms that are significantly affected---by having an observed outcome outside the 95\% credible interval for the counterfactual outcome--- most firms are negatively affected both in terms of markups and profit rates. By the last quarter of 2021, the share of firms with positive effects increased but is still lower than those with negative effects. This suggests that there were more negatively affected firms than positively affected firms in most quarters.

\begin{table}[H]
    \centering
    \caption{Heterogeneity in the effects of the pandemic on markups\\ (Yearly Specification)}
    \label{tab:het_markups_yearly}
    \begin{threeparttable}    
        \begin{tabular}{lcccccc}
           \tabularnewline\midrule\midrule
           Dependent Variable: & \multicolumn{6}{c}{Average markup rate pandemic effect in 2020--2021}\\
            & \multicolumn{2}{c}{All} & \multicolumn{2}{c}{2020} & \multicolumn{2}{c}{2021} \\ 
           Model:        & (1)            & (2)            & (3)            & (4)            & (5)           & (6)\\
           \midrule \emph{Variables} &   &   &   &   &   &  \\
          COGS   & 0.0856         & 0.0659         & 0.0490         & 0.0463         & 0.1300         & 0.0937\\
                    & (0.0851)       & (0.0912)       & (0.0893)       & (0.0965)       & (0.1497)       & (0.1591)\\
          Sales & -0.1534$^{*}$  & -0.0864        & -0.1393        & -0.0975        & -0.1689        & -0.0724\\
                    & (0.0905)       & (0.0926)       & (0.0946)       & (0.0983)       & (0.1594)       & (0.1614)\\
          Employment   & 0.0519$^{***}$ & 0.0082         & 0.0951$^{***}$ & 0.0568$^{***}$ & -0.0016        & -0.0550$^{*}$\\
                    & (0.0156)       & (0.0170)       & (0.0183)       & (0.0187)       & (0.0260)       & (0.0295)\\
          Stock-exchange tenure   & 0.0030$^{***}$ & 0.0031$^{***}$ & 0.0025$^{***}$ & 0.0018$^{**}$  & 0.0035$^{***}$ & 0.0043$^{***}$\\
                    & (0.0007)       & (0.0007)       & (0.0007)       & (0.0008)       & (0.0012)       & (0.0013)\\
          Market share & 0.8069$^{**}$  & -0.3262        & -0.0722        & -0.6241        & 1.818$^{**}$   & 0.0390\\
                    & (0.3312)       & (0.4701)       & (0.3854)       & (0.5395)       & (0.7961)       & (0.7951)\\
          \midrule \emph{Fixed-effects} &   &   &   &   &   &  \\
          2-digit NAICS industry          &                & Yes            &                & Yes            &                & Yes\\
          \midrule
          Mean  & 1.700 &  & 1.674 &  & 1.732 \\ 
          Mean effect  & -0.069 &  & -0.086 &  & -0.048 \\
          \midrule \emph{Fit statistics} &   &   &   &   &   &  \\
         Observations  & 5,771          & 5,771          & 3,139          & 3,139          & 2,632          & 2,632\\
      R$^2$         & 0.00632        & 0.05509        & 0.01250        & 0.05801        & 0.00734        & 0.07110\\
      Within R$^2$  &                & 0.00303        &                & 0.00495        &                & 0.00561\\
        \hline
        \end{tabular}
        \begin{tablenotes}
         \item \noindent \footnotesize Notes: The table presents cross-section firm-level regressions of the average pandemic effect  at the firm-level on the 2015--2019 average of the logarithm cost of goods sold, logarithm of sales, logarithm of employment, years since publicly-traded, and market shares. NAICS = North American Industrial Classification System. Heteroskedasticity-robust standard-errors in parentheses. Signif. Codes: ***: 0.01, **: 0.05, *: 0.1.
         \end{tablenotes}
    \end{threeparttable}
\end{table}

\begin{table}[H]
    \centering
    \caption{Heterogeneity in the effects of the pandemic on profit rates \\ (Yearly Specification)}
    \label{tab:het_profit_rates_yearly}
    \begin{threeparttable}
        \begin{tabular}{lcccccc}
           \tabularnewline\midrule\midrule
           Dependent Variable: & \multicolumn{6}{c}{Average profit rate pandemic effect in 2020--2021}\\
            & \multicolumn{2}{c}{All} & \multicolumn{2}{c}{2020} & \multicolumn{2}{c}{2021} \\ 
           Model:        & (1)            & (2)            & (3)            & (4)           & (5)          & (6)\\
           \midrule \emph{Variables} &   &   &   &   &   &  \\
          COGS    & -19.46$^{***}$ & -21.09$^{***}$ & -18.76$^{***}$ & -15.69$^{**}$ & -20.03$^{*}$ & -27.06$^{**}$\\
                    & (5.923)        & (7.359)        & (6.236)        & (6.714)       & (10.50)      & (13.76)\\
          Sales & 16.81$^{**}$   & 18.46$^{**}$   & 12.90          & 10.12         & 21.25        & 28.37$^{*}$\\
                    & (7.897)        & (9.153)        & (9.326)        & (9.903)       & (13.09)      & (16.04)\\
          Employment  & -0.2620        & -0.2221        & 3.092          & 2.458         & -4.328       & -3.901\\
                    & (3.919)        & (4.421)        & (4.992)        & (5.771)       & (6.238)      & (6.861)\\
          Stock-exchange tenure    & -0.0211        & -0.0288        & -0.0914        & -0.1333       & 0.0539       & 0.0761\\
                    & (0.0667)       & (0.0796)       & (0.0795)       & (0.0978)      & (0.1117)     & (0.1303)\\
          Market share  & -5.302         & -29.90         & -10.50         & 15.34         & -1.726       & -84.34\\
                    & (26.36)        & (72.81)        & (37.57)        & (84.41)       & (37.10)      & (127.3)\\
          \midrule \emph{Fixed-effects} &   &   &   &   &   &  \\
          2-digit NAICS industry         &                & Yes            &                & Yes           &              & Yes\\
        \midrule
        Mean & 2.438 &  & 1.274 &  & 3.833 \\ 
        Mean effect& 7.145 &  & 5.443 &  & 9.183 \\ 
        \\
          \midrule \emph{Fit statistics} &   &   &   &   &   &  \\
           Observations  & 5,771          & 5,771          & 3,139          & 3,139         & 2,632        & 2,632\\
      R$^2$         & 0.00534        & 0.01288        & 0.00867        & 0.02122       & 0.00411      & 0.01382\\
      Within R$^2$  &                & 0.00509        &                & 0.00664       &              & 0.00527\\ 
        \hline
        \end{tabular}
        \begin{tablenotes}
         \item \noindent \footnotesize Notes: The table presents cross-section firm-level regressions of the average pandemic effect at the firm-level on 2015--2019 average of the logarithm cost of goods sold, the logarithm of sales, logarithm of employment, years since publicly-traded, and market shares. NAICS = North American Industrial Classification System. Heteroskedasticity-robust standard errors in parentheses. Signif. Codes: ***: 0.01, **: 0.05, *: 0.1
        \end{tablenotes}
    \end{threeparttable}
\end{table}
 
\subsection{Heterogeneity by industry}

\begin{figure}[H]
\caption{Effects on Markups and Profit Rates by Industry}
\begin{subfigure}{.5\textwidth}
  \centering
  \includegraphics[width=\linewidth,page=1]{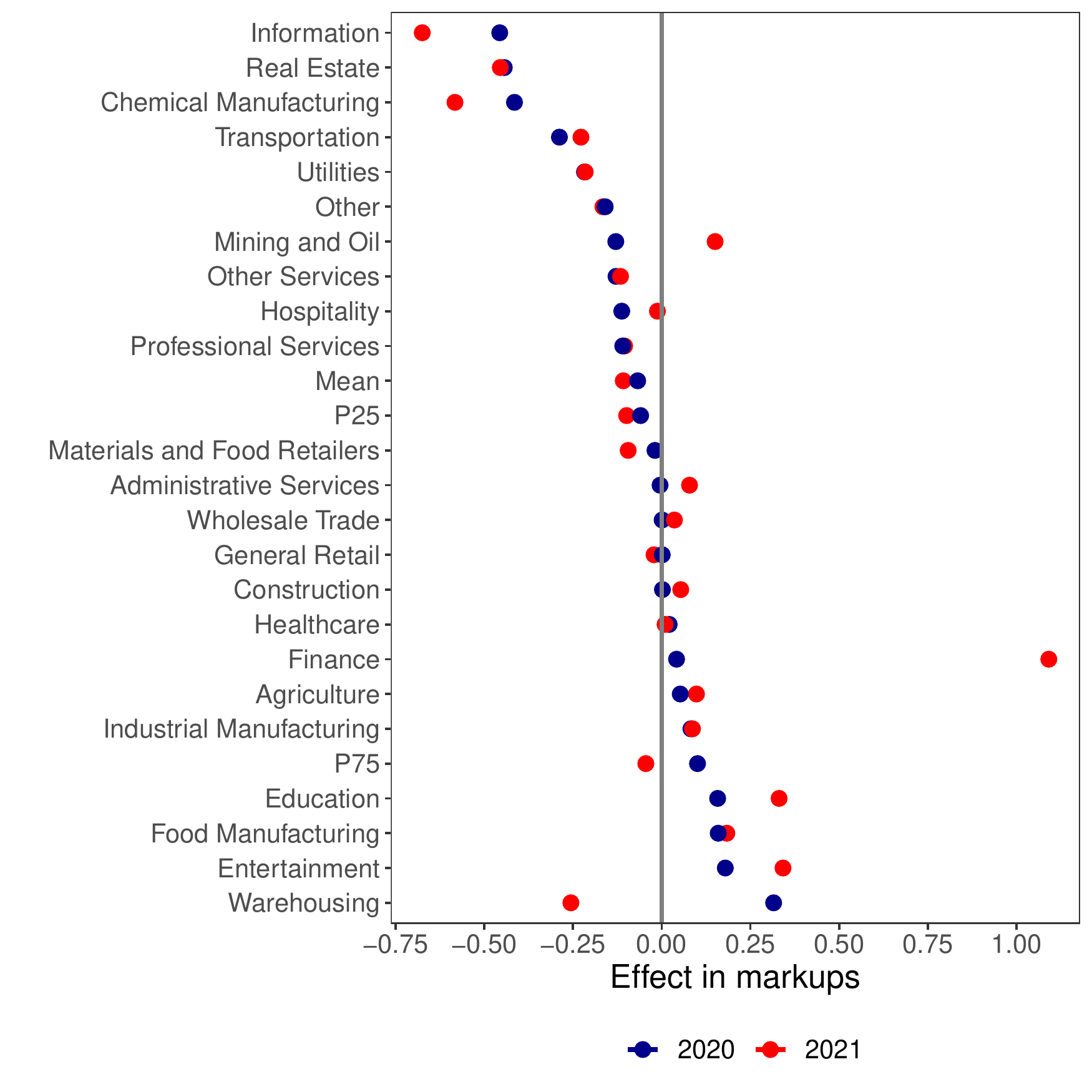}
  \caption{Markups}
  \label{fig:industry_mu}
\end{subfigure}%
\begin{subfigure}{.5\textwidth}
  \centering
  \includegraphics[width=\linewidth,page=2]{Compare_pimu_bar_ind_yearly_lollipop.pdf}
  \caption{Profit Rates}
  \label{fig:industry_pik}
\end{subfigure}
\label{fig:industry}
    \caption*{\footnotesize Note: The figure plots the sales-weighted average effects on markup and profit rates by the 2-digit NAICS industry. The mean and percentiles 25 (P25) and 75 (P75) correspond to the aggregate effects for the sample of 3,611 firms in Compustat (see also Figures \ref{fig:Trend_mu_year} and \ref{fig:Trend_pik_year}). Industries are organized in ascending order according to their corresponding effect in 2020. }
 
\end{figure}

We supplement our heterogeneity results by analyzing the effects of industry. \figurename~\ref{fig:industry} shows the industry breakdown of the effects on markup and profit rates. First, we observe that the transportation industry experienced  negative effects on profit rates in both 2020 and 2021. Following a negative demand shock due to lockdowns, we also note a fall in this industry's markups.  Interestingly, we observe similar behavior in the oil and gas industry; in 2020, lockdowns drove demand and prices down, which lead the industry to report losses. However, the commodity price surge in 2021, partly in response to a myriad of public policy responses, triggered increases in profit rates of roughly 10\%. 
 
Similarly, hospitality displays adverse effects on profit rates in 2020, with roughly no effect on markups. In contrast, information and technology, warehousing, and real estate experienced profit rates above   what could have been predicted by previous trends. Likewise, professional and administrative services, which could easily migrate to online environments, had profit rates above their historical trends. Lastly, we see modest effects on the retail industries for both markups and profit rates, which are known to have low markups \citep{Philippon2019}.

\subsection{Business cycle analysis}

One may wonder that fluctuations in economic activity may affect either markup or profit rates. To answer this question and under the assumption that the pandemic does not affect these fluctuations, we further augment Equation \eqref{eq:obs_eq} with a cyclical component.
\begin{gather}
    Y_{it} = \delta_{it} + \alpha\ cycle_{t} + \varepsilon_{it},
    \label{eq:cycle}
\end{gather}
where $cycle_t$ is taken as the Hodrick-Prescott decomposition of the US real gross domestic product in period $t$.

\begin{figure}[H]
\caption{Histogram of Coefficient of the Cycle in Models (Yearly Frequency)}
\begin{subfigure}{.5\textwidth}
  \centering
  \includegraphics[width=0.9\linewidth]{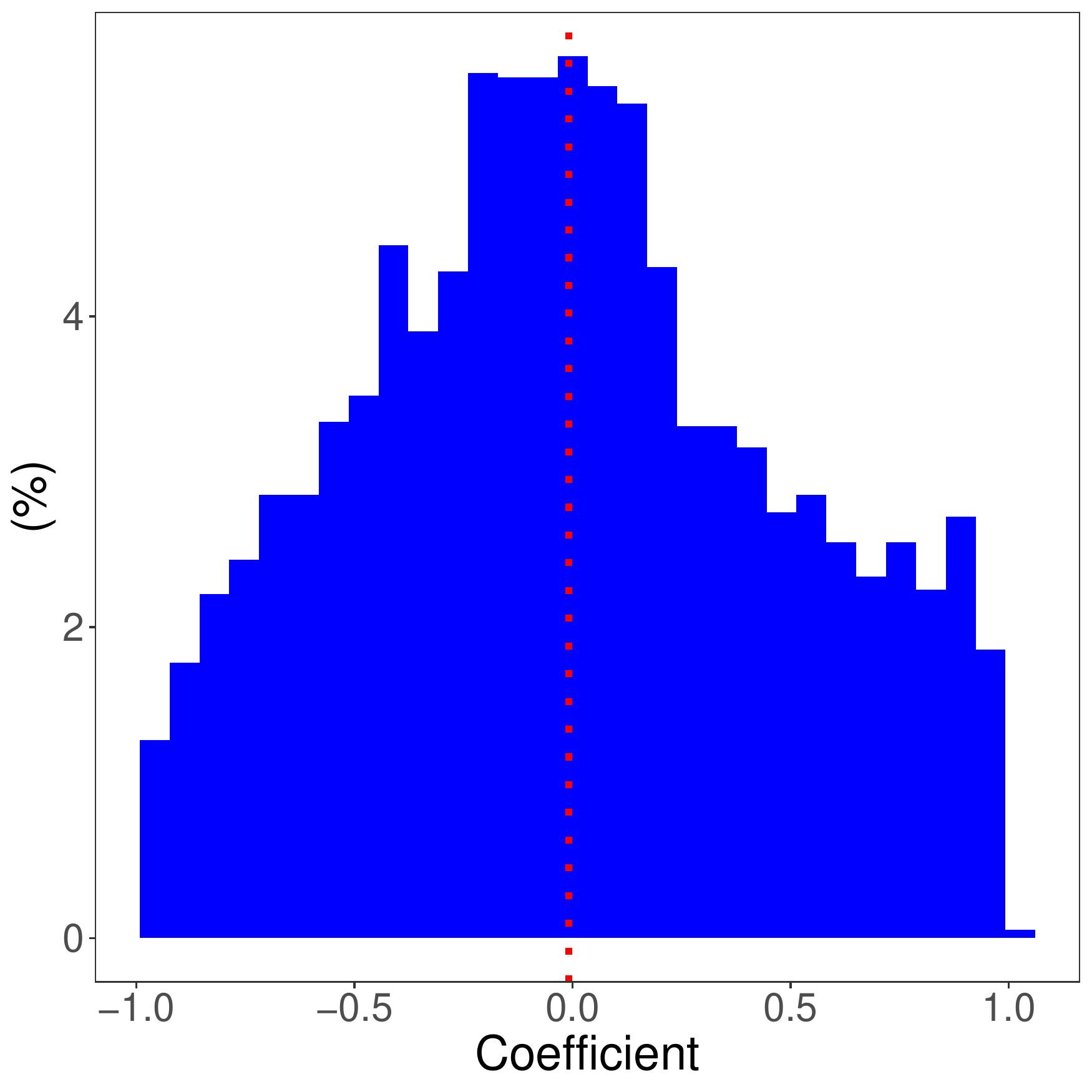}
  \caption{Markup Rates}
  \label{fig:cycle_mu}
\end{subfigure}%
\begin{subfigure}{.5\textwidth}
  \centering
  \includegraphics[width=0.9\linewidth]{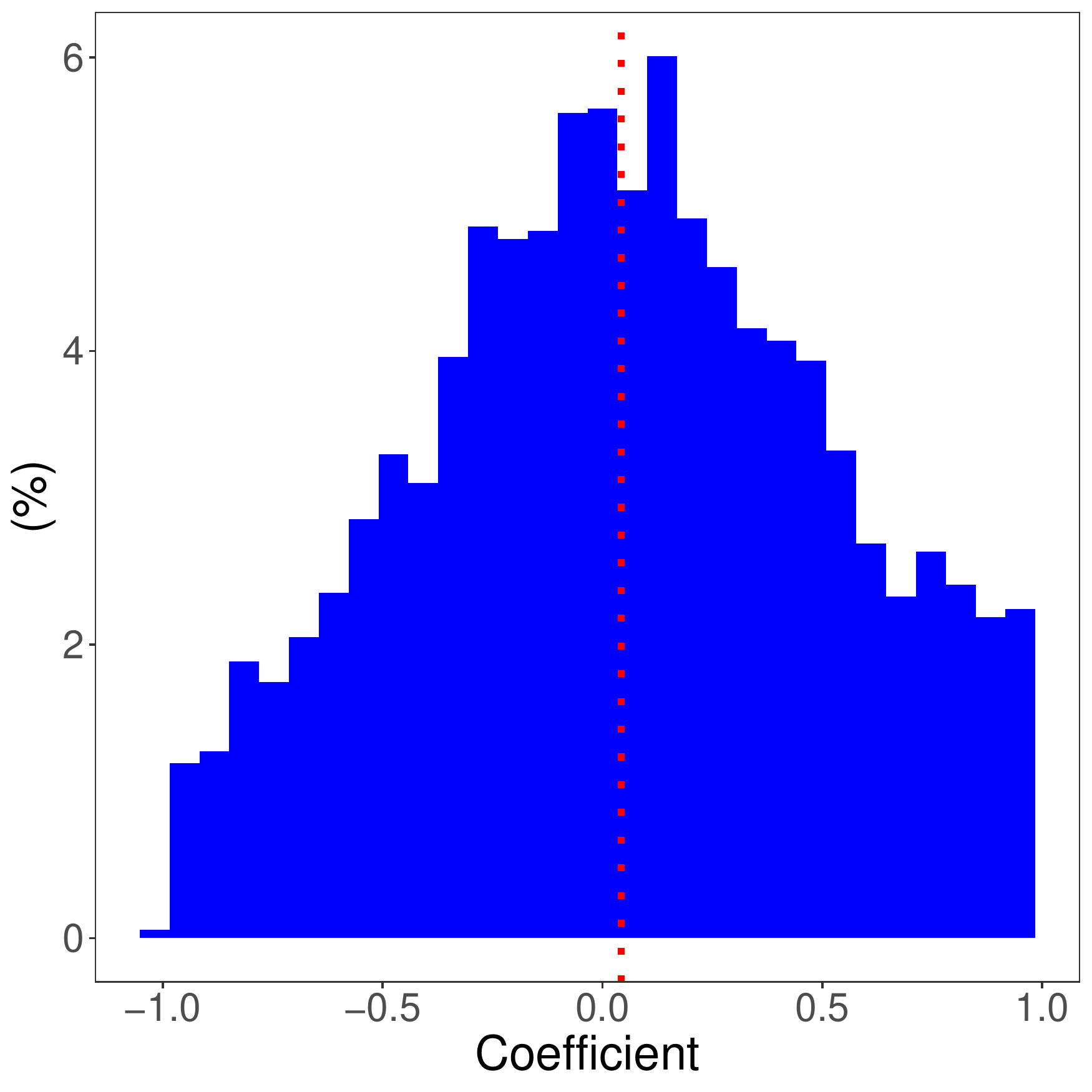}
  \caption{Profit Rates}
  \label{fig:cycle_pik}
\end{subfigure}
\label{fig:cycle_all}
\caption*{\footnotesize Note: The plots show the histograms for the average firm-specific coefficient of the cycle using a yearly specification. The red dotted line shows the average across all of these coefficients. Data were standardized before estimating the model, so the coefficients are included in the $[-1,1]$ interval.}
\end{figure}

We find evidence for acyclical markups and profit rates on average across firms. In particular, we estimate the firm-level coefficient of the GDP cycle on both markup and profit rates, $\alpha$ in Equation \eqref{eq:cycle}. Figures \ref{fig:cycle_mu} and \ref{fig:cycle_pik} show histograms of these coefficients for separate models for markup and profit rates. This distribution suggests that, on average, the cycle does not seem to affect markups and profit rates. Furthermore, we compute the percentage of significant cycle coefficients. In the case of markup rates, we find 63\% of the coefficients to be significant with a rather symmetrical share of positive (30\%) and negative (33\%) coefficients. Similarly, in the case of  profit rates 62\% of the coefficients are significant (34\% positively significant and 28\% negative significantly). 

These wide-ranging cycle coefficients may be connected to the literature on the cyclicality of markups, which has identified potential reasons for both pro- and counter-cyclical markups. In particular, the behavior of liquidity-constrained firms that adjust their customer base to stay afloat during recessions may explain counter-cyclical markups \citep{Gilchrist2017}, with larger firms being able to further smooth these liquidity constraints out \citep{hong2017customer}, whereas Keynesian demand effects may explain pro-cyclical markups \citep{Nekarda2020}. It seems that these three conjectures may have some weight in explaining the apparent average acylicality in the markup and profit rates.

\section{Robustness of findings to other modeling strategies and choices}\label{sec:robustness}

In this section, we introduce two approaches to evaluate the robustness of our main findings. In the first approach, we modify the hyperparameters associated with the priors of the Bayesian structural time series (BSTS) models, which we refer to as sensitivity analysis. In the second approach, we use the commonly-used methodology of local linear projections (LLPs) to evaluate if our main findings are robust to this fairly flexible strategy. Our sensitivity analysis corroborates our main results in Section \ref{sec:counteractual_analysis}. The LLPs approach gives similar point effects in markup rates compared to the Bayesian structural model, and a smaller effect in profit rates; however, it seems that LLPs lack power for this data as the confidence intervals seem to be too wide.

\subsection{Robustness of findings to other modeling choices: a hyper-parameter sensitivity analysis.}

We validate the findings from the BSTS model by performing a sensitivity analysis of our choice of hyperparameters. This analysis ensures that our results are not driven by this choice. As noted in Section \ref{sec:estimation}, the hyperparameters govern the prior densities of the  variances and initial values of the trend and seasonality structural components. We perform two main exercises: we either increase or decrease hyperparameters that govern the scale of the priors densities by 25\% compared to the baseline values in Table \ref{tab:priors}; by doing so, we allow for the increases in markup and profit rates to have either more or less variability \citep{Brodersen2015}. As the location hyperparameters associated with the initial values are calibrated with summary statistics of the actual series, we do not change those location hyperparameters. Table \ref{tab:Sensitivity} shows the choice of hyperparameters for each of these exercises.

\begin{table}[h]
    \centering
        \caption{Sensitivity Analysis on Hyperparameters of Bayesian Structural Model}
    \label{tab:Sensitivity}
    \begin{subtable}[t]{\linewidth} 
    \caption{Exercise 1: 25\% Increase in Hyperparameters} 
       \centering
       \label{tab:sensitivity_125}
     \begin{tabular}{cccc}
            \toprule
            Parameter & Prior density & Hyperparameter 1 & Hyperparameter 2  \\
            \midrule
              $\sigma_{\varepsilon,i}$ & Inverse-Gamma & 1.25 & 0.0125 \\
              $\sigma_{\delta,i}$ & Inverse-Gamma & 0.0125 & 40 \\
              $\sigma_{\gamma,i}$ & Inverse-Gamma &  0.0125 & 0.0125 \\
              $\delta_{i,1}$ & Normal &   $\frac{y_0-\overline{y}}{ \hat{\sigma}_y}$     & 1.25 \\ 
              $\gamma_{i,1}$ & Normal &   0 & 1.25 \\
             \bottomrule 
          \end{tabular}
 \vspace{0.3cm}
 \end{subtable}
     \begin{subtable}[t]{\linewidth} 
      \caption{Exercise 2: 25\% Decrease in Hyperparameters}    \centering
       \label{tab:sensitivity_75}
      \begin{tabular}{cccc}
            \toprule
            Parameter & Prior density & Hyperparameter 1 & Hyperparameter 2  \\
            \midrule
              $\sigma_{\varepsilon,i}$ & Inverse-Gamma & 0.75 & 0.0075 \\
              $\sigma_{\delta,i}$ & Inverse-Gamma & 0.0075 & 24 \\
              $\sigma_{\gamma,i}$ & Inverse-Gamma &  0.0075 & 0.0075 \\
              $\delta_{i,1}$ & Normal &   $\frac{y_0-\overline{y}}{ \hat{\sigma}_y}$     & 0.75 \\
              $\gamma_{i,1}$ & Normal &   0 & 0.75 \\
             \bottomrule 
          \end{tabular}
 \end{subtable}
  \begin{tablenotes}
  \item  \footnotesize   Notes:  $\sigma_{\epsilon,i}$, $\sigma_{\delta,i}$, and $\sigma_{\gamma,i}$ are the variances of the observation equation error (Equation \eqref{eq:obs_eq}), trend, and seasonal components, respectively, for a firm $i$. $\delta_{i,1}$ and $\gamma_{i,1}$ are the initial values of the trend and seasonal state components. When the prior density is inverse-gamma, hyperparameters 1 and 2 can be interpreted as the degrees of freedom and the sum of squared errors, respectively \citep[see]{Brodersen2015}. When the prior density is normal, hyperparameters 1 and 2 are the mean and standard deviation, respectively. 
  \end{tablenotes}

\end{table}

\begin{figure}[H]
\caption{Comparison of Observed and Counterfactual Markups and Profit Rates Under Different Hyperparameter choices.}
\begin{subfigure}{.5\textwidth}
  \centering 
  \includegraphics[width=0.85\linewidth]{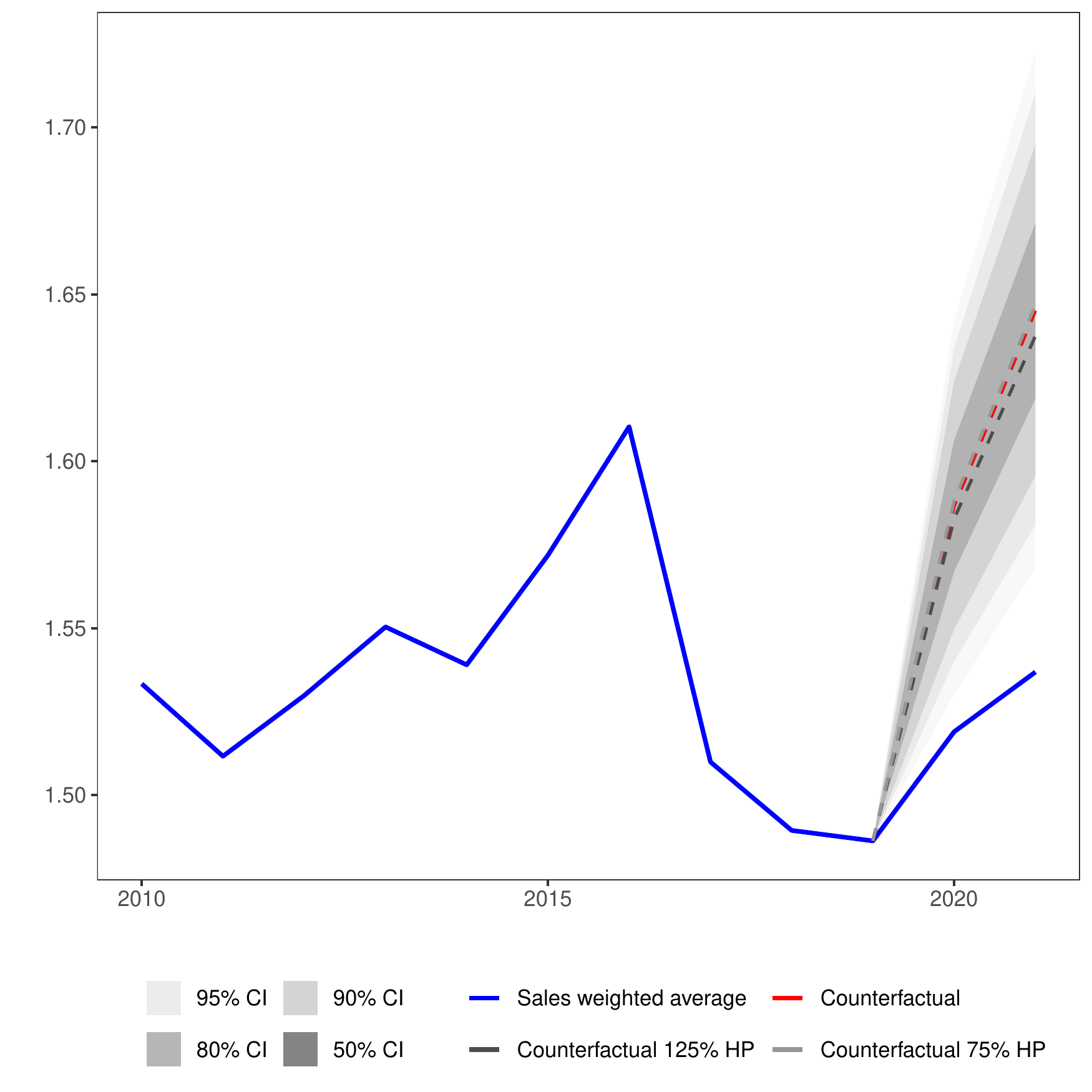}
  \caption{Markup Rate}
  \label{fig:sensitivity_mu}
\end{subfigure}
\begin{subfigure}{.5\textwidth}
  \centering 
  \includegraphics[width=0.85\linewidth]{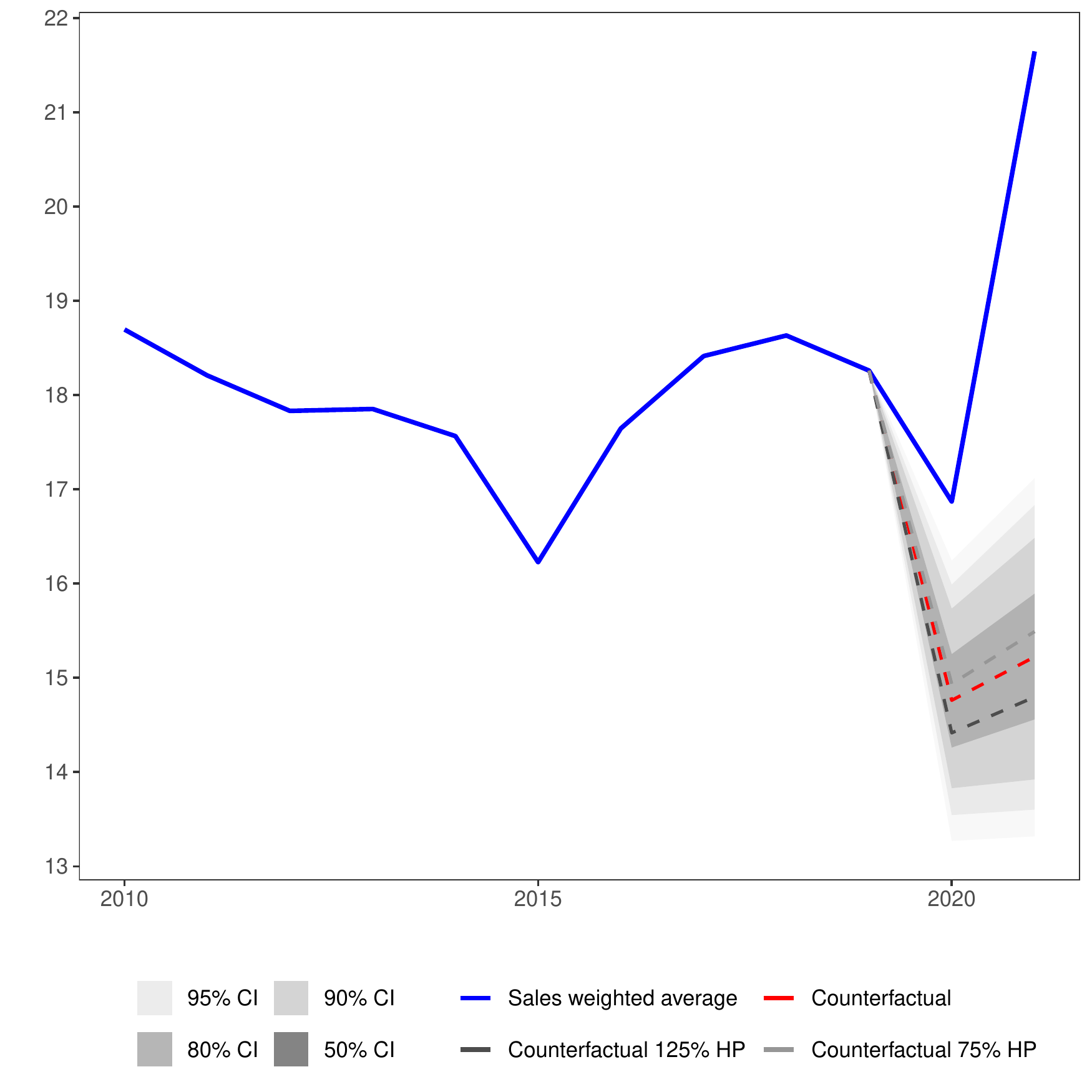}
  \caption{Profit Rate}
  \label{fig:sensitivity_pi}
\end{subfigure}
\label{fig:sensitivity}
\caption*{\footnotesize Note: The figure shows yearly sales-weighted average observed and counterfactual markup and profit rates. The solid lines represent the observed values, while the dotted lines represent the counterfactual values for 2020-2021, as calculated using firm-level local linear projection models. The shaded areas represent the 95\%, 90\%, 80\%, and 50\% (equally-tailed) credible interval for each statistic, calculated based on 5,000 posterior simulations.}
\end{figure}

Figure \ref{fig:sensitivity} shows the change in the effects on markups and profit rates under this sensitivity analysis. After either increasing or decreasing the variability of our priors, our finding that the markup rate is negatively affected for most firms does not change. Likewise, our findings for the profit rates on the impact  of the pandemic remain unchanged: on average, firms performed better than the counterfactual without the pandemic. In general, when we increase the variability of our hyperparameters, the effects are larger for most firms both in terms of markup and profit rates.  
  
\subsection{Robustness of findings to other modeling strategies: local linear projections.}

We also validate the robustness of our findings by estimating forecasts using the flexible approach of Local Linear Projections (LLPs) \citep{Jorda2005,montiel2021local}. This method creates forecasts using linear regressions of the outcome of interest on its   lags and possibly lags of other exogenous variables, which offers a flexible linear approximation of the conditional expectation of the outcome given exogenous variables. We estimate LLPs with either firm-specific projection coefficients, which we called firm-specific LLPs, or with projection coefficients constant across firms, which we called panel LLPs.

Particularly, in the case of the firm-specific LLPs, for every forecast horizon $h$ and firm $i$, we consider the linear projection of the outcome of interest, $Y_{i,t+h}$, on a vector stacking a constant and outcome lags up to period $t$, $\mathbf{X}_{it}=(1,Y_{i,t-1},Y_{i,t-2},\dots,Y_{i,t-p_i})$, with $p_i$ a positive integer,
 \begin{align}
                Y_{i,t+h} = & \ \mathbf{X}_{it}'\beta_{ih} +\xi_{ith}, \quad
                 h= \ 1,2, \ldots H;\  i=1,\dots,I, \label{eq:LLPeq}
            \end{align}
 where $\beta_{ih}$ is a $p_i+1$ vector of projection coefficients, $\xi_{ith}$ is the projection residual, and $H$ is the maximum forecast horizon, i.e., either two years or eight quarters.  Thus, Equation~\eqref{eq:LLPeq} provides the firm-specific LLPs forecast $\widehat{Y}_{i,t+h}$ for either markup or profit rates. To estimate these models, we select firm-specific lag lengths $p_i$ according to the Akaike information criterion, with a lag of up to 3 years with annual data and up to 8 quarters with quarterly data. As with BSTS, we also evaluate whether LLPs provide precise forecasts. Particularly, we find that LLPs tend to produce similarly accurate forecasts (if not slightly more accurate) compared to BSTS models according to most forecast quality statistics we consider (see Tables \ref{tab:forecast_qual} and \ref{tab:forecast_qual_LLP}).\footnote{In Appendix \ref{app:balanced_sample_forecast_quality}, we use the same sample of firms and periods to calculate forecast quality statistics for BSTS and LLPs -- we do not see relevant changes to the comparison across methods after using this sample.}

\begin{figure}[H]
\caption{Comparison of Observed and Counterfactual Profit Rates using Local Linear Projections}
\begin{subfigure}{.5\textwidth}
  \centering
  \includegraphics[width=0.85\linewidth,page=2]{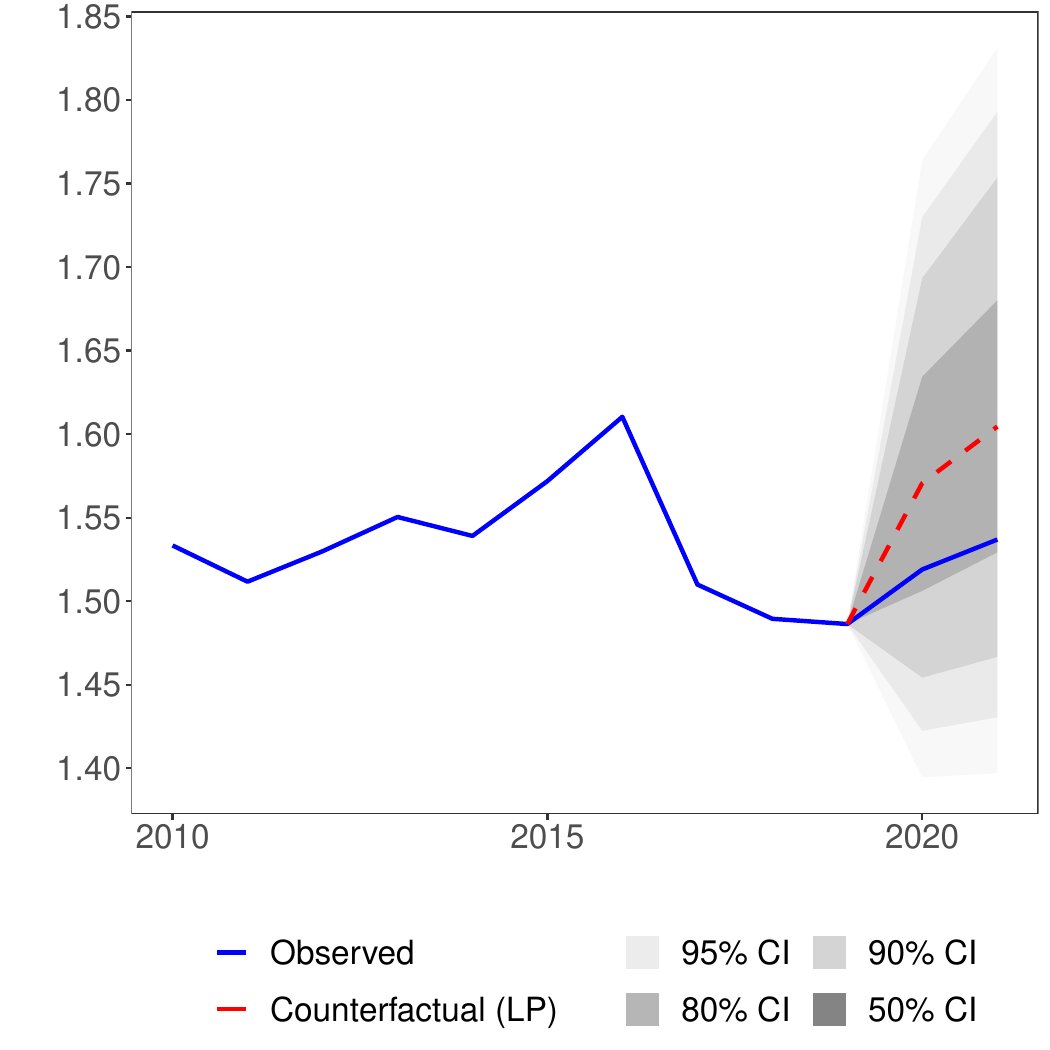}
  \caption{Markup Rate}
  \label{fig:Trend_fanchart_mu_year_lp}
\end{subfigure}%
\begin{subfigure}{.5\textwidth}
  \centering
  \includegraphics[width=0.85\linewidth,page=2]{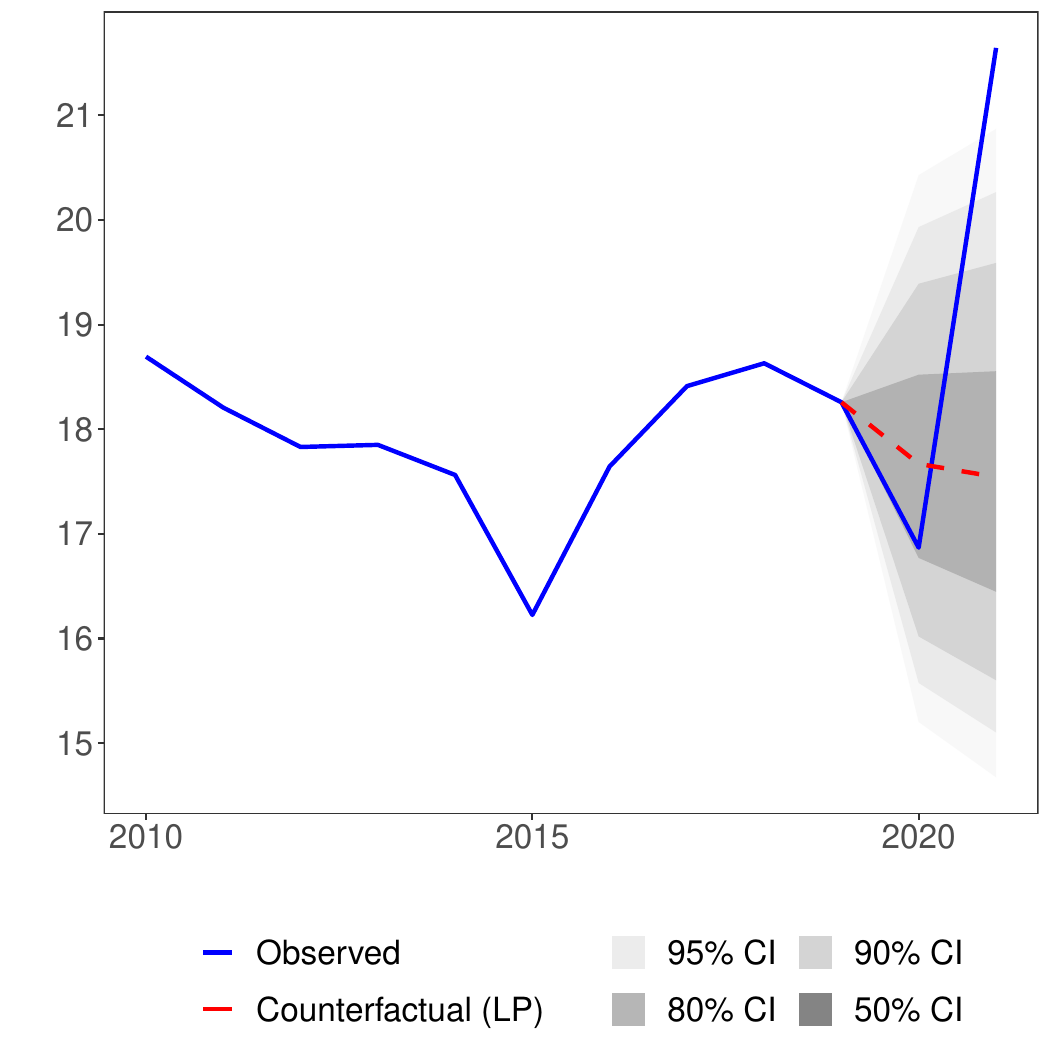}
  \caption{Profit Rate}
  \label{fig:Trend_fanchart_pik_year_lp}
\end{subfigure}%
\label{fig:Trend_fanchart_lp}
\caption*{\footnotesize Note: The figure shows yearly sales-weighted average observed and counterfactual markup and profit rates. The solid lines represent the observed values, while the dotted lines represent the counterfactual values for the period of 2020-2021, as calculated using firm-level local linear projection models. The shaded areas represent the  95\%, 90\%, 80\%, and 50\% (equally-tailed) credible interval for each statistic, calculated based on 5,000 posterior simulations.}
\end{figure}

Using LLPs, we replicate our analysis of aggregate markup and profit rates in Section \ref{sec:results_aggregate}. In particular, Figure \ref{fig:Trend_fanchart_lp} shows the counterfactual markup and profit rates estimated using firm-specific LLPs. These LLPs forecasts result in similar point estimates for markup rates when compared to the BSTS, but with higher point estimates for profit rates, which suggests that the firm-specific LLPs predict a smaller effect of the pandemic on profit rates.

To test for significant effects using the firm-specific LLPs, we compute aggregate confidence intervals using the bootstrap method. Specifically, we create 10,000 resamples with replacement by sampling firms in each year of the pandemic. We then create a 95\% confidence interval by taking the 2.5th and 97.5th percentiles of the average response across the 10,000 resamples. These confidence intervals appear to be wide, with a larger range than the BSTS's credible intervals, which only allows us to reject the null hypothesis of no significant effect on the aggregate profit rate in 2021.

\begin{table}[h]
    \centering
      \caption{\label{tab:forecast_qual_LLP} Forecast Quality Statistics of Local-Linear Projections}
    \label{tab:LLPForecast}
    \begin{subtable}[t]{\linewidth} 
    \caption{Firm-Specific Local-Linear Projection} 
       \centering
       \label{tab:LLPcomp}
     \begin{tabular}{llrrrrcp{1cm}r}
    \toprule
    Variable & Year & ME & RMSE & MAE & MAES & MAPE & Median MAPE & $n$\\
    \midrule
    Markup rate & 2018 & -0.019 & 0.697 & 0.264 & 0.152 & 0.150 & 0.058 & 2858 \\
    Markup rate & 2019 & -0.086 & 0.821 & 0.358 & 0.211 & 0.217 & 0.091 & 2751 \\
    Profit rate & 2018 & 23.490 & 488.429 & 30.526 & 3.689 & 2.294 & 0.203 & 2858 \\
    Profit rate & 2019& 16.957 & 206.812 & 29.079 & 4.244 & 3.029 & 0.313 & 2762 \\
    \bottomrule
    \end{tabular}
 \vspace{0.3cm}
 \end{subtable}
     \begin{subtable}[t]{\linewidth} 
      \caption{Panel Local-Linear Projection}    \centering
       \label{tab:LLPcomppanel}
      \begin{tabular}{llrrrrcp{1cm}r}
    \toprule
    Variable & Year & ME & RMSE & MAE & MAES & MAPE & Median MAPE & $n$\\
    \midrule
    Markup rate & 2018 & 0.017 & 0.555 & 0.223 & 0.128 & 0.120 & 0.054 & 2808\\
    Markup rate & 2019& -0.027 & 0.743 & 0.309 & 0.182 & 0.180 & 0.081 & 2808 \\
    Profit rate & 2018 & 5.809 & 96.882 & 16.257 & 1.965 & 1.590 & 0.318 & 2843 \\
    Profit rate & 2019 & 2.536 & 75.113 & 18.855 & 2.752 & 2.305 & 0.391 & 2825 \\
    \bottomrule
    \end{tabular}
 \end{subtable}
  \begin{tablenotes}
  \item  \footnotesize   Notes:  The table introduces common forecast quality measures computed for markup rates and profit rates for 2018--2019 with firm-level models estimated using information up to 2017. The measures computed are:  mean error (ME),  root mean square error (RMSE),  mean absolute error (MAE),  standardized mean absolute error  (MAES),  and the mean absolute percentage error (MAPE). Median MAPE is the median of the MAPEs across firm-level MAPEs. $n$ is the number of firms. 
   \end{tablenotes}
\end{table}

Additionally, we extend our LLPs framework to analyze whether potential unobserved heterogeneity in the projection coefficient (Equation \eqref{eq:LLPeq}) plays a role in the forecast accuracy of LLPs.\footnote{ When 
favoring either firm-specific estimation (Equation \eqref{eq:LLPeq}) instead of pooled estimation (Equation \eqref{eq:LLPeqpanel}),
\cite{pesaran2022forecasting} argues that either (i) a high level of heterogeneity in the projection coefficient, (ii) a large number of observed periods for each firm, or (iii) models with outcome lags will favor firm-specific estimation.} 
 
\begin{figure}[tp]
\caption{Comparison of Observed and Counterfactual Profit Rates using Panel Local Linear Projections}
\begin{subfigure}{.5\textwidth}
  \centering
  \includegraphics[width=0.85\linewidth,page=3]{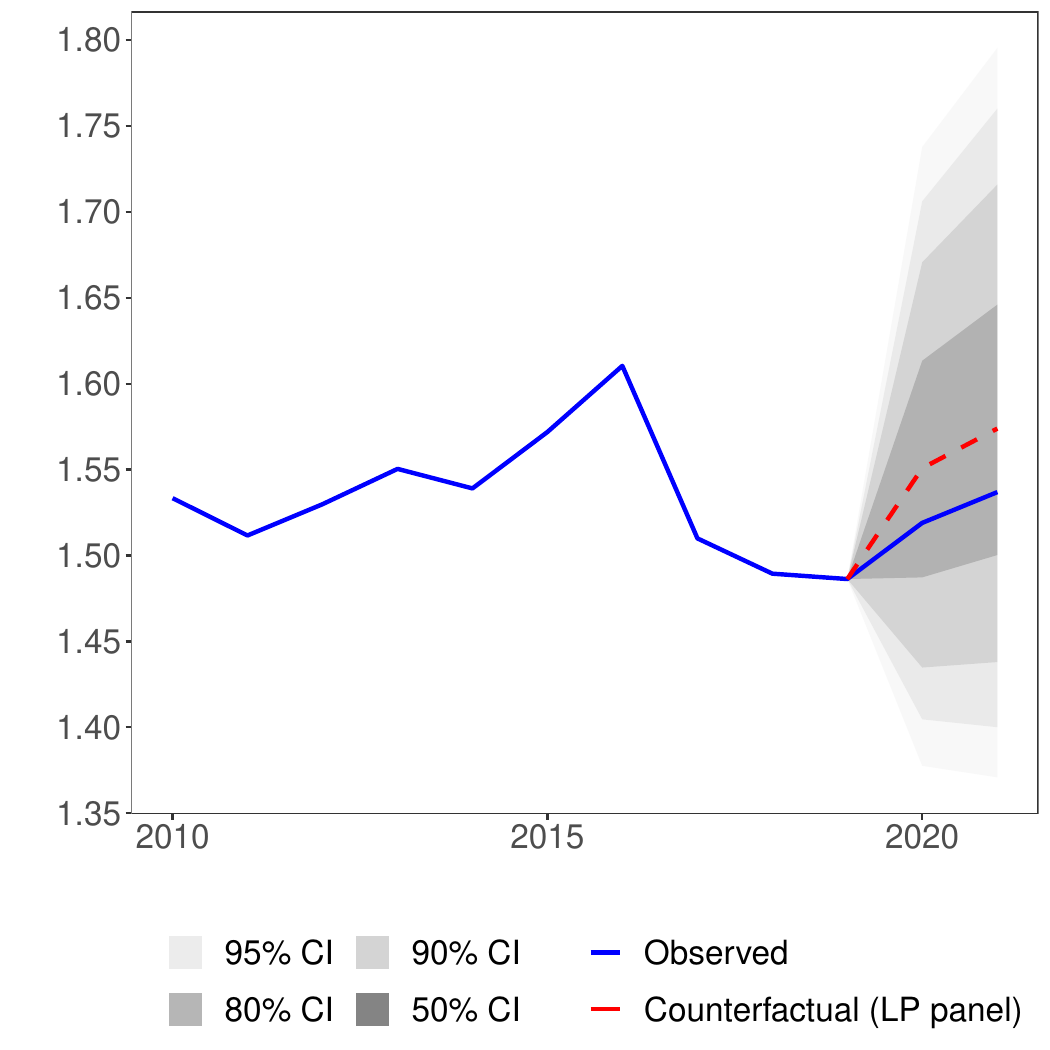}
  \caption{Markup Rate}
  \label{fig:Trend_fanchart_mu_year_lppanel}
\end{subfigure}%
\begin{subfigure}{.5\textwidth}
  \centering
  \includegraphics[width=0.85\linewidth,page=3]{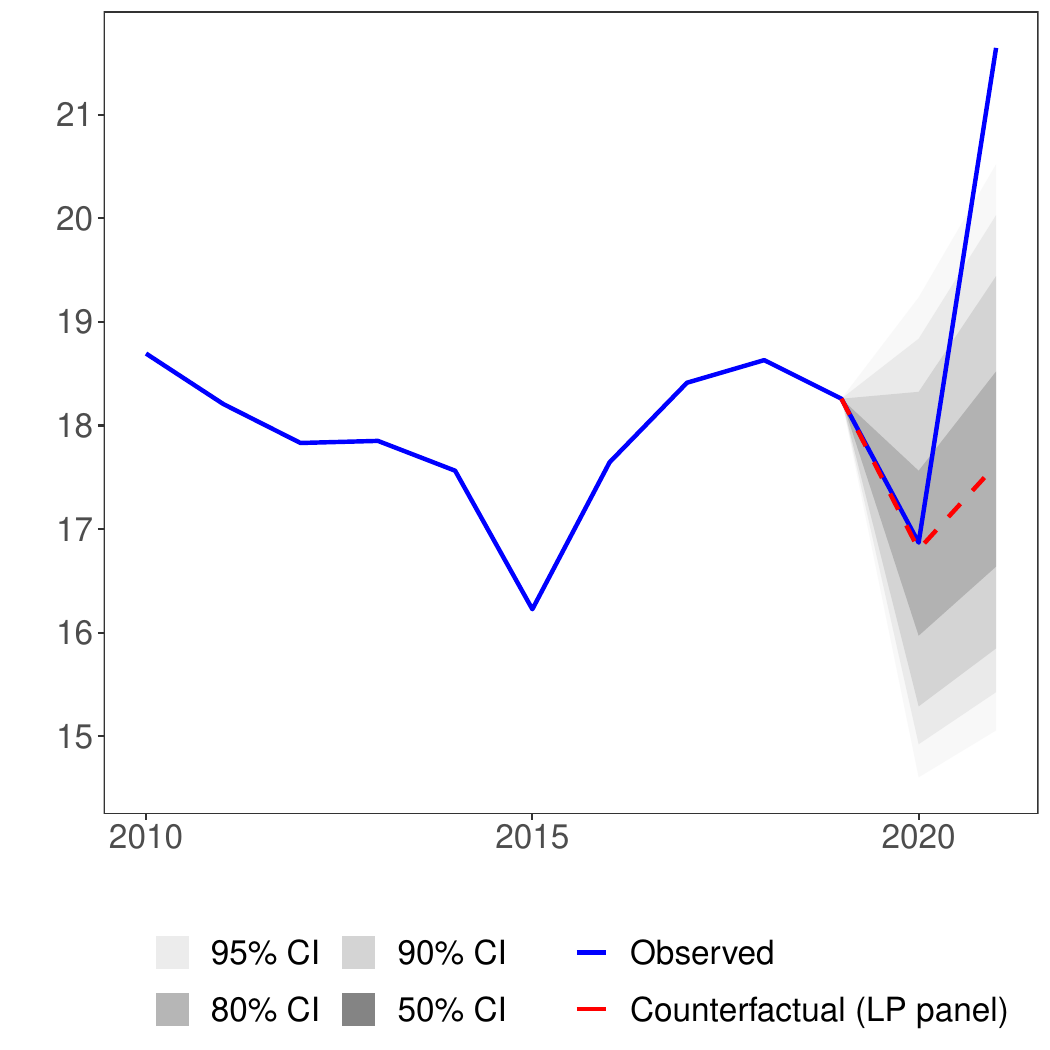}
  \caption{Profit Rate}
  \label{fig:Trend_fanchart_pik_year_lppanel}
\end{subfigure}%
\label{fig:Trend_fanchart_lppanel}
\caption*{\footnotesize Note: The figure shows yearly sales-weighted average observed and counterfactual markup and profit rates. The solid lines represent the observed values, while the dotted lines represent the counterfactual values for the period of 2020-2021, as calculated using a panel data local linear projection model. The shaded areas represent the  95\%, 90\%, 80\%, and 50\% (equally-tailed) credible interval for each statistic, calculated based on 5,000 posterior simulations.}
\end{figure}

We refer to this extension as panel LLPs.
To this purpose, we extend the linear projection in Equation \eqref{eq:LLPeq} to incorporate a constant projection coefficient across firms and firm fixed effects as follows, 
\begin{align}
                Y_{i,t+h} = & \ \mathbf{X}_{it}'\beta_{h} + u_{ih} +\zeta_{ith};               \quad    h= \ 1,2, \ldots, H;\ i=1,\dots,I; \label{eq:LLPeqpanel} 
            \end{align}
where $\mathbf{X}_{it}=(1,Y_{i,t-1},Y_{i,t-2},\dots,Y_{i,t-p})$ is a $p+1$ vector stacking a constant and lags of the outcome $Y_{it}$, $\beta_{h}$ is a panel-level projection coefficient, $u_{ih}$ is a firm fixed effect, and $\zeta_{ith}$ is a projection residual.\footnote{In Equation \eqref{eq:LLPeqpanel}, we do not include time fixed effects as we cannot forecast time fixed effects for 2020 and 2021 which makes it infeasible to compute forecasts for the outcome of interests.} Similarly to the firm-specific LLPs specification, we choose the lag length $p$ based on the Akaike information criterion.\footnote{We also estimate the following specifications not reported here: (i) Equation \eqref{eq:LLPeqpanel} replacing firm-fixed effects with industry fixed effects, (ii) Equation \eqref{eq:LLPeqpanel} without the firm fixed effect $u_{ih}$ (or pooled specification), and (iii) Equation \eqref{eq:LLPeqpanel} with industry-specific projection coefficients $\beta_h$. We find similar results in these specifications to our baseline specification in Equation~\eqref{eq:LLPeqpanel}.} Using this extended model, we compute individual forecasts for each firm and bootstrapped standard errors as before, and plot these estimates in Figure \ref{fig:Trend_fanchart_lppanel}.\footnote{It seems that panel LLPs are more accurate than firm-specific LLPs (see Panel (a) and (b) in Table \ref{tab:forecast_qual_LLP}), which suggests that there is no high unobserved heterogeneity in the projection coefficient, $\beta_{ih}$. In Appendix \ref{app:balanced_sample_forecast_quality}, we use the same sample of firms and periods to calculate forecast quality statistics for BSTS and LLPs -- we do not see relevant changes to the comparison across methods after using this sample.}

The panel LLPs results are quite similar to the firm-level LLPs, i.e., there are no effects in markups and profit rates but positive effects in profit rates in 2021 (see Figures \ref{fig:Trend_fanchart_lp} and \ref{fig:Trend_fanchart_lppanel}). In summary, it seems that the LLPs have limited power for this data, as their confidence intervals are relatively wide and their forecast quality statistics indicate they are not as precise as the BSTS.

\section{Conclusion}

We assess the impact of the COVID-19 pandemic on market power and profitability. Specifically, we compare the realized series of markup and profit rates to a forecast based on information up to the beginning of the pandemic. Essentially, we argue that this forecast represents a counterfactual of what would have happened if the pandemic had never occurred. Our counterfactual reveals that the COVID pandemic adversely affected markup rates for the average firm in 2020 and 2021. Likewise, we find that most firms had record profit rates that are significant in the sense that pre-pandemic trends cannot explain those trends. 

We estimate counterfactuals using the arguably flexible framework of Bayesian structural time series (BSTS), which can accommodate multiple sources of variation, including local trends and seasonality. On the one hand, these estimations suggest that markup rates were 4.3\% and 6.6\% lower than their counterfactuals in 2020 and 2021, respectively. On the other hand, the counterfactual suggests that had the pandemic not happened, the profit rate would have been 15.2\% in 2021 which is 6.4 percentage points lower than the observed value in 2021. Furthermore, we find that the effects of the pandemic are statistically significant.

Our findings are largely robust to changes in our modeling strategy. Particularly, we evaluate the plausibility of our assumptions and the robustness of our results by implementing three strategies. First, specification and goodness-of-fit tests support the assumptions and the quality of the model. Second, our results are largely resilient to changes to modeling choices, such as different data frequencies (either quarterly or yearly) or changing the hyperparameters of the BSTS priors. Third, similar findings follow from counterfactuals constructed from other modeling strategies, such as local linear projections, which is a commonly used and flexible method to compute forecasts. In summary, specification and goodness-of-fit tests validate our assumptions, and other modeling strategies indicate that our results are robust.   

Furthermore, this paper uncovers the heterogeneity of these effects in terms of pre-pandemic baseline characteristics such as markup rates, firm size, stock-exchange tenure, employment, profitability, and market shares. Our findings show that companies with a longer history on the stock exchange tend to have a larger impact on markup rates. Additionally, companies with high sales or low costs tend to have a significant positive impact on profit rates. 

We complement this heterogeneity analysis with a breakdown of the effects of the pandemic by industry. Particularly, industries such as information, real estate, and chemical manufacturing had lower markups than expected, while warehousing and entertainment had higher markups. A similar analysis of profit rates by industry shows that transportation, entertainment, and hospitality tended to have lower profit rates than expected, whereas warehousing and real estate had higher profit rates. 

Our results have implications for the current academic and economic policy debate on the impacts of the pandemic on businesses. For instance, this work can help guide future research on the impact of fiscal and monetary COVID relief policies. Likewise, understanding the changes, and the subsequent policy response, in markup and profit rates during the pandemic can help policymakers guide strategies for addressing the impacts of the pandemic on markets. Additionally, future research will uncover whether firms may have reacted to the increased uncertainty brought by the pandemic by reducing markups in order to maintain or expand their customer base. Finally, the pandemic-induced lockdowns may have created a reallocation within firms' input use, such as the rise of remote workers \citep{Bloom2015,Bloom2020}, which could have decreased selling, general, and administrative expenses, while increasing the value of the costs of goods sold.
 
\newpage
\bibliography{Bibliography} 
 
\appendix

\renewcommand\thefigure{\thesection.\arabic{figure}}    
\setcounter{figure}{0}  
\setcounter{table}{0}
\renewcommand{\thetable}{A\arabic{table}}

\newpage

\section{Data and summary statistics}
\label{app:data}

To calculate our measures of markup and profit rates, we remove all firms with negative sales, cost of goods sold (COGS), and selling, general and administrative expenses (SG\&A). We also remove records below the first and above the 99th yearly percentiles of the COGS and SG\&A to sales ratios per year. We removed records without either Compustat firm identification numbers or industry codes. After the previous steps, we also remove values below the first and above the 99th percentile of two cost-shares (the ratio of COGS to the total of COGS and capital expenditure, and the ratio of COGS to the total of COGS, capital expenditure, and SG\&A). Before computing these ratios, we deflate the respective series using the GDP deflator with 2010 as the base year. 

In addition to these restrictions on the sample, we further impose the following conditions to estimate counterfactuals reliably. First, we require firms with data for both the pre-pandemic and the pandemic, so we remove 18,022 (17,003) firms that exited the panel before or in 2019, of which 79\% left before 2010. Then, we eliminate 191 (431) firms that either entered the panel during or after 2020 or had less than two years of available data before this year. Finally, to compute counterfactuals for firms with at least 3 years of observations, we drop an additional 802 (925) firms. After these filtering steps, our sample comprises 3,611 (3,192) publicly-traded firms with yearly (quarterly) data, no extreme values in costs ratios, information for at least 5 years before the pandemic and operating during it.

\begin{table}[H]
    \centering
    \caption{Summary Statistics}
    \label{tab:summary_stats}
    \begin{threeparttable}
        \begin{tabular}{lcccc} \hline
            Variable & Acronym & Mean & Median & N. obs \\ \hline
            \multicolumn{5}{l}{\textit{Panel A: Sample 1955--2021}} \\
             \hspace{0.2cm}  Sales & Sale, PQ & 2,125,657 & 159,894 & 269,899\\
             \hspace{0.2cm}  Cost of goods sold & COGS, V & 1,414,280 & 94,946& 269,899\\
             \hspace{0.2cm}  Capital stock & PPEGT,K & 1,655,290 & 62,386 & 269,899 \\
             \hspace{0.2cm}  SG\&A & XSG\&A,X & 380,988  & 32,308 & 269,899 \\
             \hspace{0.2cm}  Wage bill & CXLR,WL & 1,088,658& 117,408 & 32,982 \\
             \hspace{0.2cm}  Employment & Emp,L & 8,910 & 900 & 241,083\\ 
            
            \multicolumn{5}{l}{\it Panel B: Sample 1955--2016}\\
             \hspace{0.2cm}  Sales & Sale, PQ & 1,953,210 & 149,483 & 248,378 \\
             \hspace{0.2cm}  Cost of goods sold & COGS, V &  1,306,434 & 90,116 & 248,378  \\
             \hspace{0.2cm}  Capital stock & PPEGT,K & 1,474,565 & 58,178 & 248,378  \\
             \hspace{0.2cm}  SG\&A & XSG\&A,X & 349,539& 30,043 &248,378 \\
             \hspace{0.2cm}  Wage bill & CXLR,WL & 1,106,980 & 131,148& 28,321 \\
             \hspace{0.2cm}  Employment & Emp,L & 8,426 & 866 & 221,685 \\ 
            
            \multicolumn{5}{l}{\it Panel C: Sample 1955-2016 from \citet{DeLoecker2020}} \\
            \hspace{0.2cm} Sales & Sale, PQ & 1,922,074 & 147,806 & 247,644 \\
            \hspace{0.2cm} Cost of goods sold & COGS, V & 1,016,550 & 55,384 & 247,644 \\
            \hspace{0.2cm} Capital Stock & PPEGT,K & 1,454,210 & 57,532 & 247,644 \\
            \hspace{0.2cm} SG\&A & XSG\&A,X & 342,805 & 29,682 & 247,644 \\
            \hspace{0.2cm} Wage bill & CXLR,WL & 1,093,406 & 130,486 & 28,116 \\
            \hspace{0.2cm} Employment & Emp,L & 8,363 & 863 & 221,121 \\
             \hline
        \end{tabular}
        \begin{tablenotes}
         \item \noindent \footnotesize Notes: Thousands yearly US\$ deflated using the GDP deflator with base year 2010. In column 2 of the table, we also report the Compustat acronym to keep track of the specific variables with respect to that dataset.
        \end{tablenotes}
    \end{threeparttable}
\end{table}

\newpage

\section{Counterfactual analysis using cost-of-goods weighting}

\label{app:cogsweighting}

\begin{figure}[H]
\caption{Comparison of Observed and Counterfactual Markup and Profit Rates, weighting by Cost of Goods Sold (COGS) }
\begin{subfigure}{.5\textwidth}
  \centering
  \includegraphics[width=0.85\linewidth, page=4]{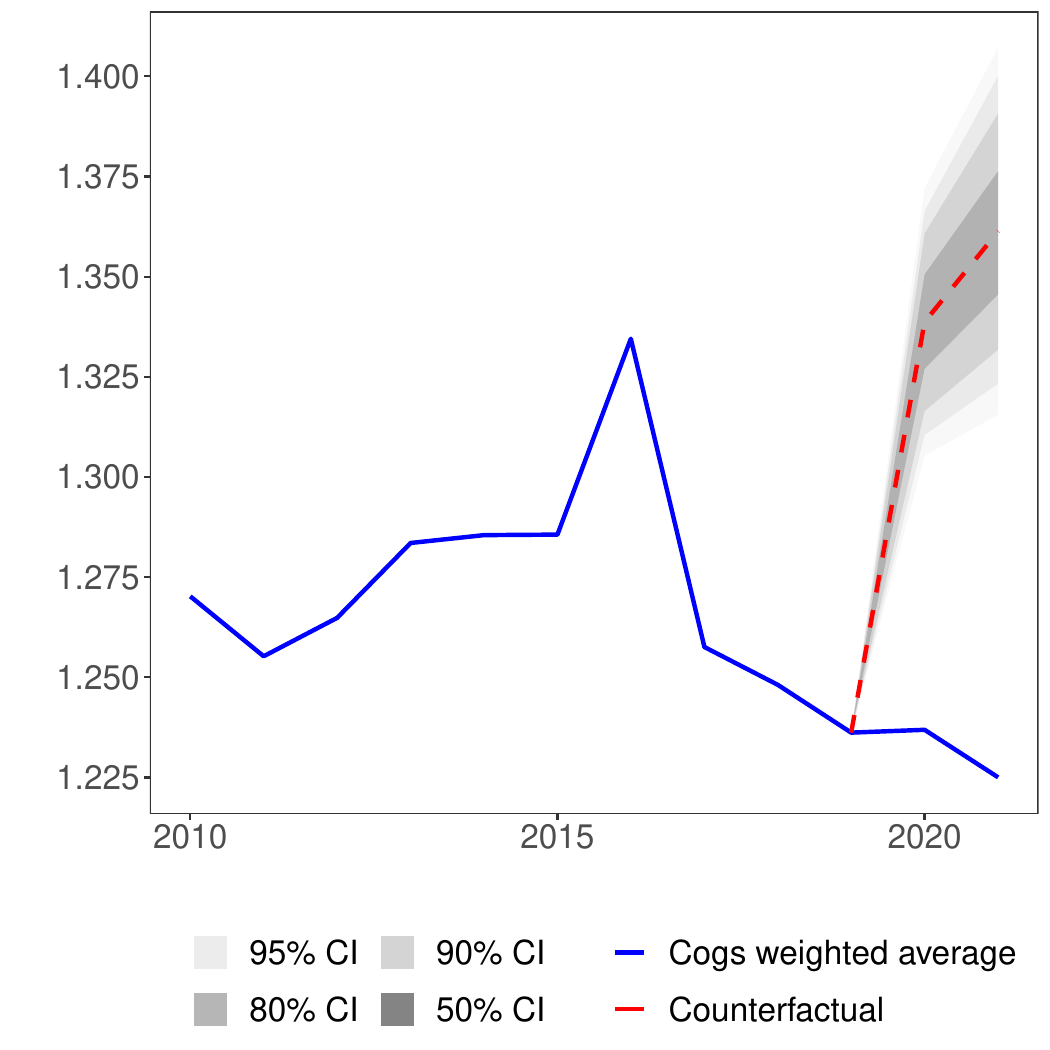}
  \caption{Markup Rate}
  \label{fig:Trend_fanchart_mu_year_cogs}
\end{subfigure}%
\begin{subfigure}{.5\textwidth}
  \centering
  \includegraphics[width=0.85\linewidth,page=4]{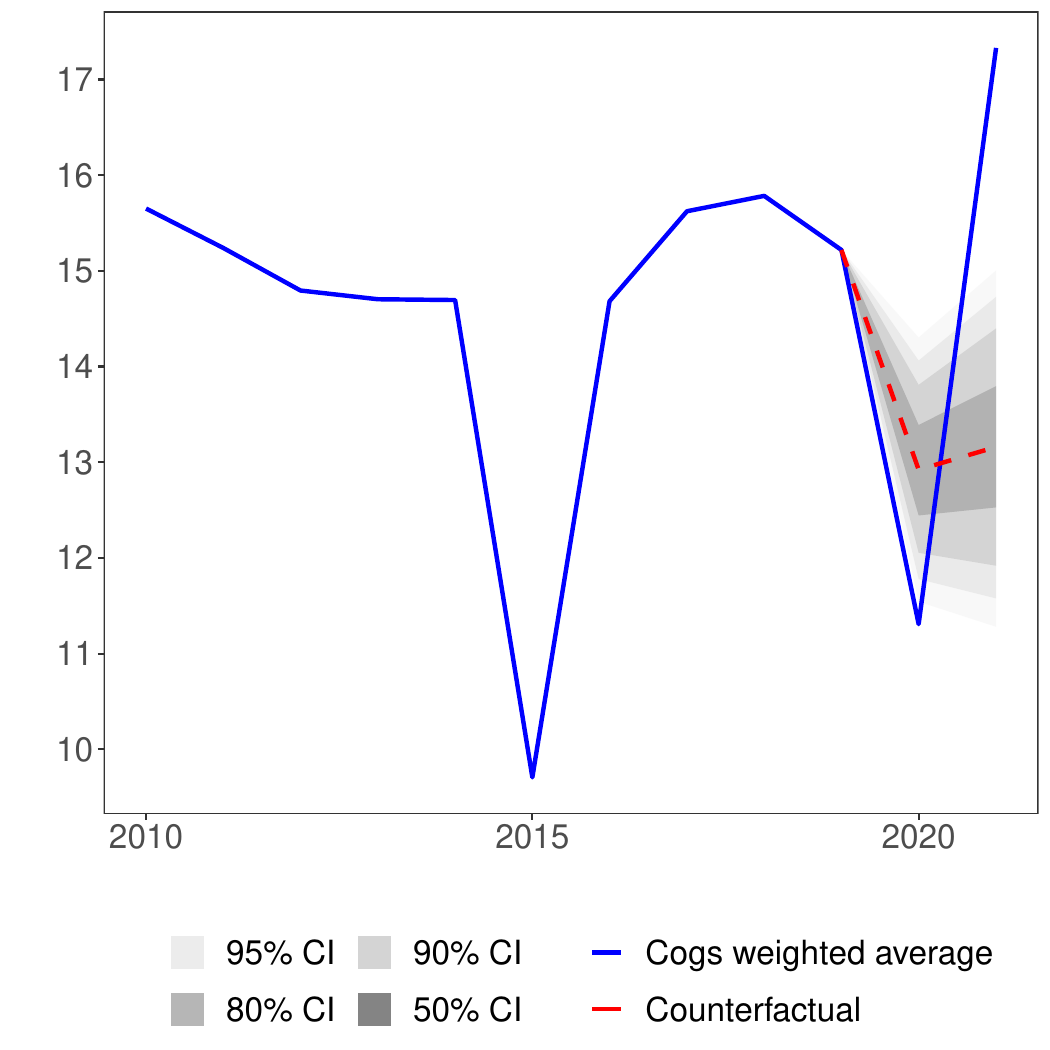}
  \caption{Profit Rate}
  \label{fig:Trend_fanchart_pik_year_cogs}
\end{subfigure}
\label{fig:Trend_year_cogs}
\caption*{\footnotesize Note: Note: The figure shows the cogs-weighted aggregate mean of the yearly observed and counterfactual markups (Panel (a)) and profit rates (Panel (b)). The solid lines represent the observed values, while the dotted lines represent the counterfactual values for 2020-2021, as calculated using firm-level Bayesian structural time series models. The shaded areas represent the  95\%, 90\%, 80\%, and 50\% (equally-tailed) credible interval for each statistic, calculated based on 5,000 posterior simulations.}
\end{figure}

In this section, we analyze the impact of our choice of weights on the results for the aggregate markup and profit rates. In Figures \ref{fig:Trend_mu_year} and \ref{fig:Trend_pik_year} we use sales weights, thus giving more importance to companies with higher sales. Alternatively, one can use costs of good sold (COGS) as weights instead of sales as presented in Figures \ref{fig:Trend_year_cogs}. Therefore, we give more importance to firms or industries with high cost structure, such as aviation, hospitality, or oil and gas. The results for markups remain similar to those using sales-weighting, i.e., the average firm has lower realized markups than its counterfactual markups. 

On the other hand, when weighting by COGS, the results for profit rate become mixed: firms had negative effects in 2020 and exhibited a strong recovery in 2021. This is more likely related to the weighting: high cost firms seem to have lost profits during 2020, whereas low cost firms increased their profits (sales-weighting case). 

\subsection{Forecast performance for BSTS and LLPs under  a balanced sample}
\label{app:balanced_sample_forecast_quality}

This subsection presents the measures on \tablename~\ref{tab:forecast_qual} and \tablename~\ref{tab:forecast_qual_LLP}, but considering a fully balanced panel. The results show that the presence of unbalanced panels does not change the relative performance of each forecasting strategy.
\begin{table}[h]
      \caption{\label{tab:forecast_qual_bal} Forecast Quality Statistics Across Models, Balanced Sample.} 
            \centering  
     \begin{subtable}[t]{\linewidth} 
    \caption{Firm-Specific Bayesian Time Series Models} 
       \centering
       \label{tab:BSTScomp_bal} 
     \begin{tabular}{llrrrrcp{1cm}r} 
    \toprule
    Variable & Year & ME & RMSE & MAE & MAES & MAPE & Median MAPE & $n$\\
    \midrule
    Markup rate & 2018 & -0.015 & 0.790 & 0.361 & 0.210 & 0.200 & 0.105 & 2808 \\
    Markup rate & 2019 & -0.061 & 0.844 & 0.384 & 0.230 & 0.233 & 0.120 & 2693 \\
    Profit rate & 2018 & 19.632 & 146.502 & 27.645 & 3.224 & 2.957 & 0.333 & 2843 \\
    Profit rate & 2019& 16.809 & 148.274 & 28.480 & 3.878 & 3.608 & 0.367 & 2718 \\
    \bottomrule
    \end{tabular} 
 \vspace{0.3cm}
 \end{subtable}
    \begin{subtable}[t]{\linewidth} 
    \caption{Firm-Specific Local-Linear Projection} 
       \centering
       \label{tab:LLPcomp_bal} 
     \begin{tabular}{llrrrrcp{1cm}r}
    \toprule
    Variable & Year & ME & RMSE & MAE & MAES & MAPE & Median MAPE & $n$\\
    \midrule
    Markup rate & 2018 & -0.023 & 0.644 & 0.248 & 0.144 & 0.144 & 0.057 & 2808 \\
    Markup rate & 2019 & -0.086 & 0.767 & 0.337 & 0.203 & 0.206 & 0.089 & 2693 \\
    Profit rate & 2018 & 22.345 & 488.279 & 29.279 & 3.414 & 2.280 & 0.201 & 2843 \\
    Profit rate & 2019& 16.425 & 206.869 & 27.893 & 3.798 & 3.032 & 0.309 & 2718 \\
    \bottomrule
    \end{tabular} 
 \vspace{0.3cm}
 \end{subtable} 
 \begin{subtable}[t]{\linewidth} 
      \caption{Panel Local-Linear Projection}    \centering
       \label{tab:LLPcomppanel_bal} 
            \begin{tabular}{llrrrrcp{1cm}r}
    \toprule
    Variable & Year & ME & RMSE & MAE & MAES & MAPE & Median MAPE & $n$\\
    \midrule
    Markup rate & 2018 & 0.017 & 0.555 & 0.223 & 0.128 & 0.120 & 0.054 & 2808\\
    Markup rate & 2019& -0.030 & 0.721 & 0.294 & 0.176 & 0.172 & 0.078 & 2693 \\
    Profit rate & 2018 & 5.809 & 96.882 & 16.257 & 1.896 & 1.590 & 0.318 & 2843 \\
    Profit rate & 2019 & 2.679 & 73.403 & 18.088 & 2.463 & 2.282 & 0.373 & 2718 \\
    \bottomrule 
    \end{tabular} 
 \end{subtable}    
        \justifying   \footnotesize
      Notes:  The table introduces common forecast quality measures computed for markup rates and profit rates for 2018--2019 with firm-level models estimated using information up to 2017. The measures computed are:  mean error (ME),  root mean square error (RMSE),  mean absolute error (MAE),  standardized mean absolute error  (MAES),  and the mean absolute percentage error (MAPE). Median MAPE is the median of the MAPEs across firm-level MAPEs. $n$ is the number of firms and the sample is balanced to consider the same observations across exercises.   
\end{table}

\section{Quarterly Analysis}
\label{sec:further_analysis}

\begin{figure}[H]
\caption{Counterfactual Analysis of Average and Quartiles of Markup Rates}
\begin{subfigure}{.5\textwidth}
  \centering
  \includegraphics[width=0.85\linewidth]{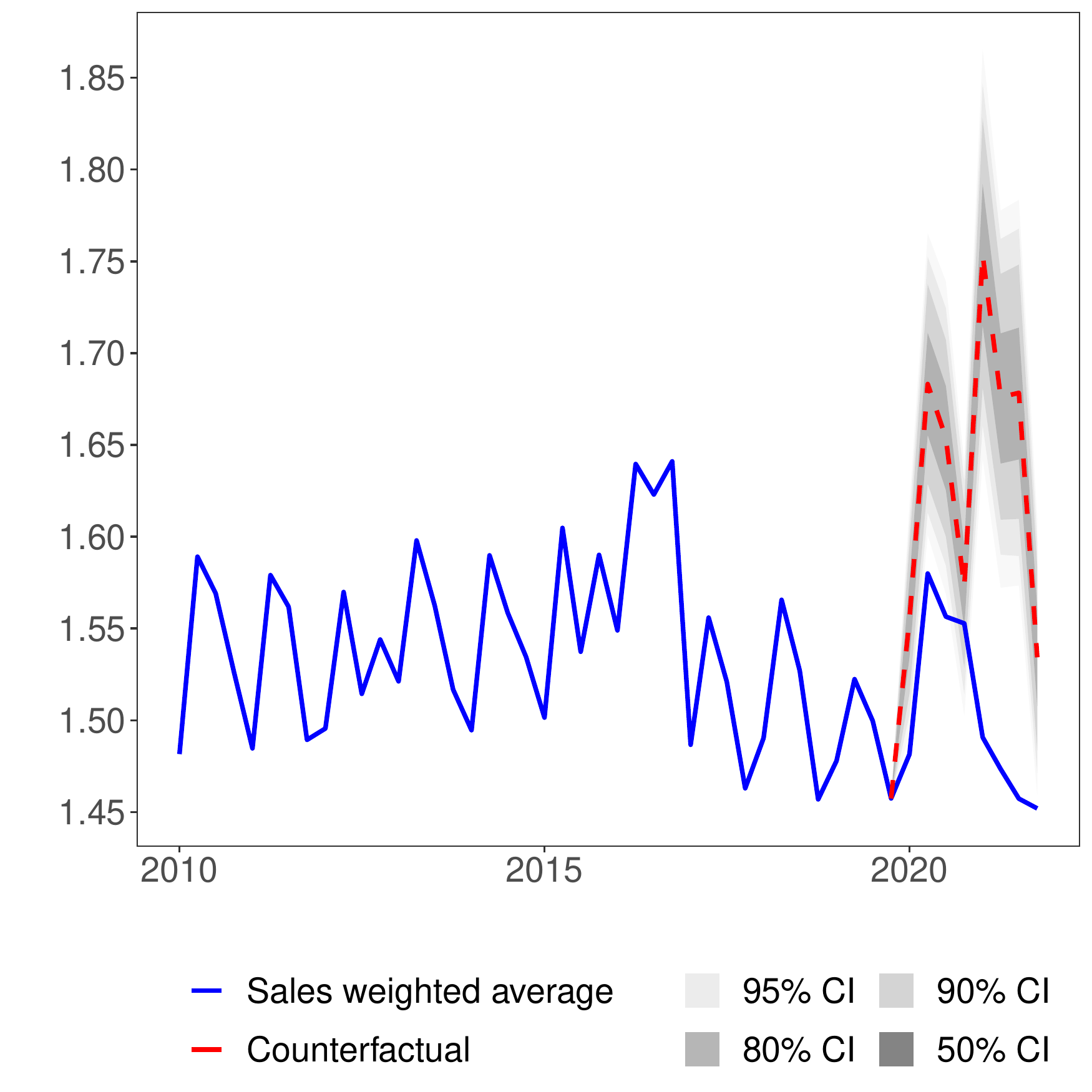}
  \caption{Average}
  \label{fig:Trend_fanchart_mu_quarter}
\end{subfigure}%
\begin{subfigure}{.5\textwidth}
  \centering
  \includegraphics[width=0.85\linewidth]{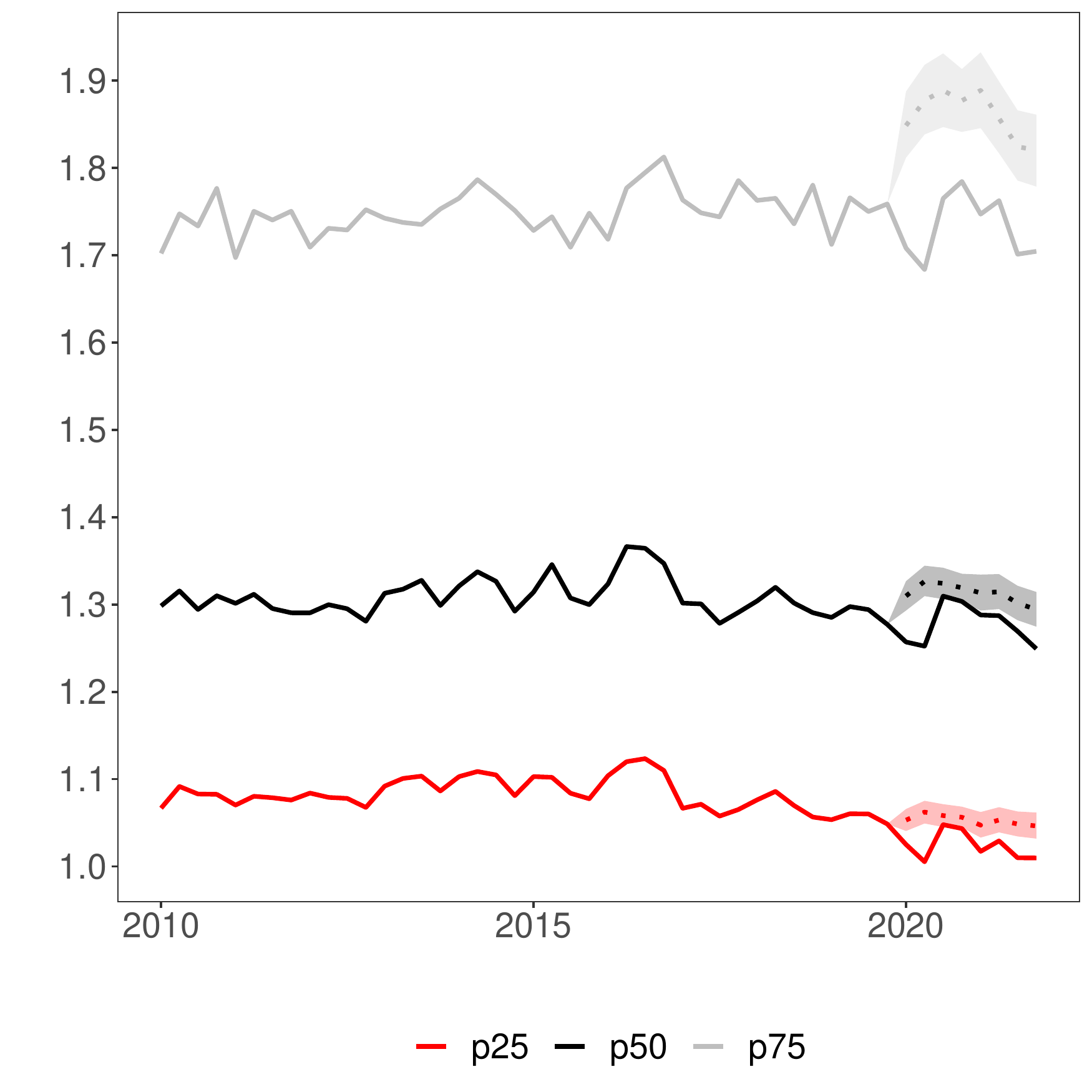}
  \caption{Quartiles}
  \label{fig:Trend_distr_mu_quarter}
\end{subfigure}
\label{fig:Trend_mu_quarter}
\caption*{\footnotesize Note: The figure shows four aggregate statistics of the quarterly observed and counterfactual markup rates. The statistics are the sales-weighted average (shown in Panel (a)), the first quartile (shown in Panel (b) in red), the second quartile (shown in Panel (b) in black), and the third quartile (shown in Panel (b) in gray). The solid lines represent the observed values, while the dotted lines represent the counterfactual values for 2020-2021, as calculated using firm-level Bayesian structural time series models. The shaded areas represent the  95\%, 90\%, 80\%, and 50\% (equally-tailed) credible interval for each statistic, calculated based on 5,000 posterior simulations on Panel (a). 95\% (equally-tailed) credible intervals are displayed for the statistics on Panel (b). }
\end{figure}

\begin{figure}[H]
\caption{Counterfactual Analysis of Average and Quartiles of Profit Rates}
\begin{subfigure}{.5\textwidth}
  \centering
  \includegraphics[width=0.85\linewidth]{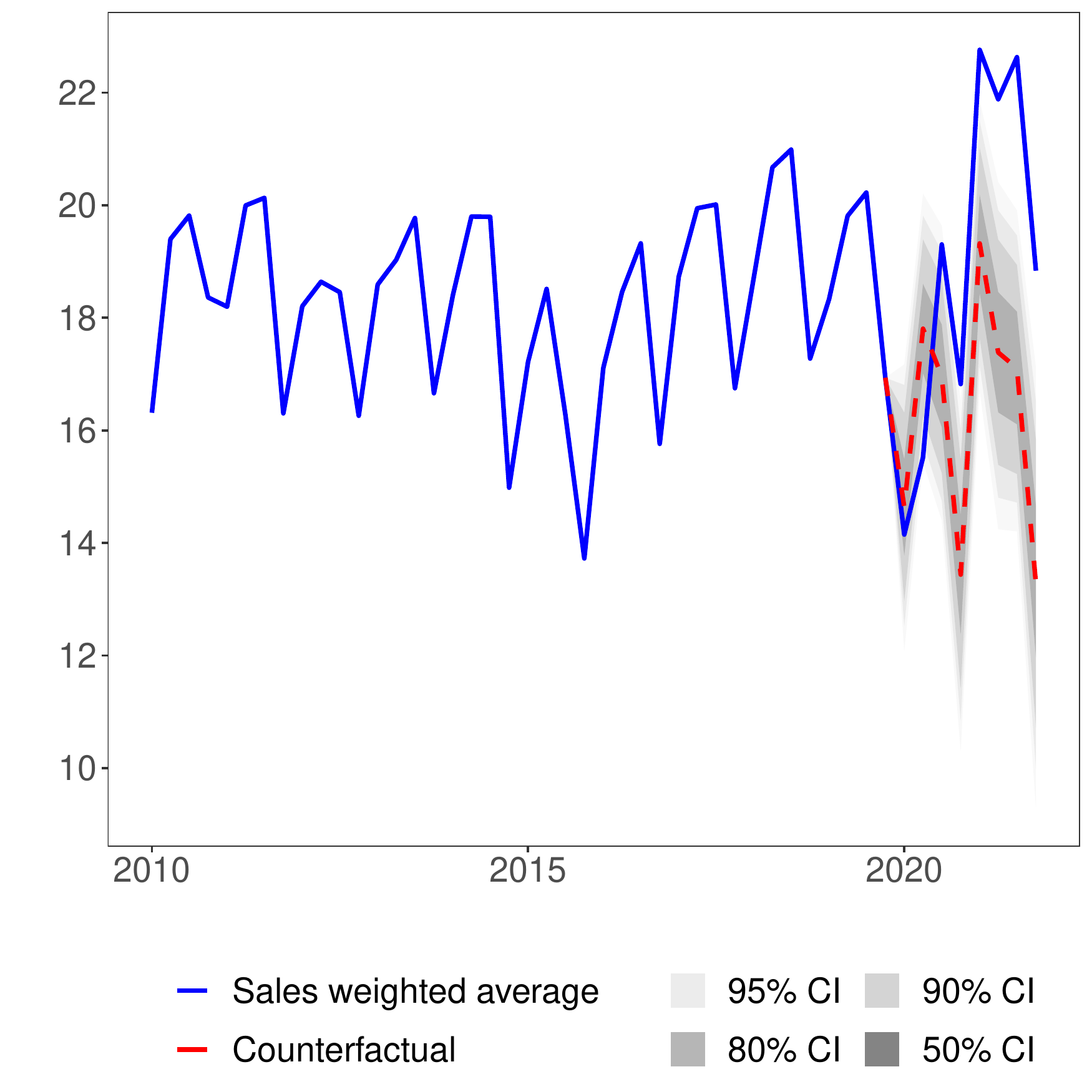}
  \caption{Average}
  \label{fig:Trend_fanchart_pik_quarter}
\end{subfigure}%
\begin{subfigure}{.5\textwidth}
  \centering
  \includegraphics[width=0.85\linewidth]{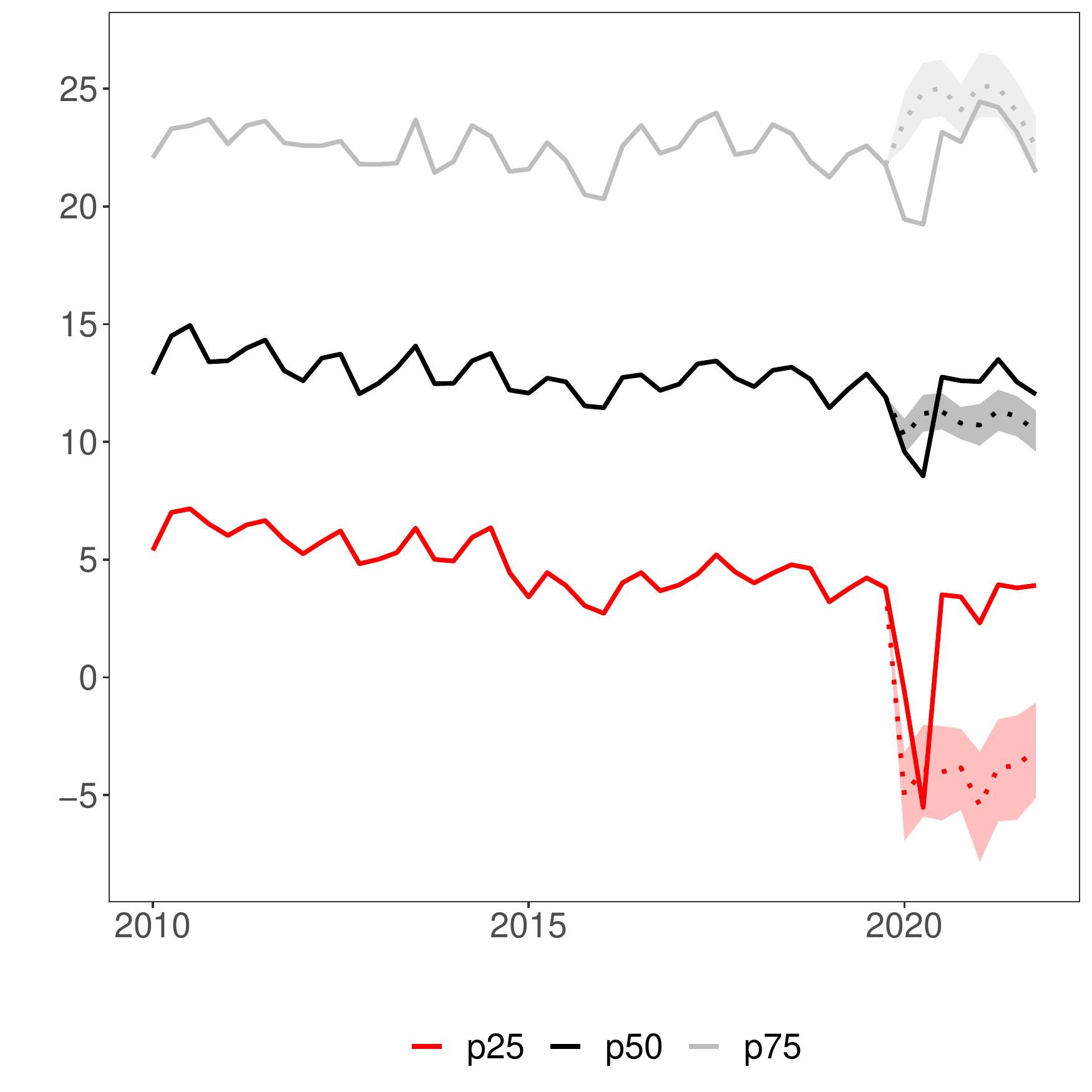}
  \caption{Quartiles}
  \label{fig:Trend_distr_pik_quarter}
\end{subfigure}
\label{fig:Trend_pik_quarter}
\caption*{\footnotesize Note: The figure shows four aggregate statistics of the quarterly observed and counterfactual profit rates. The statistics are the sales-weighted average (shown in Panel (a)), the first quartile (shown in Panel (b) in red), the second quartile (shown in Panel (b) in black), and the third quartile (shown in Panel (b) in gray). The solid lines represent the observed values, while the dotted lines represent the counterfactual values for 2020-2021, as calculated using firm-level Bayesian structural time series models. The shaded areas represent the  95\%, 90\%, 80\%, and 50\% (equally-tailed) credible interval for each statistic, calculated based on 5,000 posterior simulations on Panel (a). 95\% (equally-tailed) credible intervals are displayed for the statistics on Panel (b). }
\end{figure}

This section expands the aggregate counterfactual analysis in Figures \ref{fig:Trend_mu_year} and \ref{fig:Trend_pik_year}, but using quarterly data instead of yearly data. Figures \ref{fig:Trend_mu_quarter} and \ref{fig:Trend_pik_quarter} shows similar results for the markup rate: markups would have been higher had the pandemic not happened. Particularly, for the average firm, the annual average observed markup rate was 4.5\% and 11.6\% lower than its counterfactual value in 2020 and 2021, respectively. Interestingly, in the case of profit rates, the positive effects of the pandemic only seem to have started in the third quarter of 2020. More precisely, the average firm had a profit rate of 16.9\% and 21.6\% in 2020 and 2021, respectively, which are 0.7 and 4.75 percentage points higher than the counterfactuals.

\subsection{Quarterly heterogeneity analysis}\label{app:heterogeneity}

This section analyzes the heterogeneous effects of the pandemic  using quarterly data, which provides insights into the unfolding of the pandemic in 2020 and 2021.

\subsubsection{Effect heterogeneity regressions}

Finally, we validate the results in Section~\ref{sec:heterogeneity} by averaging the quarterly effects into the annual frequency and repeating the accounting exercise, with the results displayed on \tablename~\ref{tab:het_markup_rates_quarterly} and \ref{tab:het_profit_rates_quarterly}. The implied yearly effects backed by the high-frequency analysis seem to be more reactive to cogs and sales, with very similar results for employment, stock exchange tenure, and market shares. On the other hand, the results for profit rates exhibit very similar results for all of the variables. This suggests that the short-term effects on markups may be more reactive to cogs and sales, in a way that when averaged out, they still matter with respect to those calculated using the yearly data. This could be driven by the immediate changes in very short-term demand movements \cite{Gilchrist2017}.

\begin{table}[tp]
    \centering
    \caption{Heterogeneous Effects of the pandemic on Markup Rates  (Quarterly Specification)}
    \label{tab:het_markup_rates_quarterly}
    \begin{threeparttable}
        \begin{tabular}{lcccccc}
           \tabularnewline\midrule\midrule
           Dependent Variable: & \multicolumn{6}{c}{Average markup rate pandemic effect in 2020--2021}\\
            & \multicolumn{2}{c}{All} & \multicolumn{2}{c}{2020} & \multicolumn{2}{c}{2021} \\ 
           Model:        & (1)             & (2)             & (3)             & (4)             & (5)            & (6)\\
           \midrule \emph{Variables} &   &   &   &   &   &  \\
          COGS  & -0.244$^{***}$ & -0.263$^{***}$ & -0.296$^{***}$ & -0.313$^{***}$ & -0.181$^{**}$ & -0.197$^{**}$\\
                    & (0.054)        & (0.057)        & (0.066)        & (0.073)        & (0.086)       & (0.088)\\
          Sales & 0.197$^{***}$  & 0.248$^{***}$  & 0.230$^{***}$  & 0.278$^{***}$  & 0.160$^{*}$   & 0.212$^{**}$\\
                    & (0.056)        & (0.058)        & (0.067)        & (0.072)        & (0.093)       & (0.090)\\
          Employment  & 0.037$^{***}$  & 0.008          & 0.065$^{***}$  & 0.033$^{**}$   & 0.000         & -0.028\\
                    & (0.013)        & (0.013)        & (0.014)        & (0.015)        & (0.022)       & (0.023)\\
           Stock-exchange tenure     & 0.002$^{***}$  & 0.002$^{**}$   & 0.003$^{***}$  & 0.002$^{*}$    & 0.001         & 0.001\\
                    & (0.001)        & (0.001)        & (0.001)        & (0.001)        & (0.002)       & (0.002)\\
          Market share  & 0.837$^{***}$  & 0.002          & 0.235          & -0.012         & 1.481$^{**}$  & -0.047\\
                    & (0.301)        & (0.308)        & (0.253)        & (0.287)        & (0.680)       & (0.566)\\
          \midrule \emph{Fixed-effects} &   &   &   &   &   &  \\
          2-digit NAICS industry         &                 & Yes             &                 & Yes             &                & Yes\\
          \midrule
          Mean & 1.599 &  & 1.590 &  & 1.610 \\ 
          Mean effect & -0.040 &  & -0.045 &  & -0.033 \\ 
          \midrule \emph{Fit statistics} &   &   &   &   &   &  \\
          Observations  & 5,017           & 5,017           & 2,752           & 2,752           & 2,265          & 2,265\\
      R$^2$         & 0.021          & 0.075          & 0.044          & 0.095          & 0.010         & 0.082\\
      Within R$^2$  &                & 0.018          &                & 0.034          &               & 0.009\\
        \hline    
         \end{tabular}
        \begin{tablenotes}
         \item \noindent \footnotesize Notes: The table presents cross-section firm-level regressions of the average pandemic effect at the firm-level on the 2015--2019 average of the logarithm cost of goods sold, logarithm of sales, logarithm of employment, years since publicly-traded, and market shares. NAICS = North American Industrial Classification System. Heteroskedasticity-robust standard errors in parentheses. Signif. Codes: ***: 0.01, **: 0.05, *: 0.1
        \end{tablenotes}
    \end{threeparttable}
\end{table}

\begin{table}[tp]
    \centering
    \caption{Heterogeneous Effects of the pandemic on Profit Rates   (Quarterly Specification)}
    \label{tab:het_profit_rates_quarterly}
    \begin{threeparttable}
        \begin{tabular}{lcccccc}
           \tabularnewline\midrule\midrule
           Dependent Variable: & \multicolumn{6}{c}{Average profit rate pandemic effect in 2020--2021}\\
            & \multicolumn{2}{c}{All} & \multicolumn{2}{c}{2020} & \multicolumn{2}{c}{2021} \\ 
           Model:        & (1)            & (2)            & (3)            & (4)            & (5)      & (6)\\
           \midrule \emph{Variables} &   &   &   &   &   &  \\
          COGS  & -15.889$^{***}$ & -14.229$^{***}$ & -21.252$^{***}$ & -15.754$^{***}$ & -9.525   & -12.315\\
                    & (4.186)         & (4.621)         & (3.705)         & (3.712)         & (8.018)  & (9.062)\\
          Sales  & 16.075$^{***}$  & 15.365$^{***}$  & 20.067$^{***}$  & 15.685$^{***}$  & 11.476   & 15.221$^{*}$\\
                    & (4.605)         & (4.843)         & (5.232)         & (4.780)         & (7.842)  & (8.775)\\
          Employment & 1.140           & -0.076          & 2.610           & 0.796           & -0.795   & -1.563\\
                    & (1.871)         & (2.122)         & (2.826)         & (2.845)         & (2.296)  & (3.126)\\
          Stock-exchange tenure    & 0.149           & 0.005           & 0.140           & -0.081          & 0.143    & 0.086\\
                    & (0.093)         & (0.086)         & (0.115)         & (0.107)         & (0.153)  & (0.140)\\
          Market share & -71.887$^{*}$   & -20.439         & -100.702        & 9.398           & -43.641  & -59.176\\
                    & (39.196)        & (44.425)        & (70.899)        & (67.204)        & (27.539) & (38.484)\\
          \midrule \emph{Fixed-effects} &   &   &   &   &   &  \\
          2-digit NAICS industry         &                & Yes            &                & Yes            &          & Yes\\
                    \midrule
          Mean & 2.974 &  & -0.747 &  & 7.546 \\ 
          Mean effect & -1.070 &  & -4.573 &  & 3.235 \\ 
          \midrule \emph{Fit statistics} &   &   &   &   &   &  \\
          Observations  & 5,017          & 5,017          & 2,752          & 2,752          & 2,265    & 2,265\\
       R$^2$         & 0.009           & 0.027           & 0.015           & 0.067           & 0.004    & 0.009\\
      Within R$^2$  &                 & 0.005           &                 & 0.006           &          & 0.006\\
        \hline     
        \end{tabular}
        \begin{tablenotes}
         \item \noindent \footnotesize Notes: The table presents cross-section firm-level regressions of the average pandemic effect at the firm-level on 2015--2019 average of the logarithm cost of goods sold, the logarithm of sales, logarithm of employment, years since publicly-traded, and market shares. NAICS = North American Industrial Classification System. Heteroskedasticity-robust standard errors in parentheses. Signif. Codes: ***: 0.01, **: 0.05, *: 0.1
        \end{tablenotes}
    \end{threeparttable}
\end{table}

\newpage

\subsection{Tracking the fraction of firms with significant effects in the short-term.}

\begin{figure}[H]
\caption{Fraction of Firms with Significant effects of the pandemic, Bayesian Structural Time Series Model}
\begin{subfigure}{.5\textwidth}
  \centering
  \includegraphics[width=0.9\linewidth,page=1]{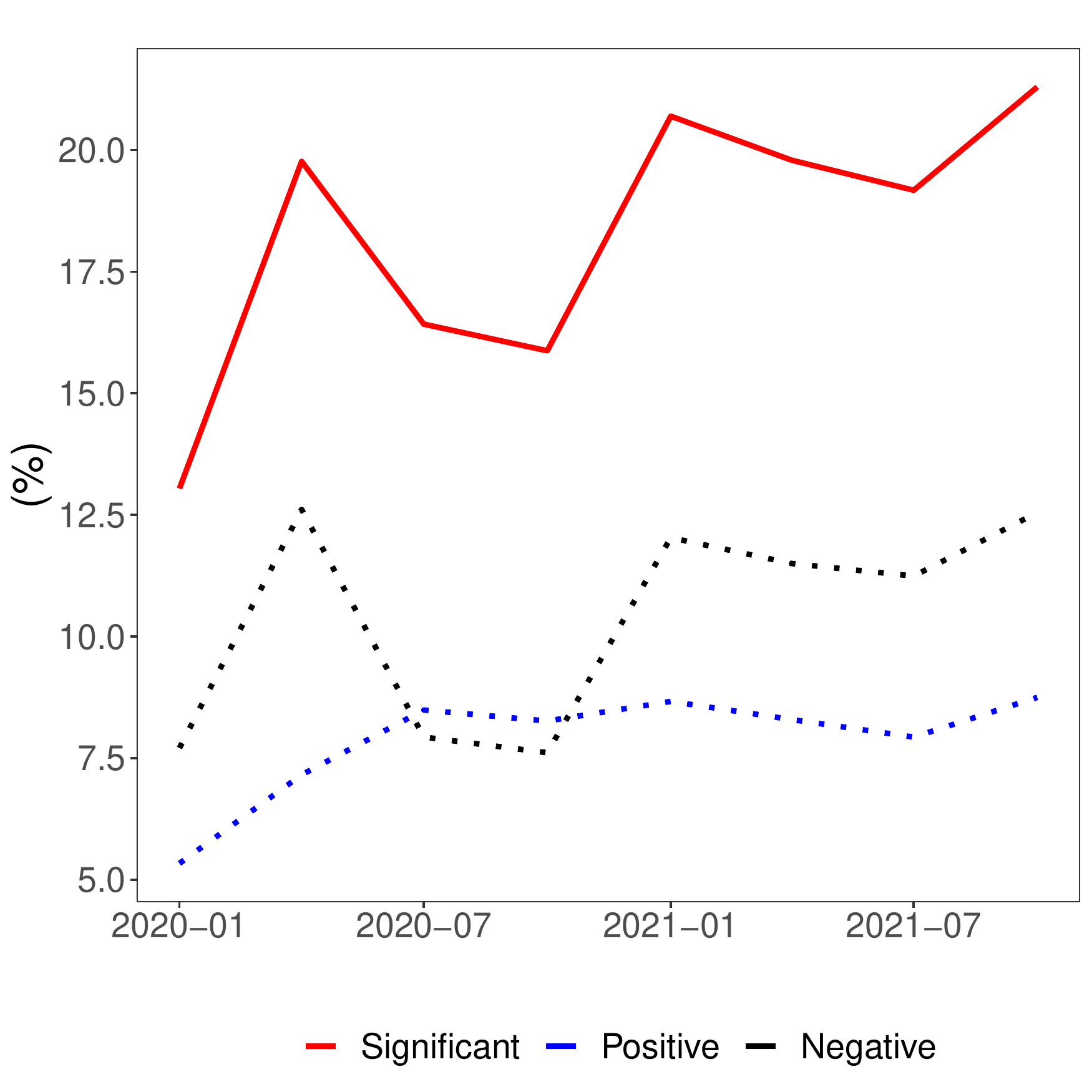}
  \caption{Markup Rates}
  \label{fig:frac_sign_mu_quarterly}
\end{subfigure}%
\begin{subfigure}{.5\textwidth}
  \centering
  \includegraphics[width=0.9\linewidth,page=1]{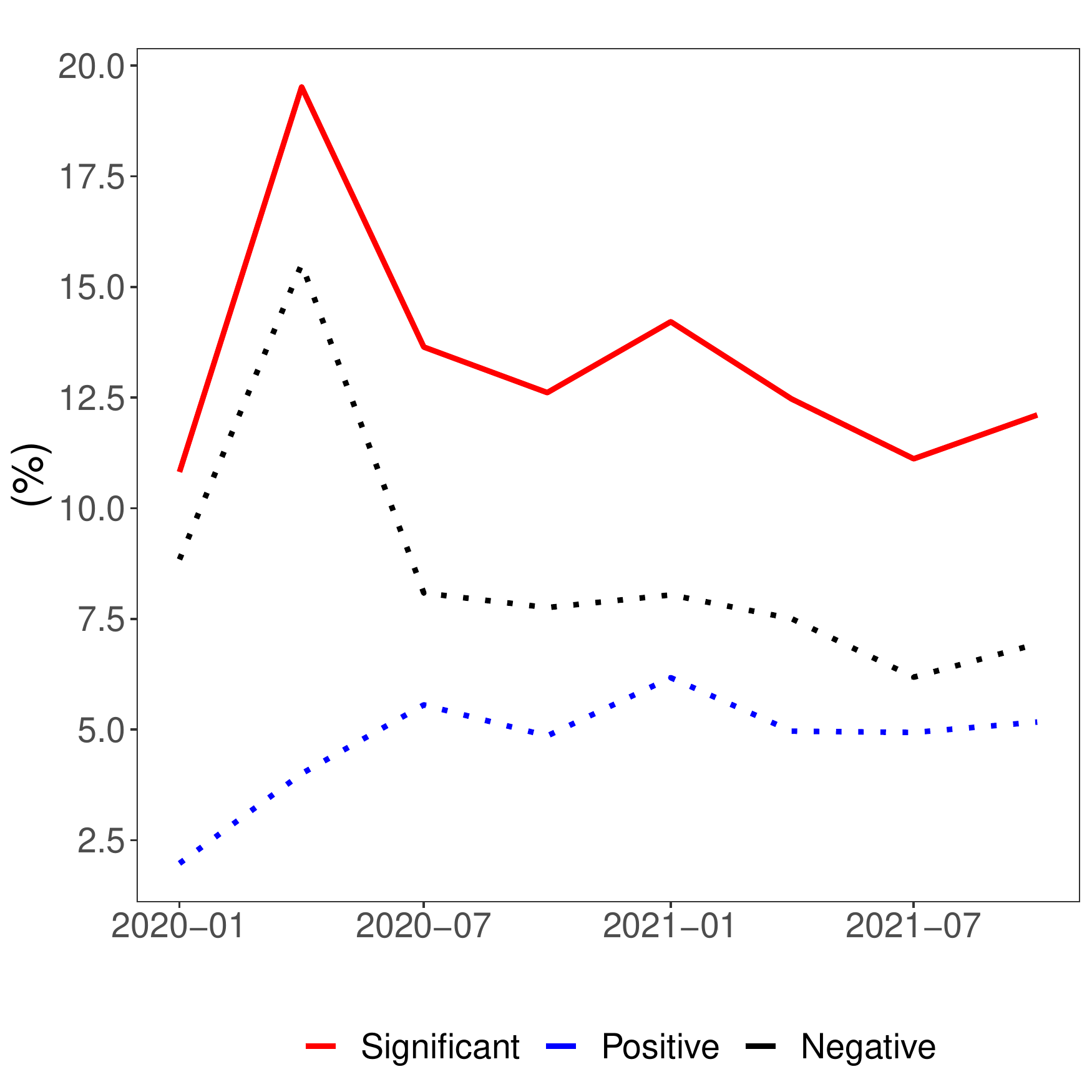}
  \caption{Profit Rates}
  \label{fig:frac_sign_pik_quarterly}
\end{subfigure}
\label{fig:frac_sign_quarterly}
    \caption*{\footnotesize Note: The figure plots the fraction of firms with significant (red solid), positive and significant (blue dotted), and negative and significant (black dotted) effects. A significant effect is defined when the corresponding 95\% posterior credible interval does not contain zero.}
\end{figure}

To elucidate whether these effects on markup rates are statistically significant, we calculated the fraction of firms with significant effects, and its decomposition into the fraction of firms with significant and positive effects, and significant and negative effects. \figurename~\ref{fig:frac_sign_mu_quarterly} shows this calculation for quarterly data in 2020--2021. We find that 19\% of the firms had a significant effect on their markups at the beginning of the pandemic, with most of those firms having a negative effect and over 7\% of the firms having a positive effect. In most of 2020--2021, the number of firms with significant, negative effects is higher than the number of firms with significant, positive, effects. Counting the number of firms with at least one quarter with a significant effect, we find that 1,122 out of 3,192 firms have a significant effect. 
 
Besides analyzing the statistical significance of the markup effects, we find that only a small fraction of the effects on profit rates are significant. Figure \ref{fig:frac_sign_pik_quarterly} shows this calculation for quarterly data in 2020--2021. Particularly, we find that 19\% of the firms had significant effects on profit rates at the beginning of the pandemic, where most of the firms have a negative effect and over 4\% of the firms had a positive effect. By the first quarter of 2021, the share of firms with positive effects increase but is still lower than those with negative effects. When we count the number of firms with at least one significant effect in one quarter, we find that 953 out of 3,152 firms have a significant effect.

To avoid multiple hypothesis testing issues and given our sample has thousands of firms corresponding to at least an equal number of hypothesis tests, we implement the Benjamini-Hochberg algorithm to control for the false discovery rate (FDR), which is the expected number of statistically significant effects that are indeed null effects.\footnote{We also implement four corrections that control the family-wise error rate, i.e., the probability of detecting at least one significant effect that is indeed null, including the Bonferroni, Hochberg, Holm, and Hommel corrections, and find no significant effects. However, such procedures are quite conservative, even more so when one has thousands of hypotheses such as in our case \citep{storey2003statistical}. In addition to these procedures, we implement the $q$-value approach of \citet{storey2003statistical} and found a similar number of tests that are significant while controlling the FDR at 5\%; however, the assumption on the uniform distribution of $p$-values in the unit interval may not hold since the highest $p$-value in our sample is just below 0.5.} 
As this algorithm relies on the computed $p$-values for each  individual test to construct a  joint test for each firm for the entirety of the pandemic in our sample, we compute the firm-level $p$-value for the 2020--2021 average effect on markup rates. When controlling the FDR at 5\% for the effects on markup rates, we find that there are 1,240 out of 3,192 statistically significant effects. If we were to follow the naive approach of counting the number of firms with statistically significant effects at 5\% using the $p$-values, we would consider 1,518 significant effects. Overall, the quarterly analysis using Figure \ref{fig:frac_sign_quarterly} and the FDR analysis of statistic significance are consistent with each other, with the former producing a slightly higher fraction of firms with significant effects, i.e., 35.2\% versus 38.8\%.

We repeat the previous multiple hypothesis correction to evaluate the statistical significance of the effects on profit rates.\footnote{Likewise, we find similar effects when using  the $q$-value correction of \citet{storey2003statistical}. We also implement four extremely conservative tests correcting for the FDR (Bonferroni, Hochberg, Holm, and Hommel) and find no significant effects.} Consequently, when controlling the FDR at 5\% for the effects on profit rates, we find that there are 832 out of 3,152 statistically significant effects. If we were to follow the naive approach of counting the number of firms with statistically significant effects at 5\% using the $p$-values, we would consider 1,200 significant effects. Overall, the quarterly analysis using \figurename~\ref{fig:frac_sign_pik_quarterly} and the FDR analysis of statistic significance are consistent with each other, with the former producing a slightly higher fraction of firms with significant effects, i.e., 29.9\% versus 26.4\%.

One concern when performing the Benjamini-Hochberg correction is that either markup or profit rates may be highly correlated across firms. To alleviate this concern, we perform both factor and correlation analyses as in \citet{bermingham2014understanding}. To have a sense of how correlated the series are across firms, that study suggests estimating the correlation between each firm's markup or profit rate with the principal component calculated across firms. To avoid losing a big fraction of our sample and given the pattern of entry and exit of firms in the stock exchange, we estimate principal components in 5-year windows. We then compute the average $R^2$ for the firm-level regressions of either the markup or profit rate on the corresponding first principal component. We obtain an average $R^2$ of just 0.07 and 0.06 for markup rates and profit rates, respectively. This result indicates that there is not a sizable correlation across firms for either the markup or the profit rates. Moreover, the average pairwise correlation in the markup and profit rates across firms is very small and equal to 0.01 and 0.016, respectively.

In addition, we conduct the corresponding robustness checks by  plotting this decomposition for both the LLPs model and using the hyperparameters in the sensitivity analysis presented in Section~\ref{sec:robustness}. 

\subsubsection{Fraction of firms with significant effects of the pandemic, LLPs Model}

\begin{figure}[H]
\caption{Fraction of Firms with Significant effects of the pandemic, LLPs Model}
\begin{subfigure}{.5\textwidth}
  \centering
  \includegraphics[width=0.9\linewidth,page=2]{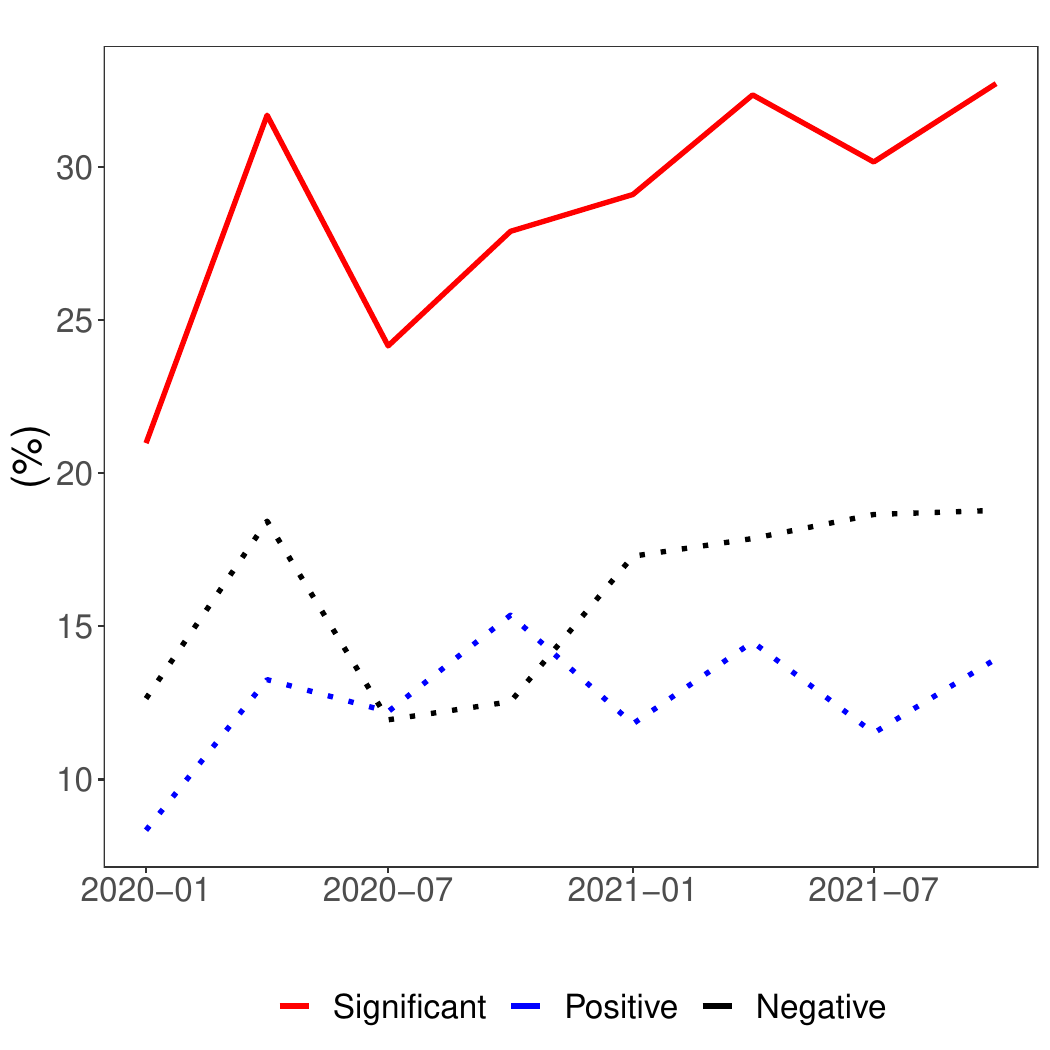}
  \caption{Markup Rates}
  \label{fig:frac_sign_mu_quarterly_LLP}
\end{subfigure}%
\begin{subfigure}{.5\textwidth}
  \centering
  \includegraphics[width=0.9\linewidth,page=2]{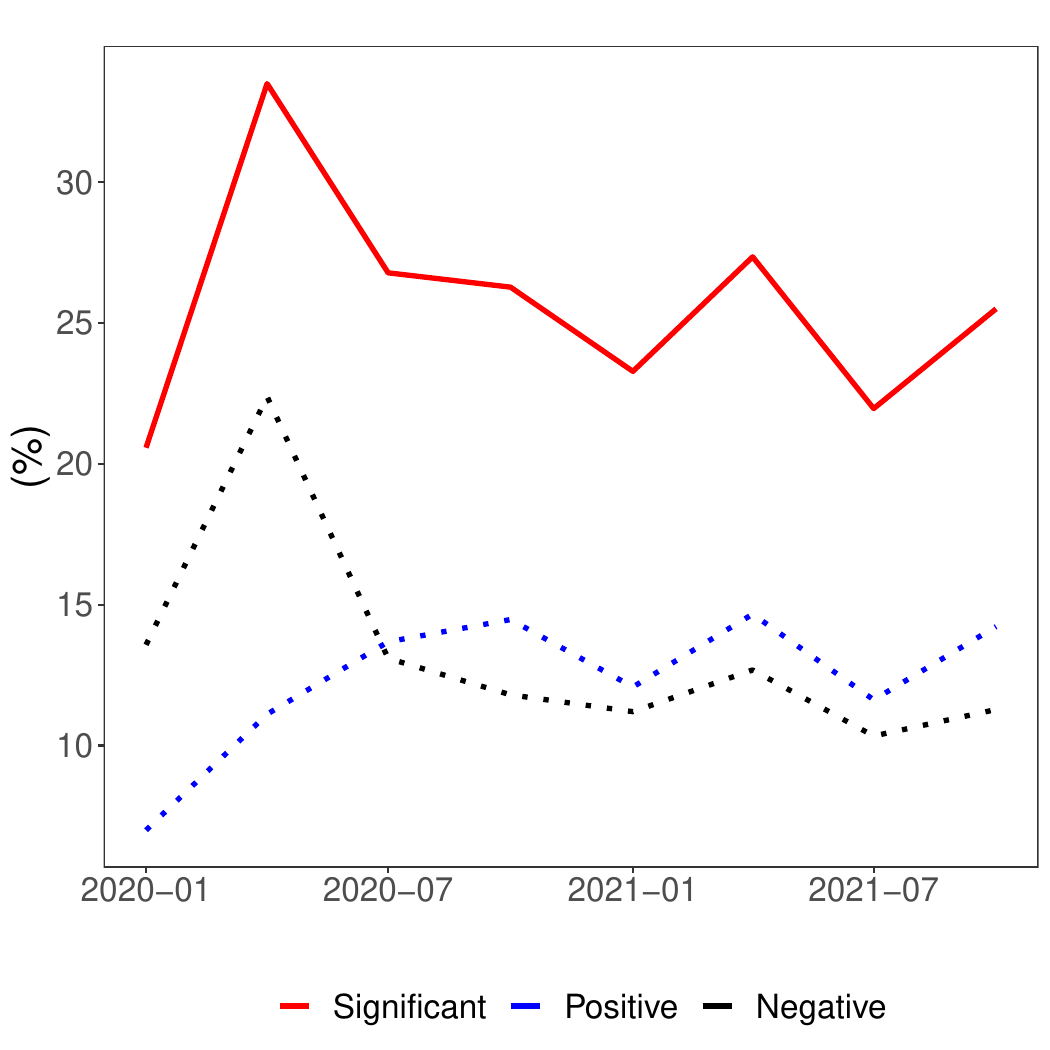}
  \caption{Profit Rates}
  \label{fig:frac_sign_pik_quarterly_LLP}
\end{subfigure}
\label{fig:frac_sign_quarterly_LLP}
    \caption*{\footnotesize Note: The figure plots the fraction of firms with significant (red solid), positive and significant (blue dotted), and negative and significant (black dotted) effects. A significant effect is defined when the corresponding 95\% posterior credible interval does not contain zero.}
\end{figure}
 
Using LLPs, we replicate our analysis for the aggregate markup and profit rates in Section \ref{sec:counteractual_analysis}. Overall, both BSTS and LLPs forecasts find that there are more negatively affected firms in the first two-quarters of the pandemic, with LLPs reporting more significantly affected firms than BSTS. (see Figures \ref{fig:frac_sign_quarterly} and \ref{fig:frac_sign_quarterly_LLP}). Additionally, both methods suggest that the share of firms benefiting from the pandemic was initially low in 2020 and grew steadily up to the end of 2021. LLPs seems to be more liberal as it tends to find more significant effects on both markup and profit rates compared to BSTS.

The results tell the same story as those in \figurename~\ref{fig:frac_sign_quarterly}, but the LLPs model shows more significant effects in both directions, with roughly 28\% significant effects in the first half of 2020. However, in Section~\ref{sec:model_fit} we observe that the Bayesian model has better out-of-sample performance, so the observed increase observed in the LLPs model might be driven by overestimated effects due to forecasting errors.  

\subsubsection{Fraction of firms with significant effects of the pandemic, Sensitivity Analysis}

In this section, we repeat the analysis in  \figurename~\ref{fig:frac_sign_quarterly} and changing the hyperparameters as detailed in Section~\ref{sec:robustness}. Interestingly, the results are quite similar, with increased (decreased) variability producing higher (fewer) significant effects, especially for those observing negative effects. Nonetheless, these changes seem to be modest.

\begin{figure*} 
        \caption{Sensitivity Analysis: Fraction of Firms with Significant effects of the pandemic (Quarterly Specification)}
        \label{fig:sen_ana_sig} 
        \centering
        \begin{subfigure}[b]{0.475\textwidth}
            \centering
            \includegraphics[width=\linewidth,page=1]{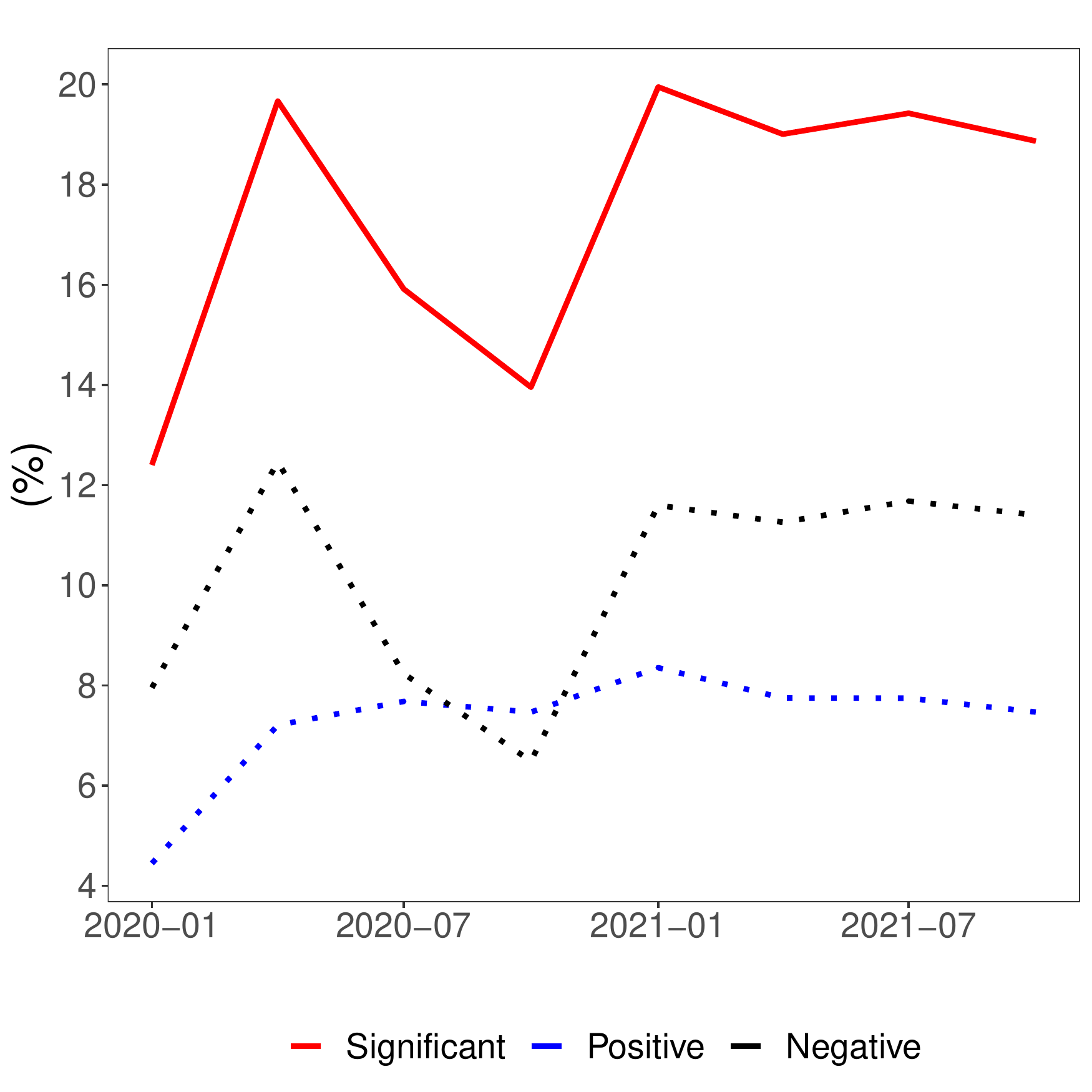}
            \caption[Network2]%
            {{\small Markups,  125\% Hyperparameters}}    
            \label{fig:frac_sign_mu_quarterly_125}
        \end{subfigure} 
        \begin{subfigure}[b]{0.475\textwidth}  
            \centering 
            \includegraphics[width=\linewidth,page=1]{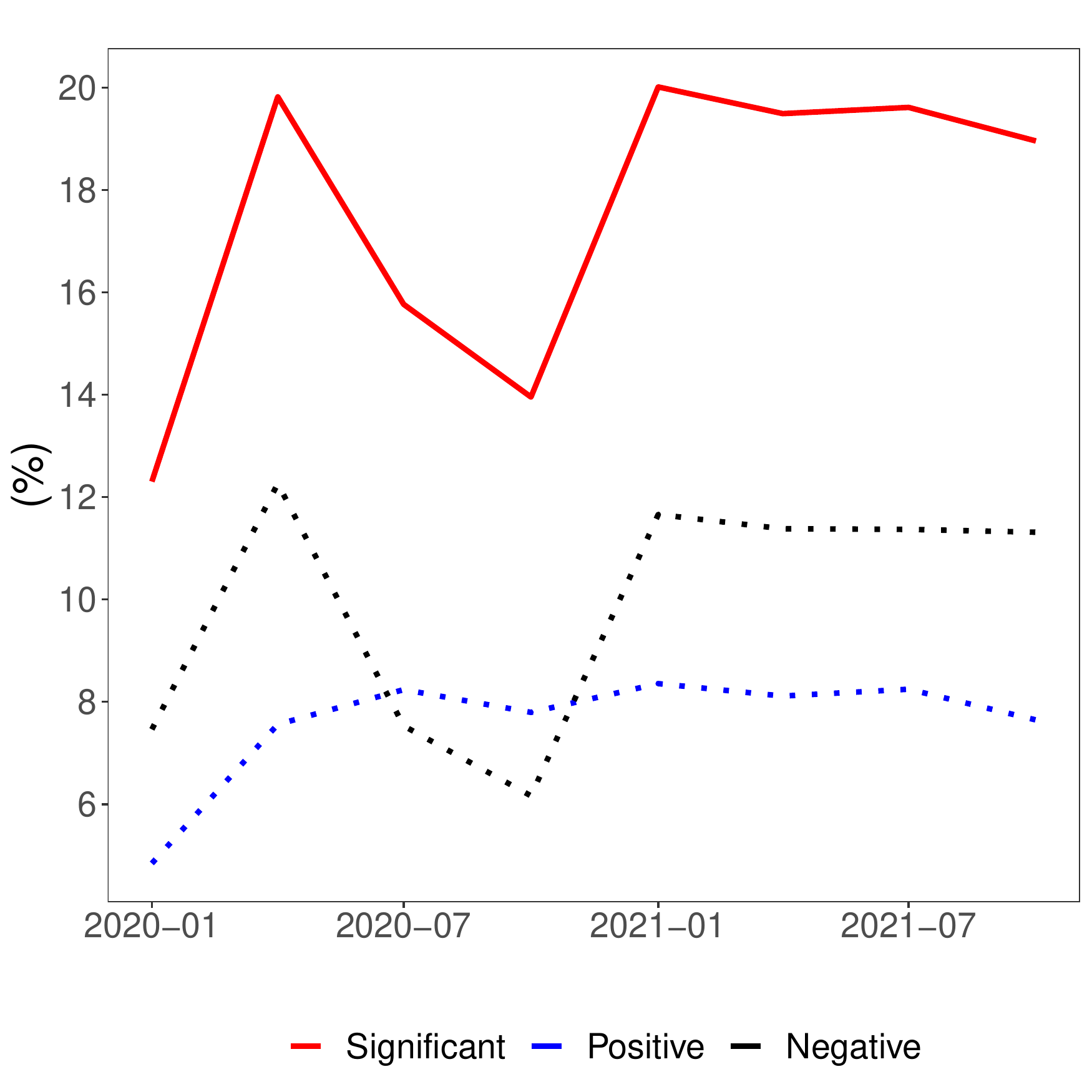}
            \caption[]%
            {{\small  Markups,  75\% Hyperparameters}}    
            \label{fig:frac_sign_mu_quarterly_075}
        \end{subfigure}
        \vskip\baselineskip
        \begin{subfigure}[b]{0.475\textwidth}
            \centering
            \includegraphics[width=\linewidth,page=1]{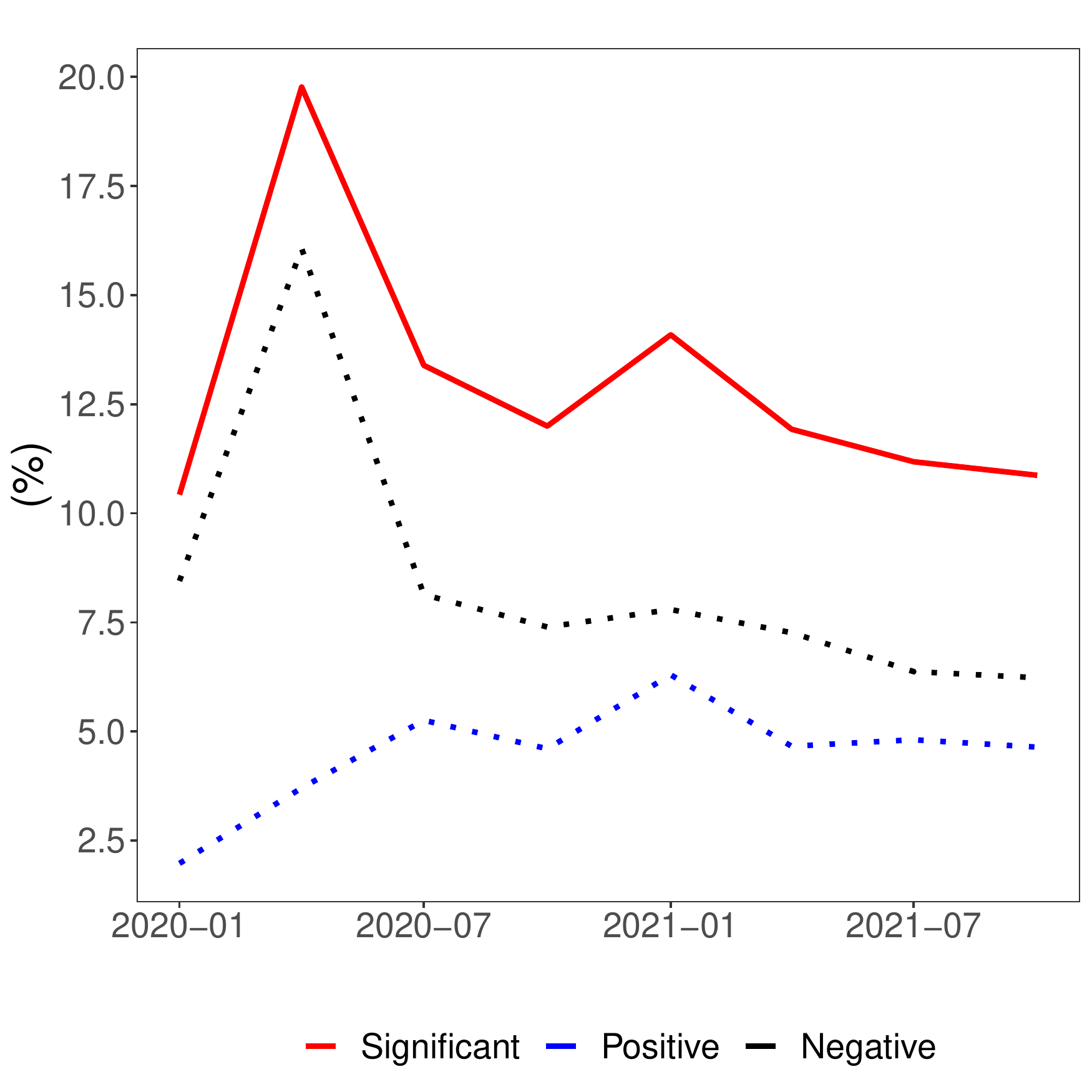}
            \caption[Network2]%
            {{\small Profit Rates, 125\% Hyperparameters}}    
            \label{fig:frac_sign_pi_quarterly_125}
        \end{subfigure}
        \hfill
        \begin{subfigure}[b]{0.475\textwidth}  
            \centering 
            \includegraphics[width=\linewidth,page=1]{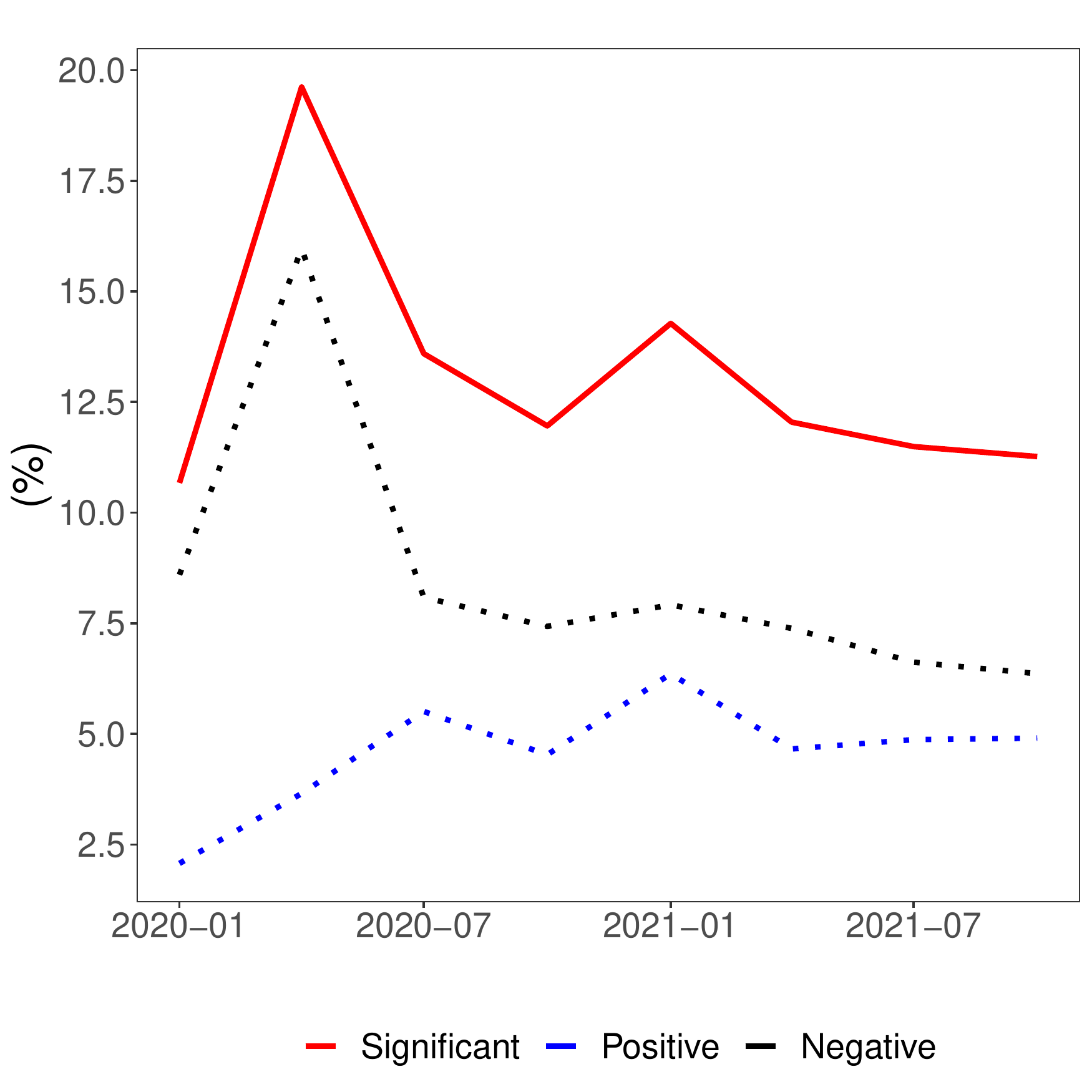}
            \caption[]%
            {{\small  Profit Rates, 75\% Hyperparameters}}    
            \label{fig:frac_sign_pi_quarterly_075}
        \end{subfigure}
   
        \caption*{\footnotesize   Note: The figure plots the fraction of firms with significant (red solid), positive and significant (blue dotted), and negative and significant (black dotted) effects. A significant effect is defined when the corresponding 95\% posterior credible interval does not contain zero. }
    \end{figure*}

After changing the variability of the hyperparameters governing the priors of the markups models, either by increasing or decreasing that variability, we validate our finding that at the beginning of the pandemic, there was a higher faction of negative firms that positively affected firms (see Panels (a) and (b) of Figure \ref{fig:sen_ana_sig} and Panel (b) of Figure \ref{fig:frac_sign_quarterly}).

\newpage

\section{Pandemic effect heterogeneity by quintiles of key firm characteristics.}

\label{app:het} 

To scale the effect across firms with different baseline markup rates, we divide the estimate of the causal impact for markup rates by the average value of markup rates in 2015--2019 as follows
\begin{gather*}
    \widehat{\tau}_{i,t+h}^s = \frac{Y_{i,t+h} - \widehat{Y}_{i,t+h}}{\overline{Y}_{i,2015\text{--}2019}},
\end{gather*}    
where $\overline{Y}_{i,2015\text{--}2019}$ is firm's $i$ average markup rate in 2015--2019. We acknowledge that choosing 2015--2019 as a scaling period is somewhat arbitrary, but it has the advantage of at least taking a period that is actually close to the pandemic years.\footnote{Since the profit rate takes negative values for some firms in some time periods, we do not scale those rates to avoid interpretation issues.}  

\begin{figure*}[p]
        \caption{Heterogeneity Analysis of Effect of Pandemic in Markups (Yearly Specification)}
        \label{fig:het_ana_markup}
        \centering
        \begin{subfigure}[b]{0.475\textwidth}
            \centering
            \includegraphics[width=\textwidth,height=5cm,page=1]{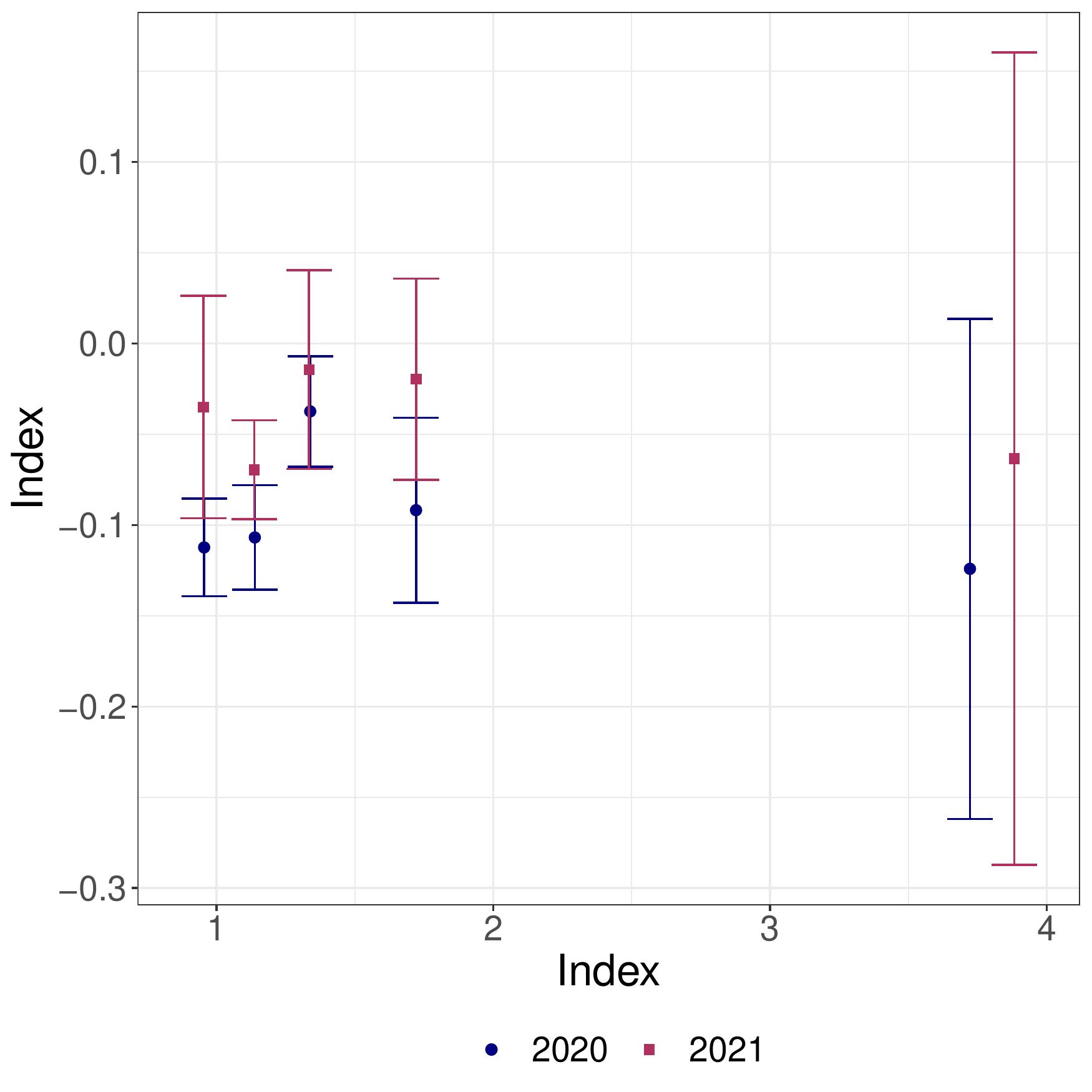}
            \caption[Network2]%
            {{\small Markup Rate}}    
            \label{fig:markup_markup}
        \end{subfigure}
        \hfill
        \begin{subfigure}[b]{0.475\textwidth}  
            \centering 
            \includegraphics[width=\textwidth,height=5cm,page=3]{Effects_year_yearly_mu_5bins_mean.pdf}
            \caption[]%
            {{\small Sales}}    
            \label{fig:markup_sales}
        \end{subfigure}
        \vskip\baselineskip
        \begin{subfigure}[b]{0.475\textwidth}   
            \centering 
            \includegraphics[width=\textwidth,height=5cm,page=8]{Effects_year_yearly_mu_5bins_mean.pdf}
            \caption[]%
            {{\small Years in stock exchange}}    
            \label{fig:markup_yearsPublic}
        \end{subfigure}
        \hfill
        \begin{subfigure}[b]{0.475\textwidth}   
            \centering 
            \includegraphics[width=\textwidth,height=5cm,page=4]{Effects_year_yearly_mu_5bins_mean.pdf}
            \caption[]%
            {{\small Profit Rate}}    
            \label{fig:markup_profit_rate}
        \end{subfigure}
        \vskip\baselineskip
        \begin{subfigure}[b]{0.475\textwidth}   
            \centering 
            \includegraphics[width=\textwidth,height=5cm,page=5]{Effects_year_yearly_mu_5bins_mean.pdf}
            \caption[]%
            {{\small Market share}}    
            \label{fig:markup_marketShare}
        \end{subfigure}
        \hfill
        \begin{subfigure}[b]{0.475\textwidth}   
            \centering 
            \includegraphics[width=\textwidth,height=5cm,page=7]{Effects_year_yearly_mu_5bins_mean.pdf}
            \caption[]%
            {{\small Employment}}    
            \label{fig:markup_employment}
        \end{subfigure}
         \caption*{\footnotesize Note: $y$-axis: effect on markup rates as $Y_{it}-\widetilde{Y}_{it}$, where $Y_{it}$ is firm $i$'s markup rate at time period $t$, and $\widetilde{Y}_{i,t}$ is the firm's counterfactual markup rate built upon a Bayesian structural time series model with seasonal and local-level stochastic trend. $x$-axis: firm's $i$ average value of either (a) markup rates, (b) sales (logarithm of constant 2010 thousand US dollars), (c) years since publicly traded, (d) profit rates, (e) market shares, and (f) employment (thousands of employees) in 2015--2019. }
    \end{figure*}

\begin{figure*}[p]
        \caption{Heterogeneity Analysis of Effect of Pandemic in Scaled Markups (Yearly Specification)} 
        \label{fig:het_ana_markup_scaled}
        \centering
        \begin{subfigure}[b]{0.475\textwidth}
            \centering
            \includegraphics[width=\textwidth,height=5cm,page=1]{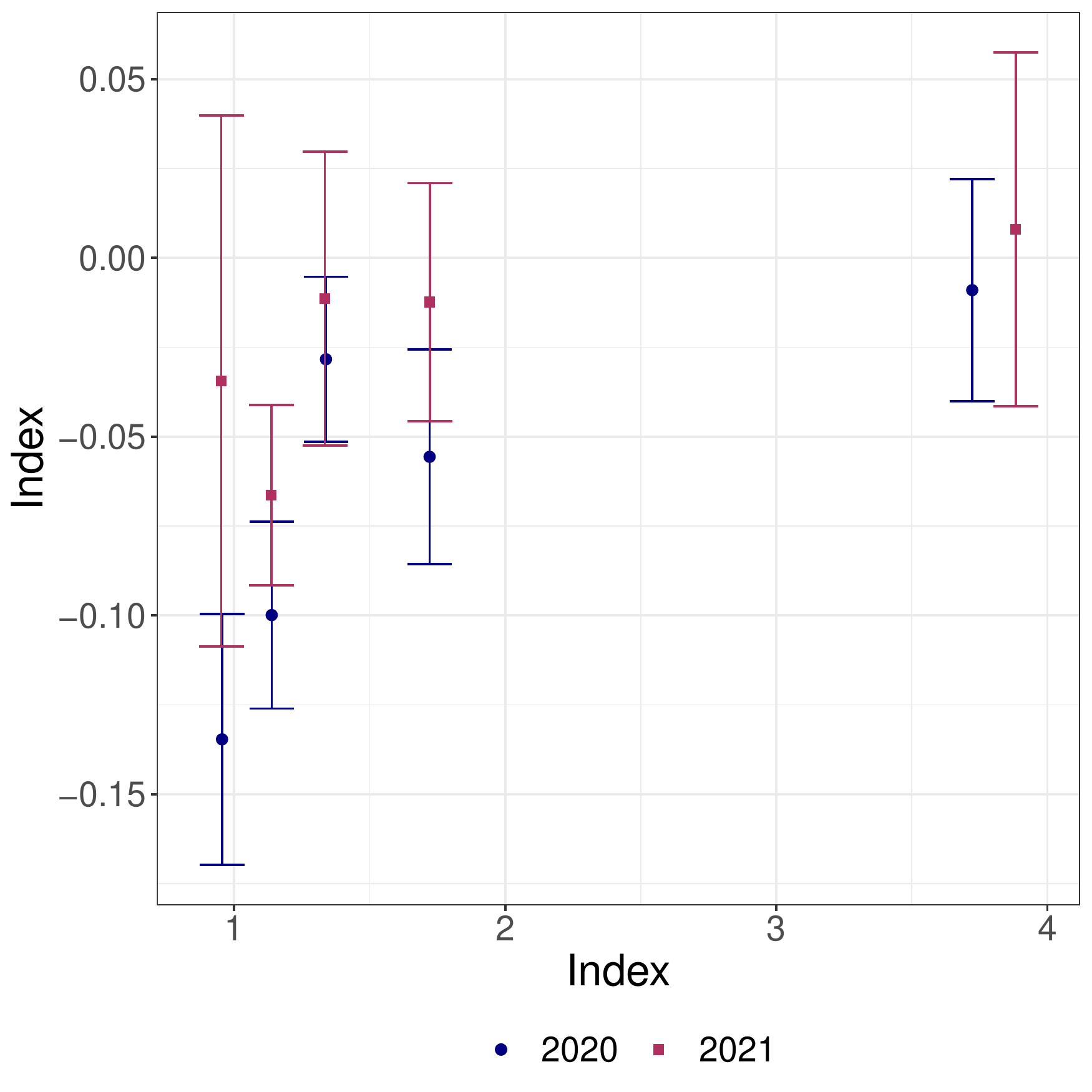}
            \caption[Network2]%
            {{\small Markup Rate}}    
            \label{fig:markup_markup_scaled}
        \end{subfigure}
        \hfill
        \begin{subfigure}[b]{0.475\textwidth}  
            \centering 
            \includegraphics[width=\textwidth,height=5cm,page=3]{Effects_year_yearly_mu_5bins_scaled.pdf}
            \caption[]%
            {{\small Sales}}    
            \label{fig:markup_sales_scaled}
        \end{subfigure}
        \vskip\baselineskip
        \begin{subfigure}[b]{0.475\textwidth}   
            \centering 
            \includegraphics[width=\textwidth,height=5cm,page=8]{Effects_year_yearly_mu_5bins_scaled.pdf}
            \caption[]%
            {{\small Years in stock exchange}}    
            \label{fig:markup_yearsPublic_scaled}
        \end{subfigure}
        \hfill
        \begin{subfigure}[b]{0.475\textwidth}   
            \centering 
            \includegraphics[width=\textwidth,height=5cm,page=4]{Effects_year_yearly_mu_5bins_scaled.pdf}
            \caption[]%
            {{\small Profit Rate}}    
            \label{fig:markup_profitRate_scaled}
        \end{subfigure}
        \vskip\baselineskip
        \begin{subfigure}[b]{0.475\textwidth}   
            \centering 
            \includegraphics[width=\textwidth,height=5cm,page=5]{Effects_year_yearly_mu_5bins_scaled.pdf}
            \caption[]%
            {{\small Market Share}}    
            \label{fig:markup_marketShare_scaled}
        \end{subfigure}
        \hfill
        \begin{subfigure}[b]{0.475\textwidth}   
            \centering 
            \includegraphics[width=\textwidth,height=5cm,page=7]{Effects_year_yearly_mu_5bins_scaled.pdf}
            \caption[]%
            {{\small Employment}}    
            \label{fig:markup_employment_scaled}
        \end{subfigure}
          \caption*{\footnotesize Note: $y$-axis: effect on markup rates as $(Y_{it}-\widetilde{Y}_{it})/\overline{Y}_{i,2015\text{--}2019}$, where $Y_{it}$ is firm $i$'s markup rate at time period $t$, $\widetilde{Y}_{i,t}$ is the firm's counterfactual markup rate built upon a Bayesian structural time series model with the seasonal and local-level stochastic trend, and $\overline{Y}_{i,2015\text{--}2019}$ is the firm's average markup rate in 2015--2019. $x$-axis: firm's $i$ average value of either (a) markup rates, (b) sales (logarithm of constant 2010 thousand US dollars), (c) years since publicly traded, (d) profit rates, (e) market shares, and (f) employment (thousands of employees) in 2015--2019. }
    \end{figure*}

\begin{figure*}[p]
        \caption{Heterogeneity Analysis of Effect of Pandemic in Markups (Quarterly Specification)}
        \label{fig:het_ana_markup_quarterly}
        \centering
        \begin{subfigure}[b]{0.475\textwidth}
            \centering
            \includegraphics[width=\textwidth,height=5cm,page=1]{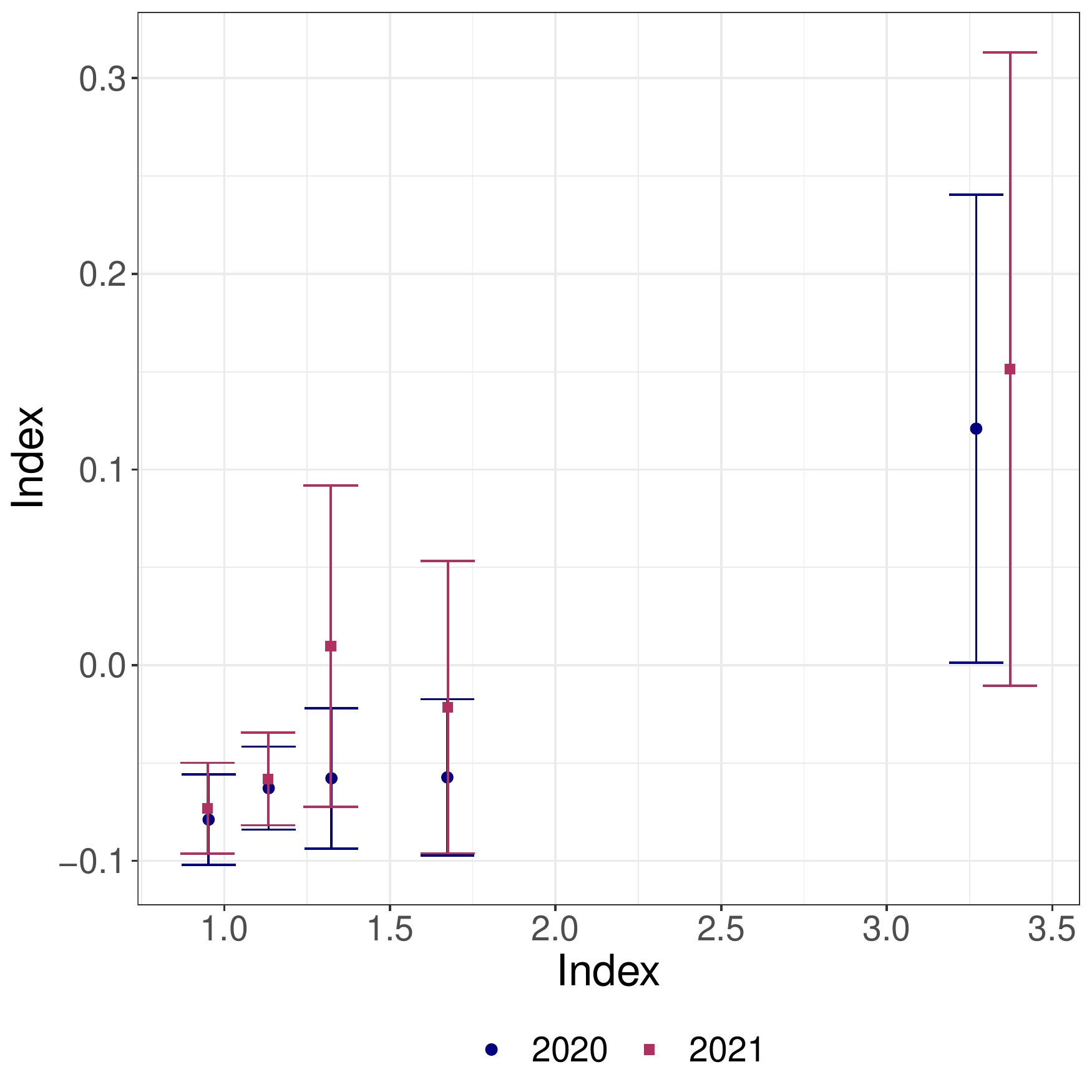}
            \caption[Network2]%
            {{\small Markup Rate}}    
            \label{fig:markup_markup_quarterly}
        \end{subfigure}
        \hfill
        \begin{subfigure}[b]{0.475\textwidth}  
            \centering 
            \includegraphics[width=\textwidth,height=5cm,page=3]{Effects_year_mu_5bins_mean.pdf}
            \caption[]%
            {{\small Sales}}    
            \label{fig:markup_sales_quarterly}
        \end{subfigure}
        \vskip\baselineskip
        \begin{subfigure}[b]{0.475\textwidth}   
            \centering 
            \includegraphics[width=\textwidth,height=5cm,page=8]{Effects_year_mu_5bins_mean.pdf}
            \caption[]%
            {{\small Years in stock exchange}}    
            \label{fig:markup_yearsPublic_quarterly}
        \end{subfigure}
        \hfill
        \begin{subfigure}[b]{0.475\textwidth}   
            \centering 
            \includegraphics[width=\textwidth,height=5cm,page=4]{Effects_year_mu_5bins_mean.pdf}
            \caption[]%
            {{\small Profit Rate}}    
            \label{fig:markup_profitRate_quarterly}
        \end{subfigure}
        \vskip\baselineskip
        \begin{subfigure}[b]{0.475\textwidth}   
            \centering 
            \includegraphics[width=\textwidth,height=5cm,page=5]{Effects_year_mu_5bins_mean.pdf}
            \caption[]%
            {{\small Market share}}    
            \label{fig:markup_marketShare_quarterly}
        \end{subfigure}
        \hfill
        \begin{subfigure}[b]{0.475\textwidth}   
            \centering 
            \includegraphics[width=\textwidth,height=5cm,page=7]{Effects_year_mu_5bins_mean.pdf}
            \caption[]%
            {{\small Employment}}    
            \label{fig:markup_employment_quarterly}
        \end{subfigure} 
        \caption*{\footnotesize Note: $y$-axis: effect on markup rates as $Y_{it}-\widetilde{Y}_{it}$, where $Y_{it}$ is firm $i$'s markup rate at time period $t$, and $\widetilde{Y}_{i,t}$ is the firm's counterfactual markup rate built upon a Bayesian structural time series model with seasonal and local-level stochastic trend. $x$-axis: firm's $i$ average value of either (a) markup rates, (b) sales (logarithm of constant 2010 thousand US dollars), (c) years since publicly traded, (d) profit rates, (e) market shares, and (f) employment (thousands of employees) in 2015--2019.}
    \end{figure*}

\begin{figure*}[p]
        \caption{Heterogeneity Analysis of Effect of Pandemic in Scaled Markups (Quarterly Specification)} 
        \label{fig:het_ana_markup_scaled_quarterly}
        \centering
        \begin{subfigure}[b]{0.475\textwidth}
            \centering
            \includegraphics[width=\textwidth,height=5cm,page=1]{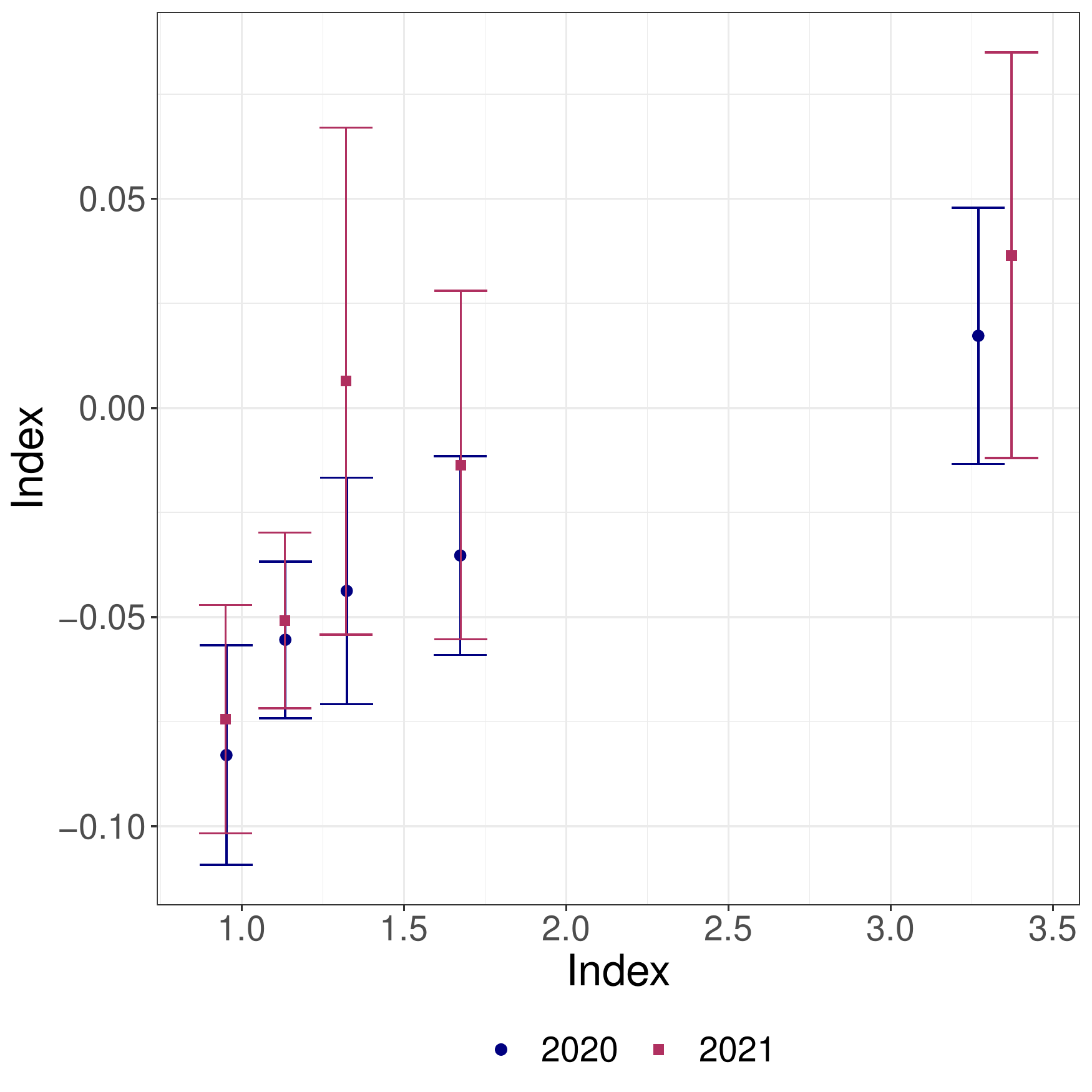}
            \caption[Network2]%
            {{\small Markup Rate}}    
            \label{fig:markup_markup_scaled_quarterly}
        \end{subfigure}
        \hfill
        \begin{subfigure}[b]{0.475\textwidth}  
            \centering 
            \includegraphics[width=\textwidth,height=5cm,page=3]{Effects_year_mu_5bins_scaled.pdf}
            \caption[]%
            {{\small Sales}}    
            \label{fig:markup_sales_scaled_quarterly}
        \end{subfigure}
        \vskip\baselineskip
        \begin{subfigure}[b]{0.475\textwidth}   
            \centering 
            \includegraphics[width=\textwidth,height=5cm,page=8]{Effects_year_mu_5bins_scaled.pdf}
            \caption[]%
            {{\small Years in stock exchange}}    
            \label{fig:markup_yearsPublic_scaled_quarterly}
        \end{subfigure}
        \hfill
        \begin{subfigure}[b]{0.475\textwidth}   
            \centering 
            \includegraphics[width=\textwidth,height=5cm,page=4]{Effects_year_mu_5bins_scaled.pdf}
            \caption[]%
            {{\small Profit Rate}}    
            \label{fig:markup_profitRate_scaled_quarterly}
        \end{subfigure}
        \vskip\baselineskip
        \begin{subfigure}[b]{0.475\textwidth}   
            \centering 
            \includegraphics[width=\textwidth,height=5cm,page=5]{Effects_year_mu_5bins_scaled.pdf}
            \caption[]%
            {{\small Market Share}}    
            \label{fig:markup_marketShare_scaled_quarterly}
        \end{subfigure}
        \hfill
        \begin{subfigure}[b]{0.475\textwidth}   
            \centering 
            \includegraphics[width=\textwidth,height=5cm,page=7]{Effects_year_mu_5bins_scaled.pdf}
            \caption[]%
            {{\small Employment}}    
            \label{fig:markup_employment_scaled_quarterly}
        \end{subfigure} 
        \caption*{\footnotesize Note: $y$-axis: effect on markup rates as $(Y_{it}-\widetilde{Y}_{it})/\overline{Y}_{i,2015\text{--}2019}$, where $Y_{it}$ is firm $i$'s markup rate at time period $t$, $\widetilde{Y}_{i,t}$ is the firm's counterfactual markup rate built upon a Bayesian structural time series model with seasonal and local-level stochastic trend, and $\overline{Y}_{i,2015\text{--}2019}$ is the firm's average markup rate in 2015--2019. $x$-axis: firm's $i$ average value of either (a) markup rates, (b) sales (logarithm of constant 2010 thousand US dollars), (c) years since publicly traded, (d) profit rates, (e) market shares, and (f) employment (thousands of employees) in 2015--2019.}
    \end{figure*}

Figures \ref{fig:het_ana_markup} and \ref{fig:het_ana_markup_scaled} report the results of this heterogeneity analysis for the markup rates, both without and with scaling by the average markup rates in 2015--2019. Firms with low markups in 2015--2019 have experienced the strongest reductions in their markup rates during the pandemic. In contrast, some of the firms in the top 20\% highest markups seem to have increased their markup rates both in the first and second years of the pandemic (Panel \ref{fig:markup_markup}). Regarding firm size heterogeneity, there are no big differences in the effects on markups (Panel \ref{fig:markup_sales}). Firms with longer tenure as public firms seem to be better at hedging the pandemic shock since they exhibit fewer negative effects in the first year of the pandemic (Panel \ref{fig:markup_yearsPublic}). Furthermore, the top 20\% firms in the profit rate tend to have positive effects on markup rates in 2021, whereas firms in the other quintiles tend to have negative effects on markups. The effects are quite heterogeneous for those firms with negative profit rates in 2015--2019. On the other hand, there are no noticeable differences in the effects on markups across market shares (Panel \ref{fig:markup_marketShare}). Finally, firms in the middle of the distribution of the number of employees tend to have lower impacts on their markups (Panel \ref{fig:markup_employment}). We find similar patterns when using the quarterly specification of the model (Figures \ref{fig:het_ana_markup_quarterly} and \ref{fig:het_ana_markup_scaled_quarterly}).

\begin{figure*}[p]
        \caption{Heterogeneity Analysis of Effect of Pandemic on Profit Rate (Yearly Specification)}
        \label{fig:het_ana_profit_rate}        
        \centering
        \begin{subfigure}[b]{0.475\textwidth}
            \centering
            \resizebox{\textwidth}{!}{\includegraphics[width=\textwidth,height=5cm,page=1]{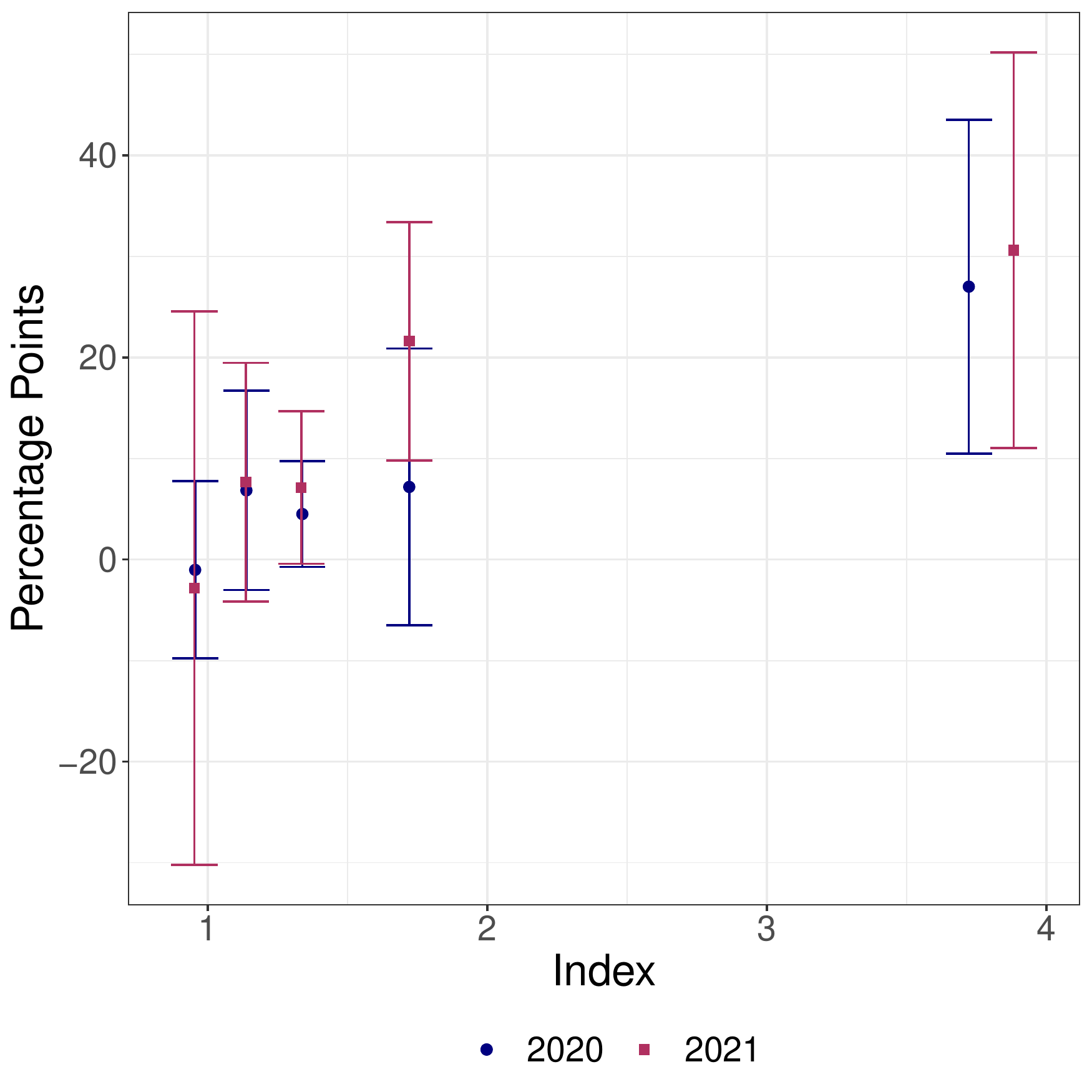}}
            \caption[Network2]%
            {{\small Markup rate}}    
            \label{fig:pik_markup}
        \end{subfigure}
        \hfill
        \begin{subfigure}[b]{0.475\textwidth}  
            \centering 
            \includegraphics[width=\textwidth,height=5cm,page=3]{Effects_year_yearly_pik_5bins_mean.pdf}
            \caption[]%
            {{\small Sales}}    
            \label{fig:pik_sales}
        \end{subfigure}
        \vskip\baselineskip
        \begin{subfigure}[b]{0.475\textwidth}   
            \centering 
            \includegraphics[width=\textwidth,height=5cm,page=8]{Effects_year_yearly_pik_5bins_mean.pdf}
            \caption[]%
            {{\small Years in stock exchange}}    
            \label{fig:pik_yearsPublic}
        \end{subfigure}
        \hfill
        \begin{subfigure}[b]{0.475\textwidth}   
            \centering 
            \includegraphics[width=\textwidth,height=5cm,page=4]{Effects_year_yearly_pik_5bins_mean.pdf}
            \caption[]%
            {{\small Profit rate}}    
            \label{fig:pik_profitRate}
        \end{subfigure}
        \vskip\baselineskip
        \begin{subfigure}[b]{0.475\textwidth}   
            \centering 
            \includegraphics[width=\textwidth,height=5.5cm,page=6]{Effects_year_yearly_pik_5bins_mean.pdf}
            \caption[]%
            {{\small Market share}}    
            \label{fig:pik_marketShare}
        \end{subfigure}
        \hfill
        \begin{subfigure}[b]{0.475\textwidth}   
            \centering 
            \includegraphics[width=\textwidth,height=5cm,page=7]{Effects_year_yearly_pik_5bins_mean.pdf}
            \caption[]%
            {{\small Employment}}    
            \label{fig:pik_employment}
        \end{subfigure} 
        \caption*{\footnotesize Note: Each plot shows the 5-bin binscatter of the pandemic effect on profit rates by the respective variable in each panel. $y$-axis: effect on profit rates as $Y_{it}-\widetilde{Y}_{it}$, where $Y_{it}$ is firm $i$'s profit rate at time period $t$, and $\widetilde{Y}_{i,t}$ is the firm's counterfactual profit rate built upon a Bayesian structural time series model with seasonal and local-level stochastic trend. $x$-axis: firm's $i$ average value of either (a) markup rates, (b) sales (logarithm of constant 2010 thousand US dollars), (c) years since publicly traded, (d) profit rates, (e) market shares, and (f) employment (thousands of employees) in 2015--2019. }
    \end{figure*}

\begin{figure*}[p]
        \caption{Heterogeneity Analysis of Effect of Pandemic in Profit Rate (Quarterly Specification)}
        \label{fig:het_ana_profit_rate_quarterly}        
        \centering
        \begin{subfigure}[b]{0.475\textwidth}
            \centering
            \resizebox{\textwidth}{!}{\includegraphics[width=\textwidth,height=5cm,page=1]{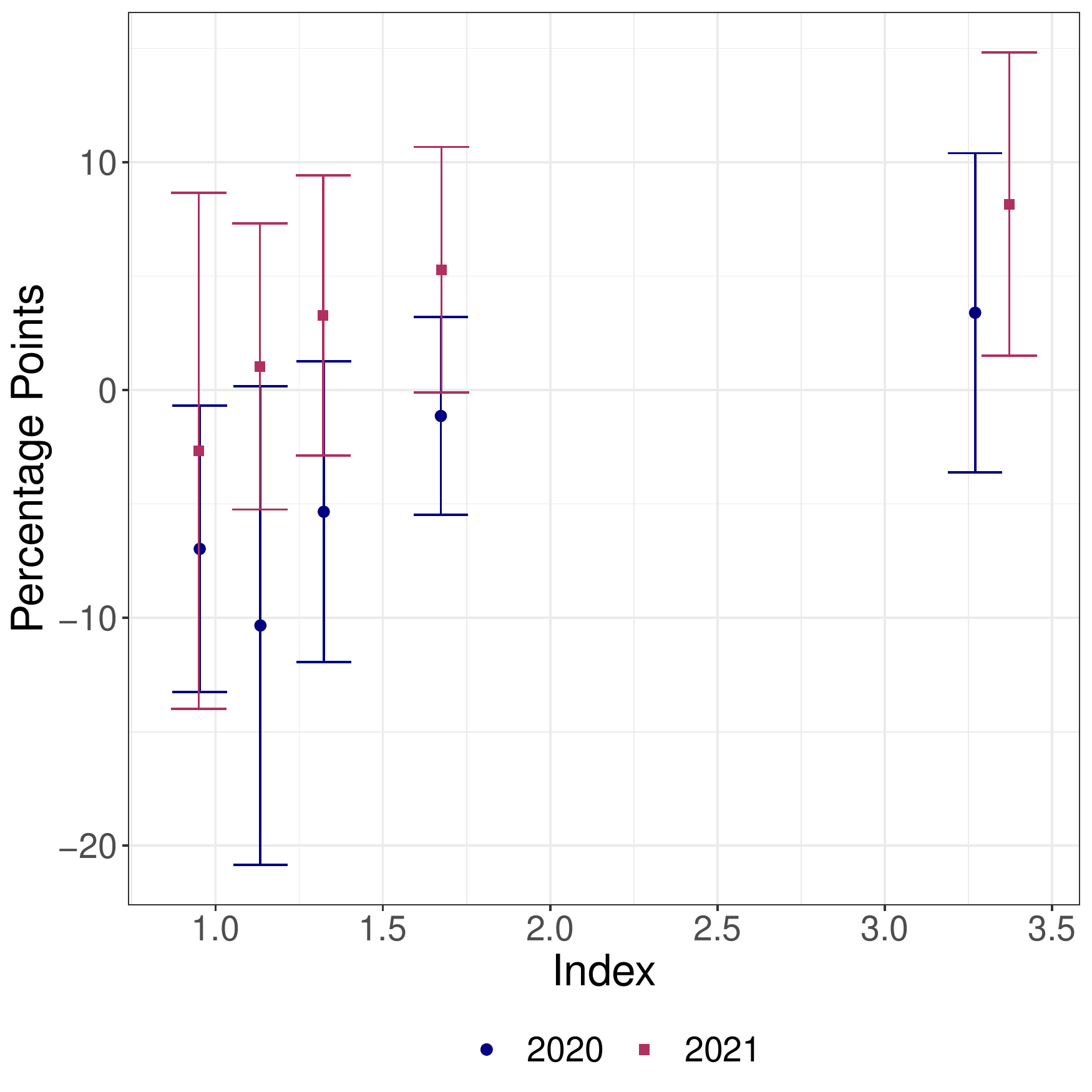}}
            \caption[Network2]%
            {{\small Markup rate}}    
            \label{fig:pik_markup_quarterly}
        \end{subfigure}
        \hfill
        \begin{subfigure}[b]{0.475\textwidth}  
            \centering 
            \includegraphics[width=\textwidth,height=5cm,page=3]{Effects_year_pik_5bins_mean.pdf}
            \caption[]%
            {{\small Sales}}    
            \label{fig:pik_sales_quarterly}
        \end{subfigure}
        \vskip\baselineskip
        \begin{subfigure}[b]{0.475\textwidth}   
            \centering 
            \includegraphics[width=\textwidth,height=5cm,page=8]{Effects_year_pik_5bins_mean.pdf}
            \caption[]%
            {{\small Years in stock exchange}}    
            \label{fig:pik_yearsPublic_quarterly}
        \end{subfigure}
        \hfill
        \begin{subfigure}[b]{0.475\textwidth}   
            \centering 
            \includegraphics[width=\textwidth,height=5cm,page=4]{Effects_year_pik_5bins_mean.pdf}
            \caption[]%
            {{\small Profit rate}}    
            \label{fig:pik_profitRate_quarterly}
        \end{subfigure}
        \vskip\baselineskip
        \begin{subfigure}[b]{0.475\textwidth}   
            \centering 
            \includegraphics[width=\textwidth,height=5cm,page=6]{Effects_year_pik_5bins_mean.pdf}
            \caption[]%
            {{\small Market share}}    
            \label{fig:pik_marketShare_quarterly}
        \end{subfigure}
        \hfill
        \begin{subfigure}[b]{0.475\textwidth}   
            \centering 
            \includegraphics[width=\textwidth,height=5cm,page=7]{Effects_year_pik_5bins_mean.pdf}
            \caption[]%
            {{\small Employment}}    
            \label{fig:pik_employment_quarterly}
        \end{subfigure}
        \caption*{\footnotesize Note: Each plot shows the 5-bin binscatter of the pandemic effect on profit rates by the respective variable in each panel. $y$-axis: effect on profit rates as $Y_{it}-\widetilde{Y}_{it}$, where $Y_{it}$ is firm $i$'s profit rate at time period $t$, and $\widetilde{Y}_{i,t}$ is the firm's counterfactual profit rate built upon a Bayesian structural time series model with seasonal and local-level stochastic trend. $x$-axis: firm's $i$ average value of either (a) markup rates, (b) sales (logarithm of constant 2010 thousand US dollars), (c) years since publicly traded, (d) profit rates, (e) market shares, and (f) employment (thousands of employees) in 2015--2019. }
    \end{figure*}

In addition to showing the heterogeneous impacts of the pandemic on firms' markup rates, we also explored heterogeneous effects on the profit rate. Figure \ref{fig:het_ana_profit_rate} shows such a heterogeneity analysis. In terms of heterogeneity for firms with different pre-pandemic markup rates, those firms with the highest markups tend to have positive and higher profit rate effects due to the pandemic (Panel \ref{fig:pik_markup}). Smaller firms tend to be less affected, yet the differences across quintiles appear to be not significant (Panel \ref{fig:pik_sales}). Firms that became public in the last half of the 2000s and in the early 1990s tend to have more drastic negative impacts (Panel \ref{fig:pik_yearsPublic}).  By profit rate, the bottom 20\% of firms in the distribution of profit rate have highly heterogeneous effects, whereas the most profitable firms in 2015--2019 tend to be better off during the pandemic (Panel \ref{fig:pik_profitRate}). Furthermore, firms with larger market shares tend to have larger negative effects on the profit rate (Panel \ref{fig:pik_marketShare}). Finally, the very large employers and the very small employers seem to have the worse profit-rate effects of the pandemic in 2020 (Panel \ref{fig:pik_employment}). When using the quarterly specification of the model, we find similar patterns in the effects on profit rates across  the 2015--2019 average levels of profit rate, market share, and employment (Figure \ref{fig:het_ana_profit_rate_quarterly}).

\end{document}